\documentclass[12pt,preprint]{aastex}

\slugcomment{Revised Version}

\shorttitle{SMBH Mass Predictors}
\shortauthors{Aller \& Richstone}

\begin{document}

\title{Host Galaxy Bulge Predictors of Supermassive Black Hole Mass}

\author{M.C. Aller and D.O. Richstone}
\affil{Department of Astronomy, University of Michigan, Ann Arbor, Michigan, 48109}
\email{maller@umich.edu}
\email{dor@umich.edu}

\begin{abstract}

A variety of host galaxy (bulge) parameters are examined in order to determine 
their predictive power in ascertaining the masses of the supermassive black holes (SMBH)
at the centers of the galaxies. Based on a sample of 23 nearby galaxies, comprised of both
elliptical galaxies and spiral/lenticular bulges, we identify a strong correlation 
between the bulge gravitational binding energy ($E_g$), as traced by the stellar light profile, and the SMBH
mass ($M_{\bullet}$), such that $M_{\bullet} \propto E_g^{0.6}$. The scatter about the 
relationship 
indicates that this is as strong a predictor of $M_{\bullet}$ as the velocity dispersion ($\sigma$),
for the elliptical galaxy subsample. Improved mass-to-light ratios, obtained with IFU spectroscopy and I-band photometry by 
the SAURON group, were used for those sample galaxies where
available, resulting in an energy predictor with the same slope, but with reduced scatter.
Alternative $M_{\bullet}$ predictors such as 
the gravitational potential and the bulge mass are also explored, but these are found to be inferior
when compared with both the bulge gravitational binding energy and bulge velocity dispersion predictors,
for the full galaxy sample.

\end{abstract}


\section{INTRODUCTION \label{Introduction}}

The presence of supermassive black holes (SMBH) ($10^6-10^{10} M_{\odot}$), in the centers of 
virtually all galaxy bulges, has become widely accepted in recent decades. With
the acceptance of the prevalence of SMBHs in the universe has come an effort to understand them:
 how they are formed, how they
evolve, and how they relate to their host galaxy. The combination of HST spectroscopic and imaging data of
the centers of nearby galaxies with ground-based imaging and spectroscopy at
large radii, has made accurate measurements of SMBH masses in nearby galaxies possible (e.g.
\citet{KormendyG,Kormendy03}) for a large enough sample to begin to understand 
the SMBH-host galaxy connection.

Given the close physical connections between the SMBH and the host galaxy bulge (where `bulge'
in Sections~\ref{Introduction} -~\ref{DetVals} refers to \textit{either} the hot, spheroidal component of a spiral/lenticular galaxy \textit{or} to a full
elliptical galaxy) it is possible to find host galaxy (bulge) characteristic parameters which may be used
to predict the SMBH mass. Previous studies have shown that the SMBH mass ($M_{\bullet}$) is well-predicted
by both the stellar bulge luminosity and by the associated stellar bulge mass; $M_{\bullet}$ is (roughly)
linearly proportional to both the bulge luminosity in the visible and NIR wavebands and to the bulge mass 
\citep{Dres89,Korm93,KR95,McLure,Marconi,Haring}, and
\citet{Graham} has shown that galaxies with more concentrated 
bulge light have larger SMBH masses.
 The SMBH mass is also well-predicted by the bulge stellar
velocity dispersion, $\sigma$, \citep{Geb,Merritt}, such that $M_{\bullet} \propto \sigma^{4.02}$ \citep{Tremaine}. 
Each of these predictors ($M_{bulge}$, $L_{bulge}$, and $\sigma$) is physically related to each other through the fundamental
plane and the virial theorem, and each is a
physical probe of the mass distribution of the bulge. 
Understanding which of these correlations is the most \textit{fundamental} would provide insight into SMBH formation and
evolution mechanisms.

Given the close physical relationship between
these predictors [$\sigma$, $M_{bulge}$, $L_{bulge}$], the uncertainties associated with both the SMBH masses and with the host galaxy parameters,
and the small ($\lesssim$ 40) number of nearby ($\lesssim$ 110 Mpc) SMBHs with dynamically measured masses available in the samples used to determine
these relationships, it is not unexpected that \textit{none} of the predictors stand out as being the clear \textit{best}, or
most \textit{fundamental}, predictor of SMBH mass. All of the predictors show a (statistically) similar scatter of SMBH mass about the
predictive relationship of $\approx$ 0.3 dex.
\citet{Novak} applied a rigorous statistical analysis to these galaxy property predictors from both the perspective of a \textit{Theorist}
(seeking the tightest correlation, i.e. the most fundamental predictor is that with the smallest residual variance)
and from the perspective of an \textit{Observer}
(seeking the galaxy property which can provide the best estimate of the SMBH mass). 
\citet{Novak} determined that neither of these questions could be fully answered with the current small sample sizes.
The probability distributions of the residual (intrinsic) variance (\textit{Theorist}) and the probability distribution 
of the uncertainties associated with the predicted SMBH masses (\textit{Observer}) both overlapped to such a degree that 
it was neither possible to state which host galaxy property has the most \textit{fundamental} connection with the SMBH mass,
nor to state which observational property would make the best prediction of the SMBH mass.
Given the limited number of bright, nearby galaxies, and
the demise of STIS, substantially increasing the sample to better address these questions, in the near future, is difficult.

In this paper, we approach the question of what is the \textit{best} predictor of the SMBH mass by examining two different questions: 
is there either a multivariate predictor or a previously unexplored host galaxy property which is better at predicting the SMBH mass
than those properties discussed in \citet{Novak}? 
\citet{Novak} found that bulge velocity dispersion, bulge mass, and bulge luminosity are all equally good at predicting
the SMBH mass. If we assume that bulge velocity dispersion is the \textit{best} predictor of SMBH mass, since it
is not dependent on bulge-disk decompositions for non-elliptical galaxies, would a multivariate fit combining $\sigma$ and a second, or third,
parameter be a better predictor of SMBH mass than $M_{\bullet}(\sigma)$, i.e. one with lower scatter? Second, \citet{Novak} compared bulge velocity dispersion,
bulge mass, bulge luminosity and the concentration index \citep{Graham}, but is there a different, physical host-galaxy property which
is a \textit{better} (or given the sample size statistically equivalent) predictor of SMBH mass? 
The motivation underpinning these questions is to gain a better understanding of the host galaxy-SMBH symbiosis (a \textit{Theorist's}
motivation), but the 
approach employed is 
that of an \textit{Observer}; the \textit{best} predictor is determined based on the relative predictive strengths as determined
by comparing the residual \textit{scatter} in the fit, not by comparing the intrinsic uncertainties.
(However, while we approach this question from an \textit{Observer's} framework, some of the galaxy properties explored, such as gravitational
binding energy, are not properties which can be directly ``observed''; geometrical assumptions and modeling are required.)
For all of these calculations we use a single set of data and modeling assumptions which allows
for direct comparisons of the relative predictive strengths.

The paper is divided into four main sections: a discussion of the data (Section~\ref{Data}),
a description of the modeling employed to determine the host galaxies properties such as gravitational
binding energy (Section~\ref{DetVals}),
an exploration of SMBH mass predictors (Section~\ref{Predictors}) and a summary and concluding
discussion (Section~\ref{summary}). 
In this final section, we assess the relative merits of
the predictive fits, and an argument is made for why the most fundamental predictor of SMBH mass, for elliptical galaxies, is the gravitational binding energy
($M_{\bullet} \propto E_g^{0.6}$), although the bulge velocity dispersion (without additional parameters) may be more easily implemented, particularly in the
case of spiral and lenticular galaxies. Additionally, there are two appendices: the first, Appendix~\ref{Ap-A}, explores the selection of the
host galaxy parameters used in Section~\ref{HostGal}, and the second, Appendix~\ref{Ap-B}, explores the need, or lack thereof, for multivariate or log-quadratic
predictive relationships for the velocity dispersion, gravitational binding energy, gravitational potential and bulge mass predictors.

\section{DATA\label{Data}}

In order to probe the connection between the host galaxy bulge and the SMBH mass, bulge properties
(gravitational potential, gravitational binding energy, and bulge mass) are determined for a sample of nearby galaxies.
The calculations use data taken from the published literature combined with geometrical assumptions and standard mathematical
formulae, as described in Section~\ref{DetVals}.
The initial sample consists of the
30 external galaxies in \citet{Tremaine}, a mix of spirals, lenticulars (S0s) and ellipticals, which have black hole
masses determined either by stellar dynamics or by gas/maser kinematics.

\subsection{Surface Brightness Profile\label{SB}}

The primary data required to calculate the stellar bulge mass, gravitational potential and gravitational binding 
energy are the galaxy major-axis stellar surface brightness (SB) profiles from which the stellar mass profiles are generated. This
is implemented by calculating the deprojected luminosity density from the SB profile, using the 
methodology (and code) from \citet{Gebhardt-dep}, and then scaling the luminosity density profile by a
mass-to-light ratio to produce the major-axis mass profile.
The ideal galaxy SB profile for these calculations would consist of a combination of 
HST photometry at small radii and ground-based photometry at large radii,
well-matched in the region of overlap, and covering the full extent of the major axis, in the same wavelength band as
the $\Upsilon_{bpC}$ mass-to-light ratio. For spiral and lenticular galaxies there is the added requirement of a well-constrained
bulge:disk decomposition. 

\begin{figure}
\plotone{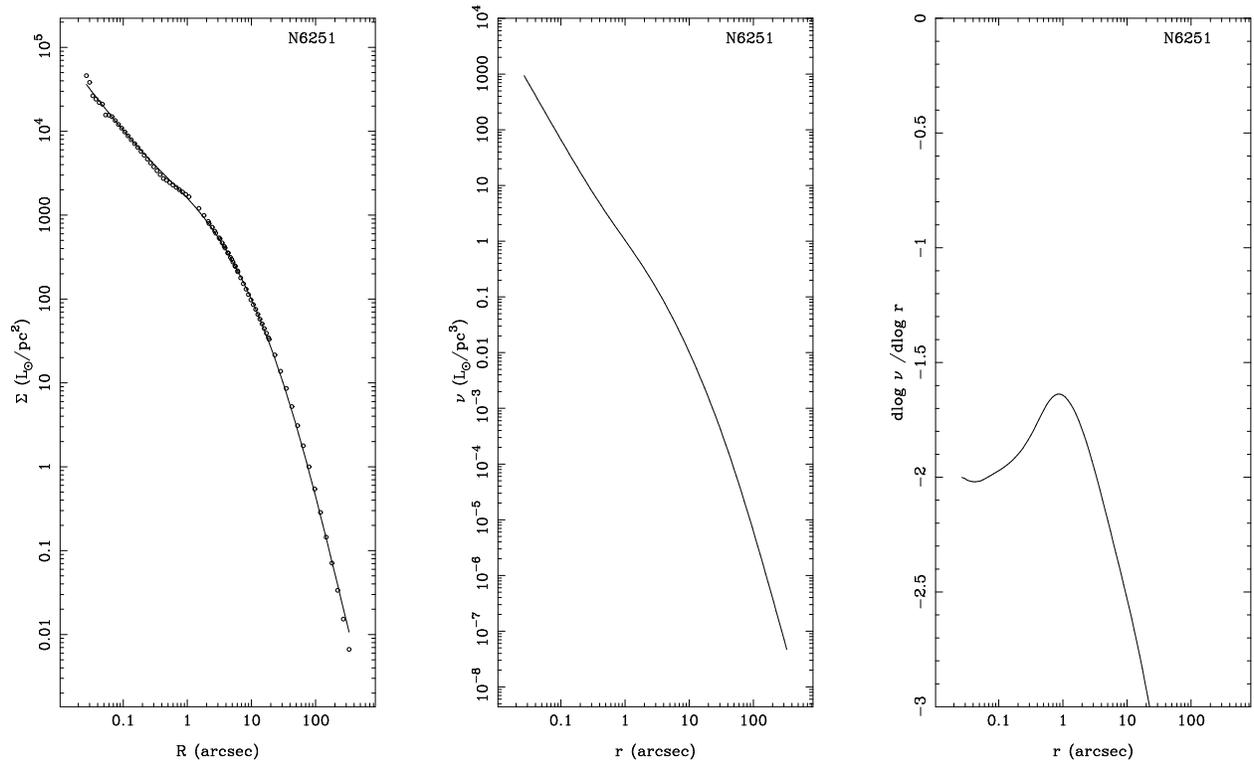}
\caption{
As a function of
radius, from left
to right, the surface brightness, the deprojected luminosity density,
and $d\log{\nu}/d\log{r}$ in the V-band along the major axis for the elliptical galaxy NGC 6251.
\label{Fig1}}
\end{figure}

The surface brightness profiles used here, given in Table~\ref{Tab1}, are a combination of data from different
literature sources, obtained using different instrumentation (both different telescopes and different detectors), different wavelength-bands,
and different reduction techniques and corrections, along different axes; 
these have been combined and homogenized here to produce a single major-axis profile in the desired
wavelength-band. 
In order to combine the data, the most compatible radial subsets of the full SB profile were selected from each literature source,
with visually discrepant points discarded, and with no new corrections (K-corrections, local and/or external galaxy
extinction, seeing effects and/or deconvolution) applied or removed. For elliptical galaxies for which both data points and a fit (Nuker-law,
S\'{e}rsic, de Vaucouleurs, etc.) were available, the data points were selected in preference to the fit; the fit is the result
of a mathematical approximation to the real data structure, and although it results in a smoother SB profile, if the discontinuities
are the result of physical structural components,
they will only be imperfectly accounted for in the fit. For the spiral and lenticular galaxies, 
the fit describing the bulge light, as obtained from a literature-based bulge-disk decomposition (denoted as Fd in Table~\ref{Tab1}, with the corresponding
reference to the decomposition), is used instead of the observed combined-light profile.
All of the decompositions assumed a de Vaucouleurs profile, except for NGC 4342, which utilized a combination of an exponential nuclear and outer disk and a
Nuker-law fit to the bulge along the minor axis.
Galaxy light was considered to come solely from a bulge or a disk; light originating 
from bars or other features were included within these two components.
An example combined SB profile, along with the deprojected luminosity density and its derivative (which indicates discontinuities in the profile), all
as a function of radius, is shown for the elliptical galaxy NGC 6251 in Figure~\ref{Fig1}; this illustrates the relatively smooth connection between data from
different origins.

There were two corrections applied to the data: a correction to 
the major axis, and a correction to a common wavelength-band.  
The first correction was to rescale the SB profile to the major axis. Published SB profiles along both the galaxy major axis
and along the isophotal major axis were taken to be equivalent, although this may not be strictly true in boxy galaxies or in 
galaxies with a strong isophotal twist, and no correction was applied to these data. For SB profiles given as a function of
geometric mean radius ($r=\sqrt{ac}$), the
corresponding major axis radii, for each point in the SB profile, are described as $r ({1-\epsilon})^{-0.5}$. 
The ellipticities were taken as a function of radius (denoted by the symbol P in Table~\ref{Tab1}) when available; otherwise
the ellipticity value at the nearest radial point in the profile, or an average value for the entire profile, was adopted.

The second correction was a conversion of the SB profile segments into a common
wavelength-band, the wavelength-band of the $\Upsilon_{bpC}$ mass-to-light ratio, prior to combining the segments. Given that optical filters can
differ from telescope-to-telescope, the published color terms resulting in the smoothest combined SB profiles were selected, 
in two cases (see Table~\ref{Tab1}) requiring additional offsets to match the individual SB profile segments more smoothly. For direct comparisons of $I_e$, the V-band was selected
as the default, and, for each galaxy, if necessary, a second (V-band) profile was generated.
The color terms used are given in Table~\ref{Tab1}. When available, color terms as a function of radius (denoted by P in the table) were
used. When this information was unavailable, color terms as a function of aperture (A in the table) were used, or an assumption of
constant color for the entire galaxy was made; most galaxies showed only a small (if any) gradient in color. For lenticular galaxies, if no separate color
terms for the bulge and disk could be located, a single value was used; in galaxies with individual bulge and disk colors (e.g. NGC 3245),
the colors in the bulge and disk were similar.

\subsection{Host Galaxy Parameters \label{HostGal}}

In addition to the SB profile, the calculation of the bulge mass, gravitational potential and gravitational binding 
energy requires the distance to the galaxy (D), the galaxy bulge mass-to-light ratio ($\Upsilon$),
the bulge effective radius ($R_e$), the galaxy inclination ($\theta$), and the bulge apparent axis ratio (q). The SMBH mass ($M_{\bullet}$)
is also required for each galaxy in order to generate the $M_{\bullet}$-predictive relationships in Section~\ref{Predictors}. 
The values
for each of these parameters are given in Table~\ref{Tab2}. Additionally,
the effective bulge velocity dispersion ($\sigma$)
is required so that $M_{\bullet}(\sigma)$ can be determined
for each galaxy subsample and used for comparative purposes in Sections~\ref{Predictors} and~\ref{summary}; the velocity dispersion values are
enumerated in Table~\ref{Tab3}, along with the computed host galaxy properties.
The distance for each of the galaxies is taken directly from \citet{Tremaine}. \citet{Tremaine} used
SBF distances from \citet{Tonry}, when available; otherwise the distance was determined from the recession velocity assuming a Hubble constant
of 80 $km s^{-1} Mpc^{-1}$. All parameters, including the SMBH mass and mass-to-light ratios, are scaled to the distance (and, 
implicitly, use this Hubble constant). The SMBH masses are taken from the individual black hole modeling papers, along with accompanying
estimates of the uncertainties.
Generally the distance-rescaled value agrees with \citet{Tremaine}; in the few
galaxies where this is not the case, a notation is made in Table~\ref{Tab2}.
The effective bulge stellar velocity dispersions ($\sigma$, defined as the  
rms dispersion within a 2$R_e$ slit aperture) are taken from 
\citet{Tremaine} for all galaxies except NGC 4258, which was taken from \citet{Siopis}.

The dynamical mass-to-light ratio ($\Upsilon$) 
is used to convert the stellar luminosity 
density into a stellar mass density. 
The strong dependence of the computed mass, gravitational potential and gravitational binding energy on the mass-to-light ratio makes it crucial to have the most
accurate $\Upsilon$-value possible. The implications of selecting an ``incorrect'' value, and the range of values
quoted in the literature, are explored in Appendix~\ref{Ap-A}.
The \textit{best} mass-to-light ratio values, available in the literature, are those determined by \citet{Cappellari-SAURON} using 
extinction-corrected NIR photometry and integral-field-spectroscopy.
These are an improvement on other mass-to-light ratios both because of the improved spectroscopy, and because
observations at redder wavebands minimize the impact of dust;
the mass-to-light ratios determined at blue wavebands are probing the star formation history as well as the mass.
However, these superior mass-to-light ratios are only available for one-third of the total galaxy sample; therefore, for this subset of the
sample galaxies, two mass-to-light ratio values are enumerated.
For \textit{every} sample galaxy, the best mass-to-light ratio available in the literature, excluding the value in \citet{Cappellari-SAURON},
is determined and denoted as the best-pre-Cappellari value, $\Upsilon_{bpC}$ (Table~\ref{Tab2}: column 6). 
For galaxies in which $\Upsilon$ was determined in the process of measuring
$M_{\bullet}$ (using well-determined spectroscopy and photometry from both HST and ground-based observatories), this value
was adopted as $\Upsilon_{bpC}$. For galaxies in which $\Upsilon$ was not provided along with $M_{\bullet}$,
$\Upsilon_{bpC}$ was taken from \citet{Tremaine}, if available, and otherwise
from another literature source (Table~\ref{Tab2}: column 7). 
The wavelength band of $\Upsilon_{bpC}$ is taken as the default for the galaxy;
the SB profiles were converted to this wavelength-band prior to calculating the luminosity density profile.
If the mass-to-light ratio for the galaxy was provided in \citet{Cappellari-SAURON} then this $\Upsilon$, rescaled
to the $\Upsilon_{bpC}$-default-wavelength band, is given as $\Upsilon_{Cap}$ in Table~\ref{Tab2} (column 8). The
\textit{best} mass-to-light ratio, $\Upsilon_{best} \equiv \Upsilon$, used in the galaxy modeling, is taken to be $\Upsilon_{Cap}$ if it exists; otherwise, $\Upsilon_{bpC}$
is utilized. As discussed  further in Section~\ref{Predictors}, mass-dependent predictors of $M_{\bullet}$ have lower residual scatter when using the $\Upsilon_{Cap}$-values 
than when using the $\Upsilon_{bpC}$-values, for galaxies in which both are available. 
It is assumed throughout that the mass-to-light ratio is constant with radius ($\Upsilon(R) = $ constant). This assumption
was also made by investigators when determining $M_{\bullet}$; authors who investigated a radial variation
in selected sample galaxies generally did not find a large variation in mass-to-light ratio.
For example, in IC 1459 \citet{IC1459} estimate a decrease in $\Upsilon$ of 15\% per decade in radius, and in NGC 3379,
\citet{G3379} estimate $\Upsilon$ may be 25\% higher in the innermost regions than in the outermost regions of the galaxy.  

The observed bulge axis ratio (q) and bulge inclination ($\theta$), in combination with several simplifying assumptions (Section~\ref{DetVals}), are
used to geometrically project the major-axis mass profile (Section~\ref{SB}) into a 3-dimensional
mass profile, assuming the bulge is an oblate spheroid.
The inclination and axis ratio are also used to correct the observed SB profile and luminosity (and
mass) density profiles to an edge-on orientation using a multiplicative ratio of the observed (q)
and intrinsic, edge-on, (p) axis ratios; 
[$(p\sin\theta)^2 = {q^2-(\cos\theta)^2}$
e.g. \citet{Richstone84,Gebhardt-dep}].
The predictive fitting functions are based on values determined from the edge-on galaxy
orientation, although the results are qualitatively unchanged if this correction is omitted. 
As in the case of the mass-to-light ratio, there is often a range of q- and $\theta$-values present in the literature;
the range of these values and the implications of selecting an ``incorrect" value for q or $\theta$ are discussed in 
Appendix~\ref{Ap-A}.

The observed axis ratios, culled from the literature, 
are assumed to be constant as a function of radius for all galaxy bulge radii.
If multiple values for the axis ratio were
quoted in the literature, the most prevalent literature-value, or an average of the literature values, as given
in Table~\ref{Tab2}: columns 10-11, was selected as the \textit{best} value.
For the elliptical galaxies, the \textit{best} axis ratios are taken from the literature either in the form of
a single quoted value representative of the galaxy as a whole, or derived from the 
radially-dependent ellipticities associated with the SB profiles. 
For the spiral/lenticular bulges the
published literature single-values are not used;
such single-values are based on a combination of both the bulge and the disk light profile, and so are flatter than for
the bulge-component alone. 
Instead, the bulge axis ratios are based on the 
SB radial profile ellipticities at representative bulge radii. 

The bulge inclinations, for all galaxy morphologies, are assumed to be constant as a function of radius (i.e. no warping) and, in 
the case of spiral/lenticular galaxies, to have the same inclination as the galaxy disks.
The inclinations used in the determination of the $M_{\bullet}$ values are adopted, when given.
If such a value was unavailable, an inclination was taken from the literature, and if this, too, was unavailable, the inclination
was determined from the observed axis ratio. 
Inclinations are generally better-determined for the spiral/lenticular galaxies, for which the inclinations can
be measured geometrically from the observed disk axis ratio, than for the disk-less elliptical galaxies.
For several of the galaxies, the $M_{\bullet}$-determination modeling implies, or assumes, an edge-on inclination for the galaxy,
while independent structural analyses of the galaxies imply a less-than edge-on inclination. For example, \citet{N1023} find that an inclination
of {90\degr} results in the best-fitting three-integral model, used to determine the SMBH mass, for NGC 1023, while \citet{Debattista} find, in an
examination of the bar of NGC 1023, that the best inclination is 66\degr.4$\pm$1\degr.2. Likewise, \citet{Busarello} claim a less-than edge-on inclination
for NGC 3384, following their structural analysis of that galaxy. These inclination discrepancies, however, do not affect our conclusions, as illustrated in Appendix~\ref{Ap-A}.

The effective (half-light) radius of the bulge ($R_e$) 
is not directly used in the calculations; it is used as a fiducial reference point for 
the gravitational potential and other radially-dependent functions. As 
with the previous bulge parameters, there were often multiple values for $R_e$ in the literature. However,
an incorrect value will not affect the intrinsic mass profile or the computed 
gravitational binding energy; it will only affect predictors which are evaluated at $nR_e$, and so have
a minimal impact on the conclusions. For the spiral/lenticular bulges the bulge effective radius was taken from the 
bulge-disk decomposition literature source; the exception is NGC 4258 which was obtained 
directly from C. Siopis (private communication). For the 10 elliptical galaxies in \citet{Gebhardt-12}, the value was taken directly from 
that paper. For the remaining elliptical galaxies, the value from the literature, or from an unweighted
$r^{1/4}$ law fit to the SB profile, which resulted in the least scatter in a fundamental plane relationship
($\log{R_e}=a\log{I_e}+b\log{\sigma}+c$)  fit to the full sample of elliptical galaxies was chosen as the \textit{best}.
For most galaxies it was assumed that the effective radius is not a function of color, i.e. that
the color is constant as a function of radius.

\section{Determination of the Stellar Mass, Gravitational Potential, and Binding Energy \label{DetVals}}

In order to determine the bulge mass, gravitational potential and gravitational binding energy,
the input data (see Section~\ref{Data}) are combined
with well-constrained mathematical functions and several simplifying assumptions about the host galaxies.
The first assumption is that the galaxies
are smooth and featureless; however, while this may be the case for NGC 221, it is certainly
not the case for the rest of the galaxies which include, among other features,
nuclear star clusters, AGNs and cores, ionized gas disks and shells, dust
arranged in clumps, filaments, rings and disks, outer and inner disks which are sometimes
warped, and bars and jets. (For details on the individual galaxies, see
the BH modeling and SB papers previously referenced.) 
A second assumption is that not only is the galaxy
uniform, but that it is an axisymmetric oblate spheroid with constant mass density along concentric,
isodensity, oblate spheroidal shells with no radial variation in ellipticity;
if galaxies are triaxial or have isophotal/isodensity twists with radial position, this will be unaccounted for in these calculations.
Third, it is assumed
that there is no contribution from inner or outer disks, bars or the
dark halo. The binding energy (as well as the mass and potential) is considered to come solely from the 
mass associated with the visible stellar light; the binding energy could be much greater if the galaxy is
immersed in a massive dark halo. While ignoring the dark halo, which can only be inferred, rather than directly measured,
will have an impact
on all galaxy types, ignoring the presence of visible disks in the spiral and lenticular galaxies
may have a substantial effect on these galaxies, weakening the predictive power of the calculated
parameters relative to an elliptical galaxy sample.

The computation of the bulge mass, gravitational potential, and gravitational binding energy is implemented
using well-characterized mathematical expressions under the assumption that the bulge is a smooth, featureless,
oblate spheroid. For an oblate spheroid ($a=b>c$) with axes a, b and c, the radii along
the major axis are projected to any arbitrary radii using the relationship that
$r(\nu) = {r_{maj} ({{1+k_0^2\nu^2}}})^{-0.5}$ where $k_0 = \sqrt{({a/c})^2 - 1}$ and $\nu = \cos\theta$.
(In this section, r, $\theta$, and $\phi$ refer to standard spherical geometry coordinates.)
This allows for the projection of the major-axis mass density profile (see Section~\ref{SB}) to any arbitrary axis: 
$\rho({r,\nu}) = \rho({r(\nu),0})$.
With this relation, the mass enclosed within a radius r can be calculated as
\begin{equation}\label{E1}
M=\int_{\phi}\int_{\nu}\int_r{\rho(r,\nu)r^2drd\nu d\phi}.
\end{equation}
The calculation of the gravitational potential, $\Phi$, is derived from equation 2-122 of \citet{BinneyT},
under the assumption that the density does not depend on $\phi$, to be
\begin{equation}\label{E2}
\Phi(r,\nu)=-2{\pi}G {\sum_l}{{{                 
  P_{l}(\cos\nu)}
[r^{-(l+1)}A_l + r^l B_l]}}
\end{equation}
where
\begin{displaymath}
A_l = 
{{\int_0}^r}a^l {\int_{-1}^1  P_{l}(\nu) \rho(a,\nu)
 d\nu} a^2 da 
\end{displaymath}
\begin{displaymath}
B_l = {{\int_r}^\infty} a^{-(l+1)}
 {\int_{-1}^1  P_{l}(\nu) \rho(a,\nu)
 d\nu} a^2 da.
\end{displaymath}
Using this relation for potential, the binding energy can be calculated, as in equation 2-19 of 
\citet{BinneyT} as
\begin{equation}\label{E3}
E={1\over2} \int_{\phi}\int_{\nu}\int_r{\rho(r,\nu)\Phi(r,\nu)r^2drd\nu d\phi}.
\end{equation}

Our computational code utilizes multivariable, Gaussian quadrature integration techniques 
(\citet{Press} routine \textit{qgaus}) combined with simplifying assumptions about the radial extent 
and structure of the mass-density profiles. 
First, the mass profile (Section~\ref{SB}) is specified at discretely sampled
points between arbitrary limits. It is assumed that the profile is intrinsically smooth, and the space
between points is interpolated using a cubic spline algorithm: \textit{spline} from \citet{Press}.
Second, the mathematical expressions (equations~\ref{E1} -~\ref{E3}) integrate over the spatial variable (r) from zero to infinity.
It is assumed that the galaxy mass physically extends only from an innermost radius ($R_{min} \neq 0$) to an outer limit ($R_{max} \neq \infty$), and outside
of these limits the mass density is taken to be zero.  
Third, the value of $R_{min}$ from the observationally-based mass profile reflects the resolution limits of the observation, not the inner
cutoff of the galaxy mass. The mass profile was extrapolated inward to 3 Schwarzschild radii,
using an $r-\log{\rho}$ unweighted quadratic fit to the innermost 10 points of
the mass profile
using a least squares fitting algorithm (\textit{lsqfit}, discussed in Section~\ref{Predictors}).
The computed energy for fitting-sample galaxies was unchanged if the extrapolation is repeated using an assumption of constant mass or if the
innermost limit is varied to other physically reasonable values such as 0.01 pc.
Fourth, the value of $R_{max}$ likewise reflects the observational limitations rather than the physical galaxy edge. The mass
profile was extrapolated outward (to an extreme limit of $50R_e$) using a $\log{r}-\log{\rho}$ unweighted quadratic fit to the outermost 10 points of the mass profile
using \textit{lsqfit}. 
The computed energy for fitting-sample galaxies was unchanged both if the outermost limit is varied to any value beyond
a few $R_e$ and if it is assumed that the observational cutoff to the mass profiles corresponds to the physical edge of the mass profile. 
Finally, the integration only goes out to a Legendre polynomial of order 4. 
Tests using models with homogeneous spheres, Plummer density profiles,
and Satoh density profiles indicate that these selections provide sufficient spatial coverage and that the expected potentials are
produced.
The final calculated values for the bulge stellar mass enclosed by radii R, the stellar gravitational potential evaluated
at radii R, and the stellar-based gravitational binding energy, for all
30 galaxies in the original sample, are given in
Table~\ref{Tab3}.

\section{PREDICTORS OF BLACK HOLE MASS \label{Predictors}}

Utilizing the host galaxy parameters described above, predictors
of the SMBH mass, primarily in the form of power laws, were determined using a variety
of algorithms. The host galaxy-SMBH mass predictors are of the 
form
\begin{equation}\label{E4}
Y = \sum_i{a_iX_i}+ d 
\end{equation}
where $x_i$ are the host galaxy parameters, $Y \equiv \log{M_{\bullet}}$, and $X_i \equiv \log{x_i}$. The host galaxy parameters
were each normalized by a value near the mean (in log-base-10 space) 
of the 30 galaxy sample (Table~\ref{Tab3}). This normalization removes the covariance between the zero-point (d) and the slope,
as discussed
in \citet{Tremaine}. The primary fitting algorithm is \textit{lsqfit} (based on formulae/methods in \citet{Bevington}); this program 
calculates the weighted-least-squares-minimized fit to a multivariable equation of the form of equation~\ref{E4}, weighting
points only by the inverse-square of the uncertainties in Y associated with each data point, not by the uncertainties in 
$X_i$. In addition
to the coefficients ($a_i$, d), the algorithm provides coefficient uncertainties $(\delta{a_i},\delta{d})$ 
using standard least-squares-fitting
formulae, e.g. \citet{Press}. 
The reduced chi-squared for this function is determined such that
\begin{equation}\label{E5}
\chi_r^2 = {1 \over DOF}{\left({\sum_{i=1}^N{{\epsilon_{i}^{-2}}}}{[Y_i-({\sum_j{a_jX_{ji}}}+d)]^2} \right)}
\end{equation}
where DOF is the number of degrees of freedom and $\epsilon_i^2 \equiv \epsilon_{yi}^2$ ($\epsilon_y$ is the total uncertainty in $\log{M_{\bullet}}$).
The second fitting algorithm is \textit{fitexy} \citep{Press}; this only allows for the fitting, via
least-squares-minimization, of one parameter ($i \equiv 1$, equation~\ref{E4}), but, unlike \textit{lsqfit}, it includes
uncertainties in both Y and $X_1$. 
The reduced chi-squared for this function is determined by equation~\ref{E5} where $j \equiv 1$ and 
$\epsilon_i^2 = \epsilon_{yi}^2+a^2\epsilon_{xi}^2$ ($\epsilon_x$ is the uncertainty in the independent variable $X_1$).
\citet{Novak} found this algorithm for determining SMBH-predictive relationships
to be the best, i.e. the most efficient and least biased among a set of seven algorithms explored. The third routine, \textit{medfit} \citep{Press}, henceforth \textit{robust},
also fits to only one parameter, but utilizes absolute-deviation-minimization, and does not
use the uncertainties in either Y or $X_1$.

The measured uncertainties on $M_{\bullet}$ are not symmetric; therefore, the calculated fits are dependent on the method
of symmetrization employed. The default procedure (referred to as \textit{Avgerr}) is to average the uncertainties such that
$\epsilon_{yo} = {0.5}[{{\epsilon_{yo_{high}} + \epsilon_{yo_{low}}}}]$, where $\epsilon_{yo}$ is the observationally-based
uncertainty. The second method (\textit{Recent}) involves
recentering $M_{\bullet}$ between the upper and lower observational limits, such that
$Y={0.5}[{{(Y+\epsilon_{yo_{high}}) + (Y-\epsilon_{yo_{low}})}}]$. 
For all fits, the \textit{Avgerr} and \textit{Recent} fits are equivalent, unless specifically noted in the text or tables.

Additionally, although the fits minimize $\chi^2$, even with the treatment of error in both variables, the resulting $\chi_r^2$ exceeds 1.0 (see
equation~\ref{E5}). Following \citet{Tremaine}, we assume that there is an additional variance ($\epsilon_{yin}$)
due to cosmic scatter and errors in the independent variable, and account for it by adjusting the error in $Y=\log{M_{\bullet}}$ to
$\epsilon_y^2 = \epsilon_{yo}^2+\epsilon_{yin}^2$.
The intrinsic uncertainty is selected in order to always obtain $\chi_r^2 = 1.0$.
The predictive fit is then calculated using three weighting methods. In the first (\textit{OBS}), it is assumed that
there is no intrinsic error ($\epsilon_y=\epsilon_{yo}$); thus $\chi_r^2 \neq 1.0$. In the second method (\textit{INT}),
the observationally measured uncertainties are ignored ($\epsilon_y=\epsilon_{yin}$). In the final, default, method (\textit{OBS+INT}),
it is assumed that the total uncertainty is a combination of the observed and intrinsic uncertainties, as given above. 
For all fits, the three methods are equivalent, unless specifically noted in the text or tables.
Throughout, the term \textit{weightcent} will be used to refer to the selection of the combination of Y-centering (\textit{Avgerr} or \textit{Recent})
and Y-weighting (\textit{OBS}, \textit{INT} or \textit{OBS+INT}).

The final weighting option
is in the adopted uncertainty in $X_1$ ($\epsilon_x$) for the \textit{fitexy} algorithm. For the bulge velocity dispersion this is set to the value 
adopted in \citet{Tremaine}: $\epsilon_x=0.021$ (5\%).
For the gravitational binding energy, gravitational potential, and bulge mass, values of $\epsilon_x=0.1$ (26\%)
and $\epsilon_x=0.3$ (100\%) are explored to see if the fits are substantially different from the default case, $\epsilon_x=0.0$.

Throughout the paper the terms \textit{strength}, \textit{stability}, and \textit{best} are used to describe the fits. 
The \textit{strength} refers to the amount
of \textit{scatter} ($\sigma_{fy}$) about the predictive relationship,
where 
\begin{equation} \label{E6} 
\sigma_{fy}^2=\left({{\sum_i{[{Y_i-({\sum_j{a_jX_{ji}}}+d)}]}^2}} \right) {DOF}^{-1}.
\end{equation}
Less \textit{scatter} implies a \textit{stronger}
fit or predictor.
A \textit{stable} slope/fit is defined to be one for which the variations in the slope/fit, with changes in \textit{weightcent}, 
fitting algorithm and/or morphological selection, are statistically insignificant. The 
\textit{strongest} and the most \textit{stable} predictor is referred to as being the \textit{best} predictor.

The \textit{stability} of the fits is further examined for each sample using a bootstrap calculation.
A bootstrap calculation, e.g. \citet{Press}, randomly
draws N points, with replacement, from an initial set of N data points, resulting in a `new' data sample with $\approx$ 63\% unique data points.
This
method indicates whether the fit based on the original sample is being unduly influenced by the presence of specific galaxies, and provides
insight into the underlying distribution of the best-fit coefficients. These tests are only intended as a consistency check on
the slopes and uncertainties based on the full, original sample of N data points, and to further illustrate the robustness of
the results.
For each galaxy sample, the
bootstrap is run for 1000 `new' samples using both the default \textit{weightcent}, \textit{lsqfit} fitting
algorithm and the \textit{robust} algorithm. The value of $\epsilon_{yin}$ for the \textit{lsqfit} fitting is selected to produce $\chi_r^2=1.0$, assuming that only unique data points in
the `new' sample count towards the DOF (counting duplicated galaxies in the DOF
as individual galaxies had a minimal impact on the resulting slope except for the \textit{8B} sample.)
The mean, median, average deviation, standard deviation, skew and kurtosis of the distribution of values
are determined using the
\textit{selip} and \textit{moment} algorithms from \citet{Press}, for the slope (a), zero-point (d), \textit{scatter} ($\sigma_{fy}$)
and required intrinsic uncertainty in $Y=\log{M_{\bullet}}$ ($\epsilon_{yin}$).

The \textit{scatter} ($\sigma_{fy}$)
and the Snedecor F-test \citep{Snedecor,Dixon} are
used to statistically assess the relative \textit{goodness-of-fit} for the predictive relationships (for a given sample).
Using $\chi^2$ as a comparison
is not feasible since $\chi_r^2 \equiv 1$ for the least square fitting.
The F-test ratio \citep{Snedecor,Dixon}, 
\begin{equation}\label{E7}
F_{\sigma_y} = \left({{\sigma_{fy_{\sigma}}}\over{\sigma_{fy_{fit}}}}\right)^2 
\end{equation}
[where $F_{\sigma_y} > 1.0$ implies a better fit than $M_{\bullet}(\sigma)$],
 compares the square
of the residual \textit{scatter} of points around each of the two predictive relationships and determines whether the difference in \textit{scatter} is significant
given the number of degrees of freedom (DOF) in each of the relationships. 
This determination is made using the published tables in \citet{Dixon} and results are abbreviated as follows:
NSS-the fits are not different at the 75\% level (approximately
1-$\sigma$), 75SS - the fits are different at the 75\%-90\% level, 90SS - the fits are different
at the 90\%-95\% level, 95SS - the fits are different at the 95\%-97.5\% level and 99SS - the
fits are different at greater than the 99\% level.
(For ease in using published tables, assumptions that 13 DOF$\sim$12 DOF (\textit{15E}) and 21 DOF$\sim$20 DOF$\sim$19 DOF (\textit{23gal})
were made; these assumptions do not change the stated results.)
Throughout,
a mention of `minimal improvement' implies that the \textit{scatter} is lower than $M_{\bullet}(\sigma)$, but not at the 75\% probability level.
The abbreviations and symbols defined in this section are summarized in Table~\ref{Tab4}.

The remainder of this section will be structured as follows. First, the selection of galaxies used in the four fitting samples
will be explained, along with a caveat related to using the spiral/lenticular galaxies for fitting (Section~\ref{sample}).
This will be followed by the fits for the bulge velocity dispersion predictor (Section~\ref{sigma}), the multivariate
fits constructed from a combination of $\sigma$, $I_e$ and $R_e$ (Section ~\ref{sigma-other}), the gravitational binding energy
predictor (Section~\ref{energy}), the gravitational potential predictor (Section ~\ref{potential}) and the 
bulge mass predictor (Section~\ref{mass}). Each of the single-variable predictors (Sections~\ref{sigma},~\ref{energy} -~\ref{mass}), 
will begin with a discussion of results from the literature (if available) and the motivation/methodology for fitting this galaxy property. This will
be followed by the fits for each of the samples and any notable fitting caveats, along with a discussion of the \textit{stability} of the fit in terms of
the galaxy morphology,
the fitting algorithm, and the \textit{weightcent} selection, and in terms of the bootstrap results. 
Finally, plots of the fits and their residuals will be examined, and comments on the necessity, or lack thereof, 
for either a multivariate fit or a log-quadratic fit will be made. Additionally, for all but the $M_{\bullet}(\sigma)$ predictor, the
\textit{strength} of the fit compared with the $M_{\bullet}(\sigma)$ fit is assessed by looking at the F-test ratios.

\subsection{Sample Selection\label{sample}}

The initial sample of 30 external galaxies \citep{Tremaine} contains 17 elliptical galaxies, 10 lenticular galaxies and 3 spiral galaxies; of 
which 2 elliptical, 2 spiral and 3 lenticular galaxies are excluded from future discussion and fitting for the following reasons. The elliptical galaxy
NGC 2778 is rejected because it is an outlier in most relations, has been noted to contain a stellar disk
\citep{Graham,Rix} like a lenticular, and has a relatively large observational uncertainty (although its inclusion
would not change the final fits because of its associated low relative weight). The elliptical galaxy NGC 4564 and the
lenticular galaxy NGC 3384 are both rejected because of their relatively small observational uncertainties (high weight) combined
with the fact that they are marginal outliers in some of the relations (particularly those involving $R_e$ for NGC 4564); their
inclusion results in noticeably different fits. The spiral galaxies NGC 1068 and NGC 4258
are rejected because they are complicated spiral galaxies containing AGNs; obtaining a correct bulge-light-only
surface brightness profile, equivalent to elliptical galaxies, is difficult, and these galaxies are usually
outliers in calculated fits derived from the SB profiles.
The lenticular galaxy NGC 4342 is rejected because it, too, has a complicated surface brightness
profile resulting from the combination of a prominent inner and outer disk; furthermore, this galaxy has been
noted to be an outlier in other relationships, such as the bulge mass predictor by \citet{Haring}.
The lenticular galaxy NGC 4742 is rejected because the SMBH mass, tabulated in \citet{Tremaine},
has not been described in a separate, detailed analysis paper. 
The Milky Way, the remaining galaxy in \citet{Tremaine}, is not used, despite its precise SMBH mass
measurement, because there is no way to reproduce the surface-brightness-profile-driven calculations in a manner
identical to the external galaxies.
\citet{Tremaine} identifies 9 of the galaxies (the Milky Way, M31, NGC 1068, NGC 2778, NGC 3115, NGC 3379, NGC 5845, NGC 4459 and
NGC 6251) in the full-31 galaxy sample as being `questionable'; however, not all of these
are excluded from this analysis, based on our adopted selection criteria. The use of even more stringent selection criteria would 
render the samples 
too small to make any statistically meaningful comments.
However, galaxies rejected from inclusion in the initial \citet{Tremaine}
31-galaxy sample as having unreliable
$M_{\bullet}$ are excluded from the fitting here. 

The final-fitting-sample consists of 23 elliptical, spiral and lenticular bulges (\textit{23gal} sample); this full
sample is then subdivided into 3 final-fitting-subsamples based on morphology and availability of data.
The full, \textit{23gal} sample is first subdivided into a sample containing the 15 pure ellipticals
(\textit{15E}) and a sample containing the 8 remaining spiral/lenticular bulges (\textit{8B}). Finally,
there is a subsample of the \textit{15E} galaxies consisting of 8 elliptical galaxies (\textit{Cap8}) for which mass-to-light ratios from \citet{Cappellari-SAURON}
are available (see Section~\ref{HostGal}). 
These samples are summarized in Table~\ref{Tab4}.

A caveat related to fitting with the spiral/lenticular galaxies (present in both the \textit{8B} and \textit{23gal} samples)
will be noted before explicitly discussing individual predictors in the following sections.
The \textit{8B} sample exhibits problems when weighting (in Y) only by the observational uncertainty (\textit{OBS})
which are not found for the elliptical sample. These problems likely stem from either incorrect $\epsilon_{yo}$ values for some of the galaxies
(which results in incorrect weights)
or from incorrect host galaxy parameters. The calculated \textit{OBS}-weighted fits for the spiral/lenticular galaxies
have large $\chi^2$ and \textit{scatter} and have significantly different
slopes than when \textit{INT}-weighting is used. For example, the \textit{OBS}-weighted $M_{\bullet}(\sigma)$ relationship for \textit{8B}: 
$\log{(M_{\bullet}/ M_{\odot})} = (1.82 \pm 0.88)\log{(\sigma / (200 km s^{-1}))}+(7.81 \pm 0.11)$
(\textit{scatter} = 0.53) bears no resemblance to the accepted slope (which is near 4.0) and is an extremely poor fit for
the elliptical galaxies. This problem is prevalent for all of the predictive relationships examined; adding the excluded
5 bulges does not solve the problem, and removing additional bulges results in too small a sample to make
any statistically meaningful comments. 
Although the \textit{OBS} weighting
is not being used for the final fits, the observational uncertainties do factor into the \textit{OBS+INT} final fits, and so 
the predictors based on both the \textit{8B} and the \textit{23gal} samples of galaxies 
may be less reliable 
than fits based solely on the elliptical galaxies. 

\subsection{Bulge Velocity Dispersion\label{sigma}}

The first $M_{\bullet}$ predictor examined is the bulge stellar velocity dispersion ($\sigma$); this
relationship is used as a point of comparison for the predictors in the future sections.
Table~\ref{Tab5} presents the predictive fits (equation~\ref{E4})
as follows: for each sample (\textit{15E, 23gal, 8B, Cap8}) and fitting algorithm used (\textit{lsqfit, fitexy, robust}),
using the previously specified default \textit{weightcent} options (\textit{Avgerr, OBS+INT}), the slope (a) and the zero-point (d) of the fit,
along with the intrinsic uncertainty in $Y=\log{M_{\bullet}}$ ($\epsilon_{yin}$) required to produce
$\chi_r^2 = 1$, the uncertainty in $X=\log{\sigma}$ ($\epsilon_x$) applied, the maximum ($a_{max}$) and 
minimum ($a_{min}$) slope obtained with any of the viable \textit{weightcent} combinations, and the
\textit{scatter} (equation~\ref{E6}) are given. The
quantities/abbreviations are as previously specified (see Table~\ref{Tab4}). 

The fits for the four samples are, as one would expect, statistically equivalent to the fit from the
\citet{Tremaine} 31-galaxy sample; the four samples are subsets of the 31-galaxy sample and the velocity dispersion values
for all fitting-sample galaxies are taken directly from \citet{Tremaine}.
\citet{Tremaine} finds a best-fit relationship for the 31-galaxy sample of 
$\log{\left({M_{\bullet} / M_{\odot}}\right)} = (4.02\pm 0.32)\log{\left({\sigma / (200 km s^{-1})}\right)}+(8.13\pm 0.06)$,
with an intrinsic uncertainty of 0.27 (or $0.23\pm 0.05$ dex using a maximum likelihood estimate).
The slopes (Table~\ref{Tab5}: column 3) for the four (sub)samples 
are all compatible with this result,
within the stated uncertainties, as are the zero-points and requisite intrinsic uncertainties in Y ($\epsilon_{yin}$).
It may be noted that the fitted slopes are consistently lower than the \citet{Tremaine} slope of $4.02\pm 0.32$. This is not unexpected; \citet{Tremaine}
states that the 4.02 slope ``may slightly overestimate the true slope by 0.1-0.3''. 
\citet{Tremaine} made this comment after examining 
4 Tremaine-subsamples which (1) excluded the Milky Way ($a = 3.88 \pm 0.32$), (2) excluded high ($\sigma \geq 250 km s^{-1}$)
dispersion galaxies ($a = 3.77 \pm 0.49$), (3) excluded all but the 10 galaxies in \citet{Pinkney,Gebhardt-12} ($a = 3.67 \pm 0.70$)
and (4) removed 9 `questionable' galaxies: the Milky Way, M31, NGC 1068, NGC 2778, NGC 3115, NGC 3379, NGC 5845, NGC 4459 and
NGC 6251 ($a = 3.79 \pm 0.32$). 

The morphological division of the full (\textit{23gal}) sample into the elliptical (\textit{15E, Cap8}) and spiral/lenticular bulge (\textit{8B})
subsamples illustrates a difference: the elliptical samples consistently have a larger slope than the bulge sample.
This implies that altering the ratio of the spiral/lenticular:elliptical galaxies in a combined sample of galaxies will alter the slope
of the combined sample. It should also be noted, however, that while the bulge (\textit{8B}) slopes are lower, they are consistent with the
elliptical slopes within the (large) stated uncertainties of the slope. 

The morphological division of the \textit{23gal} sample also illustrates that the (statistically)
strongest fitting sample is the full-elliptical (\textit{15E}) sample (also see Section~\ref{sample}). This is evident both in the resultant \textit{scatter} from the fits ($\sigma_{fy}$) which
is lowest for the \textit{15E} sample, even when compared with the full-galaxy sample (\textit{23gal}), and in the required
intrinsic uncertainty in $\log{M_{\bullet}}$ ($\epsilon_{yin}$) which is consistently lower than for any other sample. The \textit{Cap8} is 
(statistically) weaker because it has only about half of the number of points. The \textit{8B} (bulge) sample is weaker both
because of the small number of points, and because it is intrinsically weaker (as illustrated in comparing the \textit{scatter} with the same-sized \textit{Cap8} sample). 
This is likely due to complications related to the increased amount of galaxy structure. Finally, the \textit{23gal} full-galaxy sample
is (statistically) weaker because of the presence of the \textit{8B} galaxies which are either fundamentally different (as illustrated by their
lower slopes) and/or appear different because of imperfections in the data.

The replacement of the SMBH mass for NGC 224 (4.5x$10^7 M_{\odot}$) with the higher value of 1.4x$10^8 M_{\odot}$ given in \citet{Bender-M31} (see Table~\ref{Tab2}), and the
replacement of the velocity dispersion for NGC 2787 (140 $km s^{-1}$) with the higher value of 200 $km s^{-1}$ suggested by literature measurements of the central velocity dispersion
(see Table~\ref{Tab3}), does not change the conclusions drawn from the fits, as shown in Table~\ref{Tab5}. The slope of the \textit{23gal} least-squared (\textit{lsqfit})
fit is essentially unchanged, while the slope of the \textit{23gal} \textit{robust} fit is substantially lowered (3.49 compared with 3.70), 
but is consistent with the originally stated slope-range.
The slope of the \textit{8B} \textit{lsqfit} fit is even lower than originally stated, $3.10\pm0.96$ compared with $3.27\pm0.77$, while the slope of the \textit{8B} \textit{robust}
fit is higher, 3.66 compared with 3.34, but still falls below the elliptical galaxy sample slope. This illustrates 
both the relative instability of the fitted slopes for the \textit{8B} sample, and the 
dichotomy between the elliptical and spiral/lenticular bulge sample slopes.
For all of these fits, the revised parameters result in larger values for the \textit{scatter}, and in larger requisite uncertainties in $\log{M_{\bullet}}$ ($\epsilon_{yin}$)
for the least-squared fitting, and so the revised parameters are not generally preferred over the original values.

In addition to providing information about the predictive fits based on morphological divisions, Table~\ref{Tab5} also provides information about the
importance of the fitting algorithm. First, it is evident that, for a given sample, including (or excluding)
the uncertainty in $X=\log{\sigma}$ ($\epsilon_x \equiv 0.021$) has little impact on the resulting fits. The slope increases by only 
0.04 (1 \%) when it is included, well within the stated uncertainties, and $\epsilon_{yin}$, the requisite intrinsic error in $Y=\log{M_{\bullet}}$,
decreases by only 0.01.
Second, the slopes and zero-points resulting from the least-squares-fitting algorithms are consistent with the results from the robust fitting algorithm; there
is no change in the conclusions or in the stated morphological trends. 

Altering the \textit{weightcent} selections, illustrated in the stated maximum ($a_{max}$) and minimum ($a_{min}$) slopes, does not
change the stated conclusions related to morphological type; elliptical galaxy samples persistently exhibit higher slopes than the spiral/lenticular
bulge galaxy sample. The \textit{weightcent} selections can result in slope variations, relative to the default, of up to 11.5\% (\textit{fitexy, 8B}), but these
variations are well-within the stated default-slope uncertainties. The examination by morphological type also indicates that the default slope
for ellipticals is at the high end of the range suggested by $a_{max}$ and $a_{min}$, while the bulges are at the low end, maximizing the apparent
discrepancy in the slopes. However, while the full ranges bring their slopes closer together, the elliptical galaxy sample slopes are still, on
average, higher, and higher for every specific \textit{weightcent} selection.

The \textit{stability} of the fits, indicated by the statistically negligible variation in slope for each sample with changes in fitting algorithm and 
\textit{weightcent}, is reinforced through the bootstrap calculations (Table~\ref{Tab6}).
The value of $N_{\chi}$ is the number of `new' data sets (out of 1000) for which an $\epsilon_{yin}$ resulting in 
$\chi_r^2=1.0$ could be determined; samples in which there were too few (unique) points for this to occur, even when 
$\epsilon_{yin}=0$, were excluded. 
The bootstrap exhibits large uncertainties (both \textit{lsqfit} and \textit{robust})
associated with both the \textit{Cap8} and \textit{8B} sample slopes and large values for the skew and kurtosis; this underscores the problems inherent in making
statistical analyses based on small samples. The 8-galaxy samples are reduced to (typically) 5 unique galaxies here, and with this
number of galaxies the slopes are very dependent on individual galaxy effects (and selection) rather than on a general, broad
description of the morphological group. Given this caveat, it is notable that the bulge (\textit{8B}) slopes are much higher in the bootstrap 
results; higher than even the elliptical and combined samples. This dramatic change is not seen in the similarly small \textit{Cap8}
sample, further illustrating that the bulge-only sample is weaker when making statistical claims, likely owing to the strong
influence of several (discrepant) galaxies on the slope. Removing these galaxies would result in further problems from an even more reduced sample
size, however. The results for the larger samples, \textit{15E} and \textit{23gal}, are generally consistent with the original
results, both for the robust and for the least-squares fitting results. 

\begin{figure}
\plotone{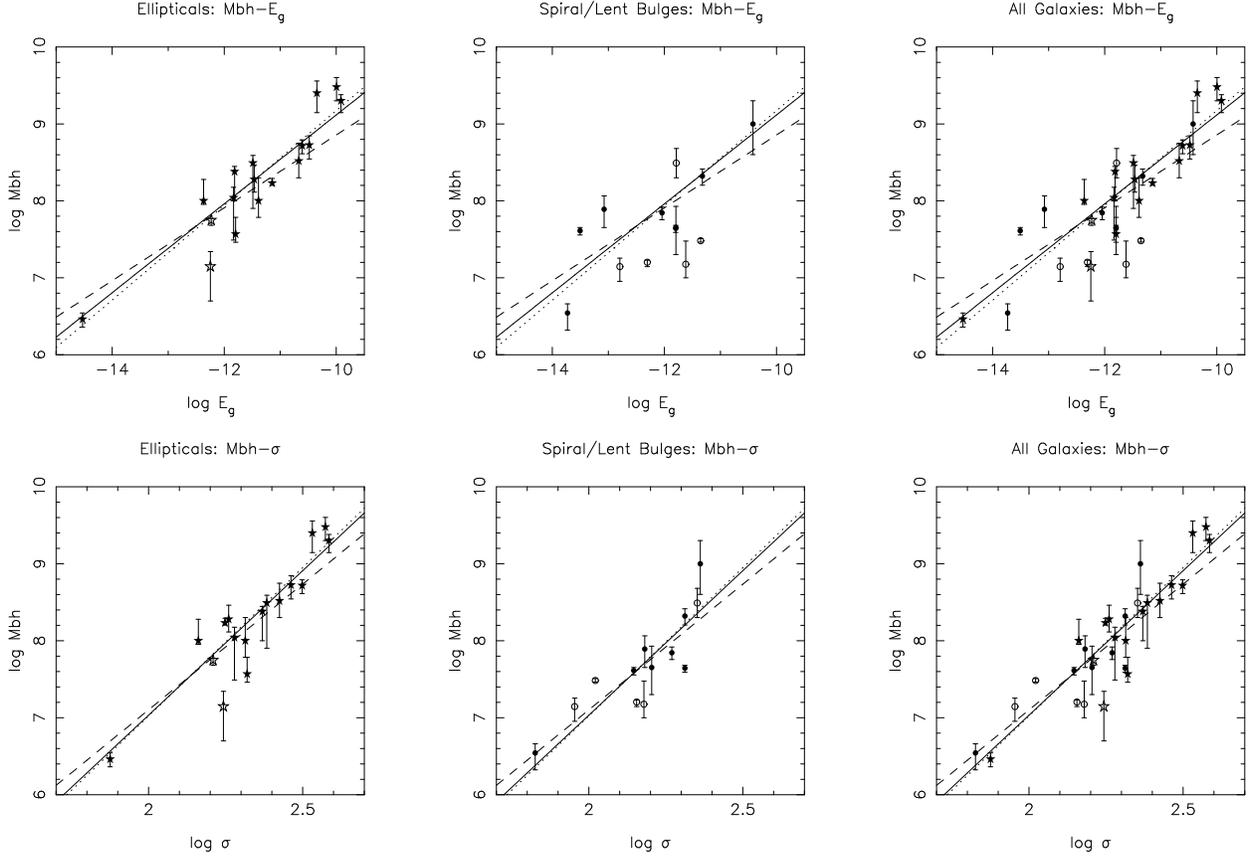}
\caption{Top three panels illustrate the relationship between the gravitational binding energy and
the black hole mass, shown with observational uncertainties on the black hole mass, and bottom
three panels illustrate the relationship between velocity dispersion ($\sigma$) and the black hole mass
for the same galaxies. From left to right, the panels show ellipticals only, spiral/lenticular bulges
only and, finally, all galaxies. Stars indicate elliptical galaxies, with filled stars denoting those
galaxies in the \textit{15E} sample used for the fitting, and the circles indicate bulges, with filled circles
denoting those galaxies in the \textit{8B} sample used for fitting. The fit for the combined galaxy sample
(\textit{23gal}) is denoted by a solid line, the fit for ellipticals only (\textit{15E}) is denoted by a dotted line,
and the fit for the bulges only (\textit{8B}) is denoted by a dashed line, in all panels.
\label{Fig2}}
\end{figure}

When the best fits, using the \textit{lsqfit} algorithm, for each sample (Table~\ref{Tab5}) are plotted, further information about the predictive relationships can be gleaned.
Looking at the bottom half of Figure~\ref{Fig2} ($M_{\bullet}$ versus $\sigma$) for ellipticals only (left), spiral/lenticular bulges only (center) and
the combined sample (right), the bulges are moderately fit by the elliptical-only relationship (dotted), while the ellipticals
are not well-fit by the (lower) bulge slope (dashed line). 
Figure~\ref{Fig3} (ellipticals in top panel, bulges in middle panel and combined sample in bottom panel) 
plots the (\textit{lsqfit}) fit residuals against the  bulge velocity dispersion (left),
the bulge effective radius (center), and the intensity at the bulge effective radius (right). 
Neither in
Figure~\ref{Fig2} nor in Figure~\ref{Fig3} does there appear to be strong evidence for the necessity of a quadratic, as opposed to a linear, parameterization;
this is further explored in Appendix~\ref{Ap-B} where formal fits of a quadratic are shown to provide no improvement in the quality of the fit (as indicated by $\sigma_{fy}$).
There also appears to be no evidence of a pattern in the residuals with 
$\sigma$, $R_e$, or $I_e$ for either the combined (\textit{23gal}) or for the elliptical (\textit{15E}) sample. (There is a minimal
\textit{8B}-fit-residual correlation with $\sigma$ when looking at only the elliptical galaxies, indicative of the previously mentioned poor lower slope fit
for the elliptical sample). 
The bulge sample (\textit{8B}-filled circles), however, shows a clear correlation between the fit-residuals and $R_e$ for all three (\textit{15E, 8B, 23gal})
of the predictive fits.

\subsection{The Combination of Bulge Velocity Dispersion, Radius and/or Intensity\label{sigma-other}} 

\begin{figure}
\plotone{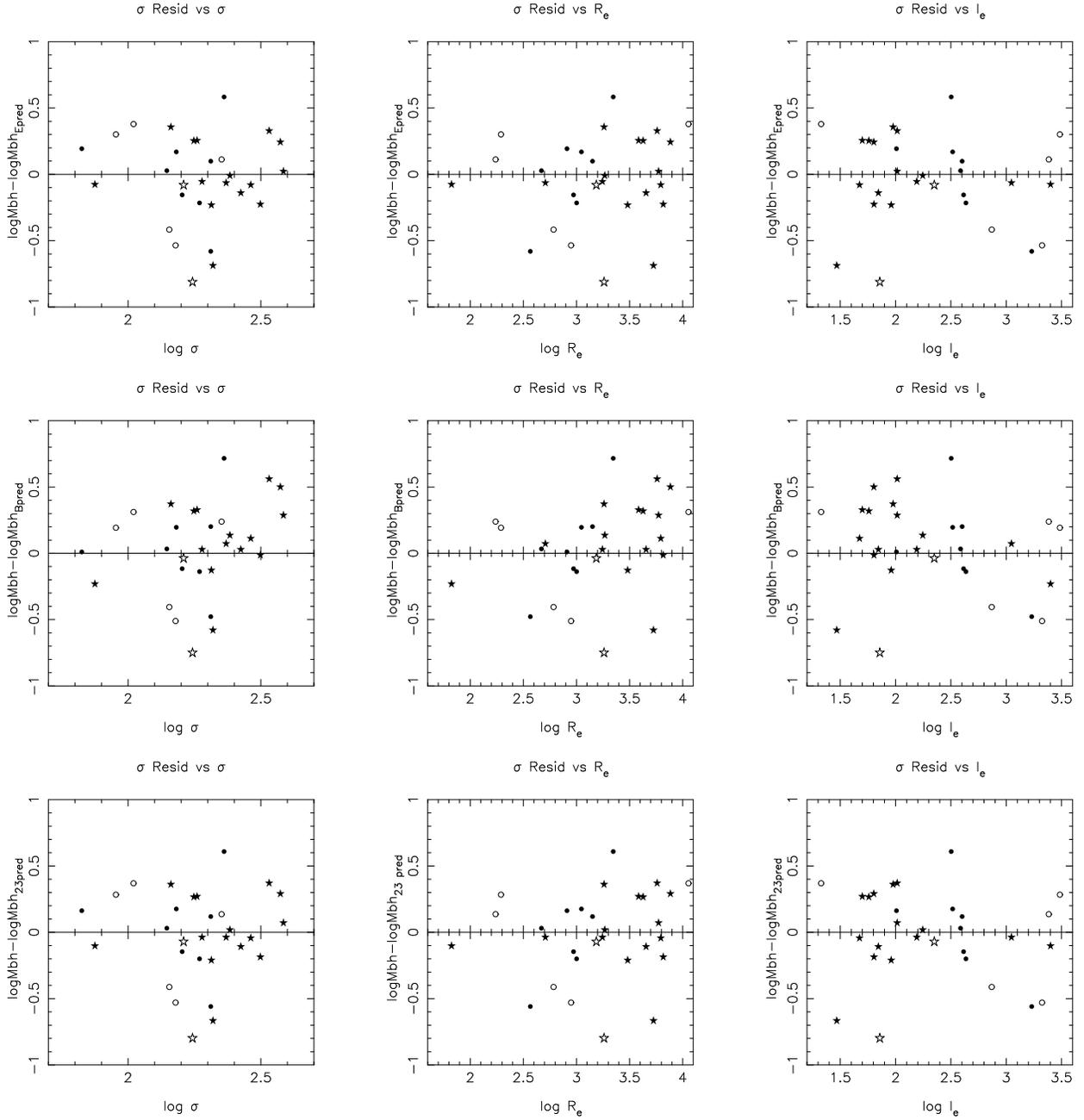}
\caption{Illustration of the residuals from the dispersion fit ($M_{\bullet obs}-M_{\bullet pred}$);
the top panel is the \textit{15E} elliptical fit, the middle panel is the \textit{8B} spiral/lenticular bulge fit and the bottom panel is the \textit{23gal} sample fit.
The left column plots the residual as a function of velocity dispersion, the middle column plots the residual as a function
of bulge effective radius, and the right column plots the residual as a function of V-band $I_e$.
The symbols are as in Figure~\ref{Fig2}. 
The galaxies with $|residuals|  \geq 0.6$  
in the top panel are NGC 821 (\textit{15E,Cap8}; filled star)
and NGC 2778 (open star), in the middle panel are
NGC 2778  and NGC 3115 (\textit{8B}, filled circle)
and in the bottom panel are
NGC 821, NGC 2778 and NGC 3115.
\label{Fig3}}
\end{figure}

The $M_{\bullet}(\sigma)$ predictor is successful for all galaxy morphologies, but the residual scatter prompts
the question of whether a multivariate fit, or an alternative single-variable fit, would reduce this scatter.
\citet{Marconi} claim
that there is ``a weak, but significant, correlation of [$M_{\bullet}-\sigma_e$] residuals
with $R_e$'' and that a ``combination of both $\sigma_e$ and $R_e$ is necessary to drive the correlations between
$M_{\bullet}$ and other bulge properties'': the so-called `fundamental plane of SMBHs'. 
The $M_{\bullet}-M_{\bullet}(\sigma)$ residuals for the elliptical (\textit{15E}) and combined (\textit{23gal}) samples,
illustrated in Figure~\ref{Fig3}, do not immediately indicate such a combination.
For the bulge-only (\textit{8B}) sample, however, there is good evidence of a correlation between the residuals and $R_e$. Examining 
the \citet{Marconi} data, it is the lenticular/spiral galaxies which are generating the appearance of
a significant correlation. If only their spiral/lenticular galaxies are considered, 
there is a relatively strong correlation present, but looking at only
their elliptical galaxies there is, at best, a minimal correlation between the residuals and $R_e$. Thus, $\sigma$ is a weaker
predictor of $M_{\bullet}$ for spiral/lenticular galaxies than for elliptical galaxies. 

\begin{figure}
\plotone{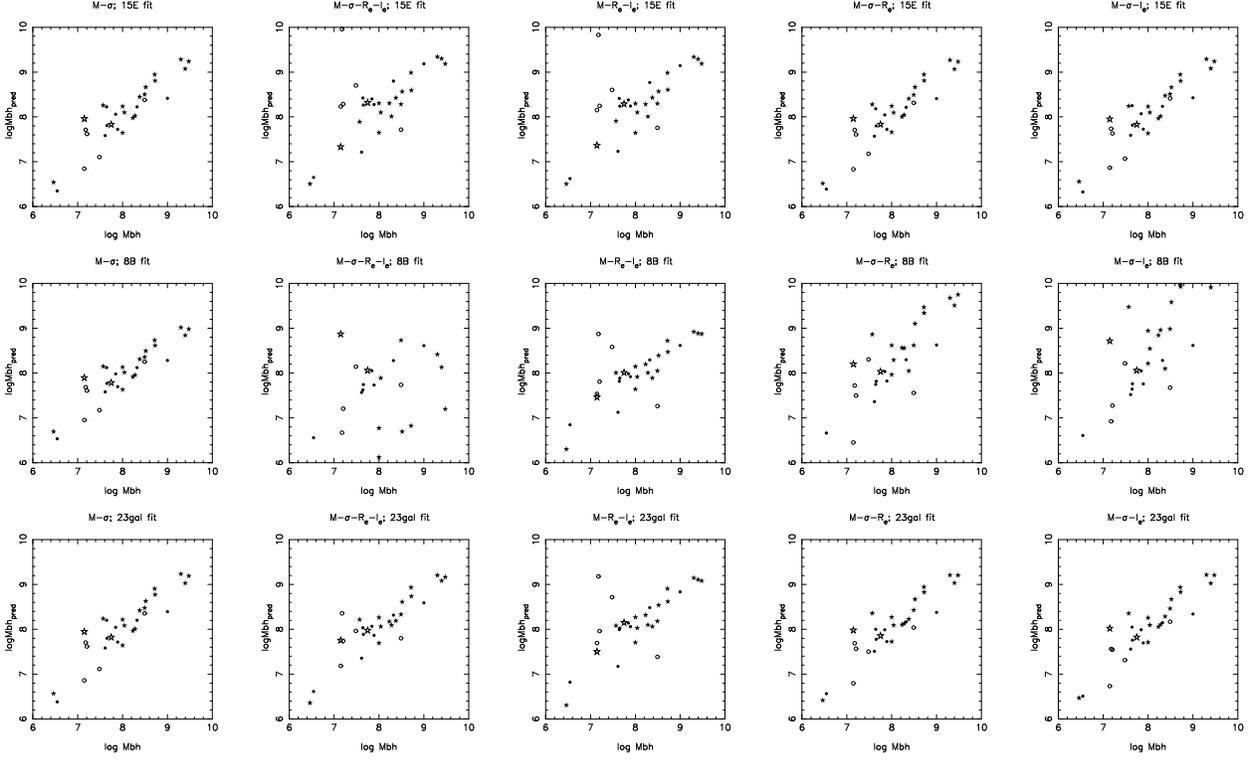}
\caption{Observed black hole mass plotted against the predicted black hole mass for multivariate fits. The top row is for the
fit based on ellipticals (\textit{15E}), the middle row is for the fit based on bulges (\textit{8B}) and the bottom row is for the fit
based on ``all'' galaxies (\textit{23gal}). From left to right the columns illustrate the $M_{\bullet}(\sigma)$ fit, the $M_{\bullet}(I_e,R_e,\sigma)$ fit,
the $M_{\bullet}(I_e,R_e)$ fit, the $M_{\bullet}(R_e,\sigma)$ fit and the $M_{\bullet}(I_e,\sigma)$ fit.
The symbols are as in Figure~\ref{Fig2}.
\label{Fig4}}
\end{figure}

Multivariate fits combining $\sigma$, $I_e$ and $R_e$ are examined, using the \textit{lsqfit} algorithm (default \textit{weightcent}), to determine if there is 
an obvious, physically-motivated, predictor of $M_{\bullet}$, other than $M_{\bullet}(\sigma)$.
The multivariate fits (equation~\ref{E4}, where $x_1=I_e / I_{e0}$, $x_2=R_e / R_{e0}$, and $x_3=\sigma / \sigma_{0}$) for the combined
(\textit{23gal}), elliptical-only (\textit{15E}) and spiral/lenticular bulge-only (\textit{8B}) samples are presented in Table~\ref{Tab7}. In
addition to the best-fit coefficients ($a_i$, d), the intrinsic uncertainty in $\log{M_{\bullet}}$ ($\epsilon_{yin}$) required to 
obtain $\chi_r^2=1.0$, the \textit{scatter} (equation~\ref{E6}), the F-test ratio ($F_{\sigma_y}$) relative to the 
$M_{\bullet}(\sigma)$ fit (equation~\ref{E7}), and the significance (Sig) of the difference relative to the $M_{\bullet}(\sigma)$ fit are given.
The fits, excluding $M_{\bullet}(I_e)$ and $M_{\bullet}(R_e)$, are illustrated
in Figure~\ref{Fig4} wherein the SMBH mass predicted by the fit (based on the top panel: \textit{15E} sample, 
central panel: \textit{8B} sample, and bottom panel: \textit{23gal} sample) is plotted against the measured SMBH mass for, 
from left to right, the $M_{\bullet}(\sigma)$, $M_{\bullet}(I_e,R_e,\sigma)$, $M_{\bullet}(I_e,R_e)$, $M_{\bullet}(R_e,\sigma)$,
and $M_{\bullet}(I_e,\sigma)$ relationships. 

The addition of $I_e$ or $R_e$ as a second parameter [$M_{\bullet}(I_e,\sigma)$ or $M_{\bullet}(R_e,\sigma)$] does not improve the quality
of the fit (based on the \textit{scatter}) for the elliptical or combined samples, despite the additional free parameter.
Furthermore, the coefficient of the additional parameter ($a_1$ and $a_2$ respectively)
is consistent with zero for the elliptical sample and is minimal (in comparison with the $\sigma$-coefficient) for the combined
sample. The bulge (\textit{8B}) sample, which was indicated to require a $M_{\bullet}(R_e,\sigma)$ fit by the Figure~\ref{Fig3} residuals,
does result in predictive fits that are significantly better than the $M_{\bullet}(\sigma)$ predictor:
$M_{\bullet} \propto I_e^{-1.02}\sigma^{4.68}$ and $M_{\bullet} \propto R_e^{0.94}\sigma^{2.90}$.
However, as illustrated in Figure~\ref{Fig4}, neither of these predictive relationships is as successful at predicting $M_{\bullet}$ for
elliptical galaxies as the $M_{\bullet}(\sigma)$ fit. Given that the bulge sample is statistically weaker and less \textit{stable} (more subject to variation with fitting
selections), as previously discussed, these two fits will not be further considered.
The $M_{\bullet}(R_e,\sigma)$ fit does not indicate $M_{\bullet} \propto \sigma^2R_e$, the
bulge mass surrogate, for any galaxy sample, nor does the $M_{\bullet}(R_e,\sigma)$ fit appear superior to 
$M_{\bullet}(\sigma)$ for elliptical galaxies either from the fits, or from examining
the $M_{\bullet}(\sigma)$ residuals versus $R_e$ (Figure~\ref{Fig3}). This is in contrast
to the \citet{Marconi} argument, based on an examination of the residuals,
that $M_{\bullet}(R_e,\sigma)$ is superior to $M_{\bullet}(\sigma)$ for all galaxies. 
Furthermore, neither $M_{\bullet}(R_e)$ nor $M_{\bullet}(I_e)$ 
are better predictors of $M_{\bullet}$; for all galaxy samples they are worse
(at a $\geq$ 1-sigma level) than $M_{\bullet}(\sigma)$. 

The three-parameter predictor, $M_{\bullet}(I_e,R_e,\sigma)$, is better (at the 1-sigma level) for bulges, and minimally
better for ellipticals and the combined sample, when compared with
$M_{\bullet}(\sigma)$. The coefficients of the fit (Table~\ref{Tab7}), however, vary strongly with galaxy sample; there is no clear relationship
applicable for all galaxies.
The bulge sample indicates a fit with minimal $I_e$ and $R_e$ and a high $\sigma$ slope;
this is a poor predictor for elliptical galaxies, as illustrated in Figure~\ref{Fig4}.
However, in both the elliptical and combined samples, the velocity dispersion is no longer
the dominant term as it was in the $M_{\bullet}(I_e,\sigma)$ and $M_{\bullet}(R_e,\sigma)$ projections of the multivariate surface.
For elliptical galaxies where $M_{\bullet} \propto \sigma^{-0.25}I_e^{2.48}R_e^{3.31}$, the preferred projection
appears to be $M_{\bullet}(I_e,R_e)$ instead of $M_{\bullet}(\sigma)$.

The $M_{\bullet}(I_e,R_e)$ fit results in a fit which is at least as
(statistically) good as $M_{\bullet}(\sigma)$.
The form of the fit, for elliptical galaxies,  
($M_{\bullet} \propto I_e^{2.34}R_e^{3.12}$) is similar to the fit for the combined sample
($M_{\bullet} \propto I_e^{1.72}R_e^{2.68}$); both suggest $M_{\bullet} \propto I_e^{2}R_e^{3}$. 
The bulge fit predicts a different relationship, $M_{\bullet} \propto I_e^{1.47}R_e^{2.39}$, with a lower dependence
on both variables. 

\begin{figure}
\plotone{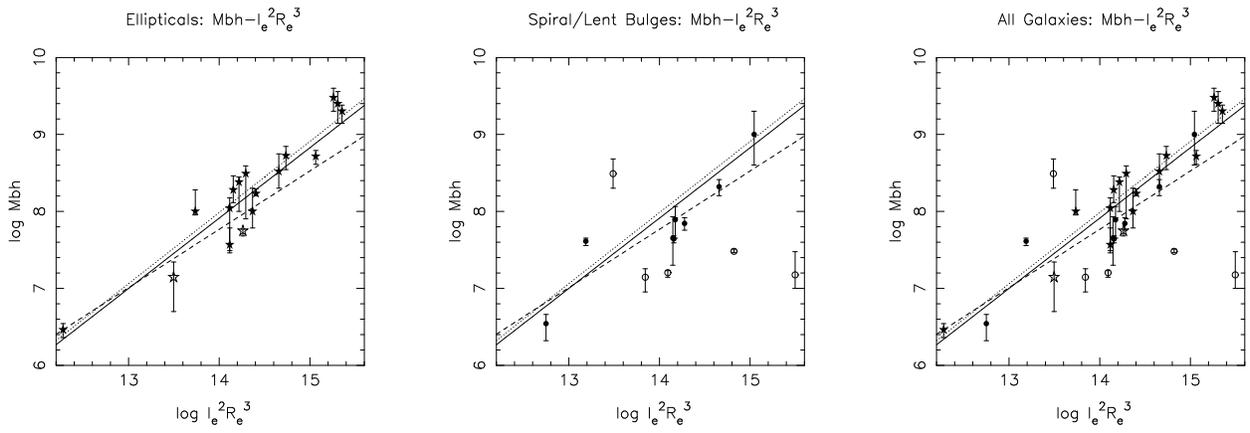}
\caption{Illustration of the relationship between $I_e^2R_e^3$ and black hole mass. The left panel
shows the ellipticals, the central panel shows spiral/lenticular bulges, and the right panel shows all galaxies.
Lines and symbols are as in Figure~\ref{Fig2}.
\label{Fig5}}
\end{figure}

Examining the suggested $M_{\bullet}(I_e^{2}R_e^{3})$ fit (Table~\ref{Tab8}, Figure~\ref{Fig5}),
the \textit{scatter} is minimally better than $M_{\bullet}(\sigma)$ for all samples. 
The lower
bulge-only slope is, within the stated uncertainties, consistent with the higher, elliptical galaxy slope (0.76 versus 0.91-0.95); this
behavior is reminiscent of the $M_{\bullet}(\sigma)$ results. This correlation is also physically significant;
the SMBH mass is proportional to the \textit{energy} (E) per $\Upsilon^2$, at the bulge effective radius:
$I_e^{2}R_e^{3} \propto E\Upsilon^{-2}$ (given $E \propto {M_e^2 / R_e}$ and
$L_e \propto I_eR_e^2$). 

\begin{figure}
\includegraphics[angle=0.,scale=0.60]{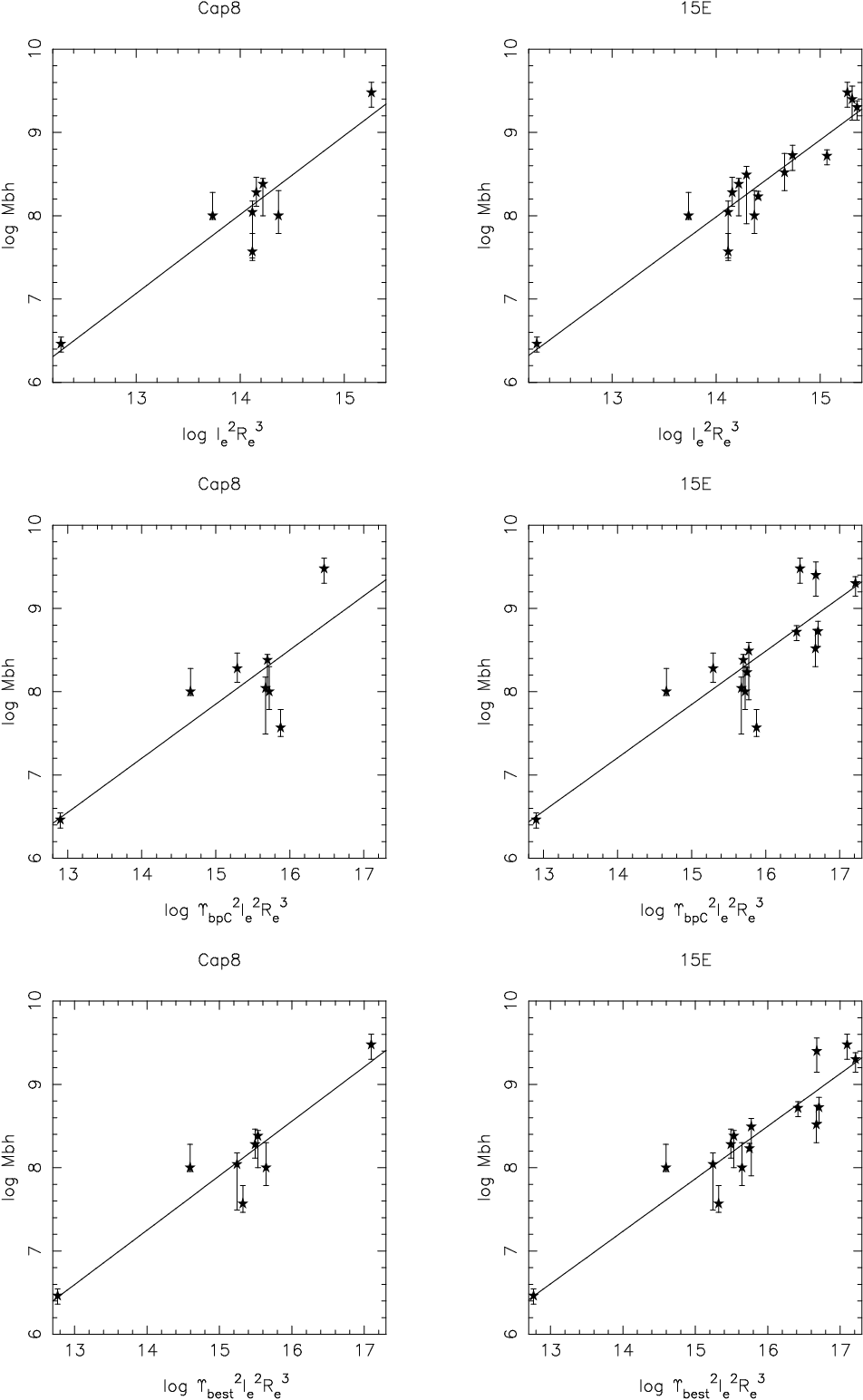}
\caption{Top panel illustrates the $I_e^2R_e^3$ fit, the middle panel the $\Upsilon_{bpC}^2I_e^2R_e^3$ fit using
the $\Upsilon_{bpC}$ values from Table~\ref{Tab2}, and the bottom panel the $\Upsilon_{best}^2I_e^2R_e^3$ fit using the
revised $\Upsilon_{Cap}$ values from \citet{Cappellari-SAURON}, where available. The left column is for the \textit{Cap8} elliptical galaxy
(sub)sample and the right column is for the \textit{15E} elliptical galaxy sample. The lines illustrate the best-fit relationship for the
galaxies in each panel. The symbols are as in Figure~\ref{Fig2}. \label{Fig6}}
\end{figure}

\begin{figure}
\plotone{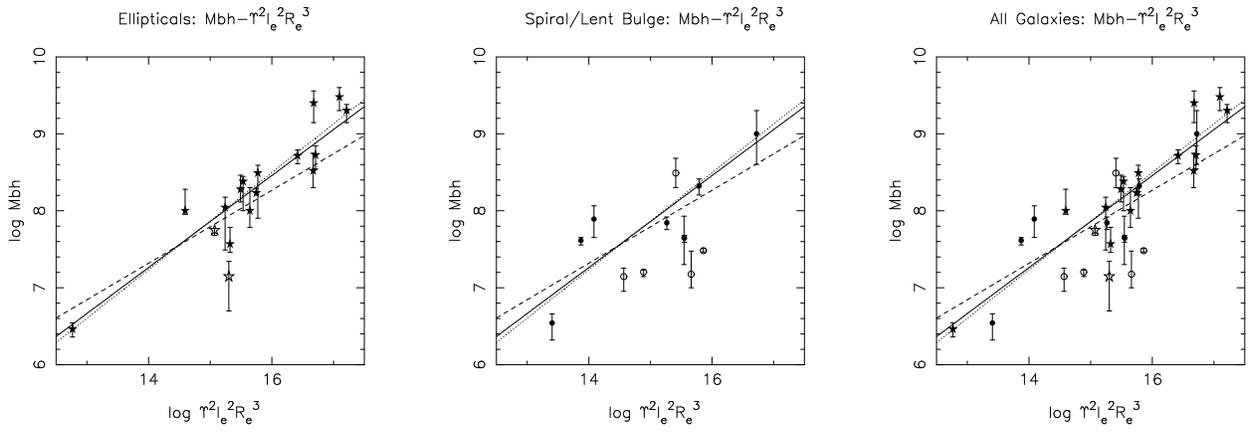}
\caption{Same as figure~\ref{Fig5}, but for $\Upsilon^2I_e^2R_e^3$ \label{Fig7}}
\end{figure}

The correlation of $M_{\bullet}$ with \textit{energy}, $M_{\bullet} \propto \Upsilon^2I_e^{2}R_e^{3}$
illuminates a problem with the $\Upsilon_{bpC}$ mass-to-light ratios. The 
$\Upsilon_{bpC}$ values (see Section~\ref{HostGal}) result in $M_{\bullet}(\Upsilon^2I_e^{2}R_e^{3})$ fits
which are worse (statistically) than $M_{\bullet}(\sigma)$. 
This is illustrated in Table~\ref{Tab8} as the $\Upsilon_{bpC}^2I_e^2R_e^3$ predictor.
However, when these $\Upsilon_{bpC}$ values are replaced with the (SAURON integral-field-spectroscopy) values from 
\citet{Cappellari-SAURON}, $\Upsilon_{Cap}$, when available, the fit improves
significantly, as illustrated in Figure~\ref{Fig6} and in Table~\ref{Tab8} ($\Upsilon_{best}^2I_e^2R_e^3$). For the sample of 8 elliptical galaxies
for which the \citet{Cappellari-SAURON} mass-to-light ratios are available (\textit{Cap8}), $M_{\bullet}(\Upsilon^2I_e^{2}R_e^{3})$
is minimally better than both $M_{\bullet}(I_e^{2}R_e^{3})$ and $M_{\bullet}(\sigma)$. It is anticipated that
if \citet{Cappellari-SAURON} ($\Upsilon_{Cap}$) values were available for the full sample of galaxies, the improvement in \textit{scatter} would be
significant; possibly enough to show a clear preference for $M_{\bullet}(E)$ over $M_{\bullet}(\sigma)$.
It is also interesting to note that although utilizing \citet{Cappellari-SAURON} ($\Upsilon_{Cap}$) values improved
the quality of the fit, the actual slope of the fit changed little or not at all. The elliptical and combined
samples show a slope of 0.60-0.65, and the lower bulge slope (0.47) is consistent with this range within the
stated uncertainties. Figure~\ref{Fig7} illustrates the higher, elliptical slope is a better descriptor of galaxies as a whole.
Allowing for a multivariate fit $M_{\bullet}(E,R_e)$, tabulated in the bottom panel of Table~\ref{Tab8} does not improve the quality of the fits.

\begin{figure}
\plotone{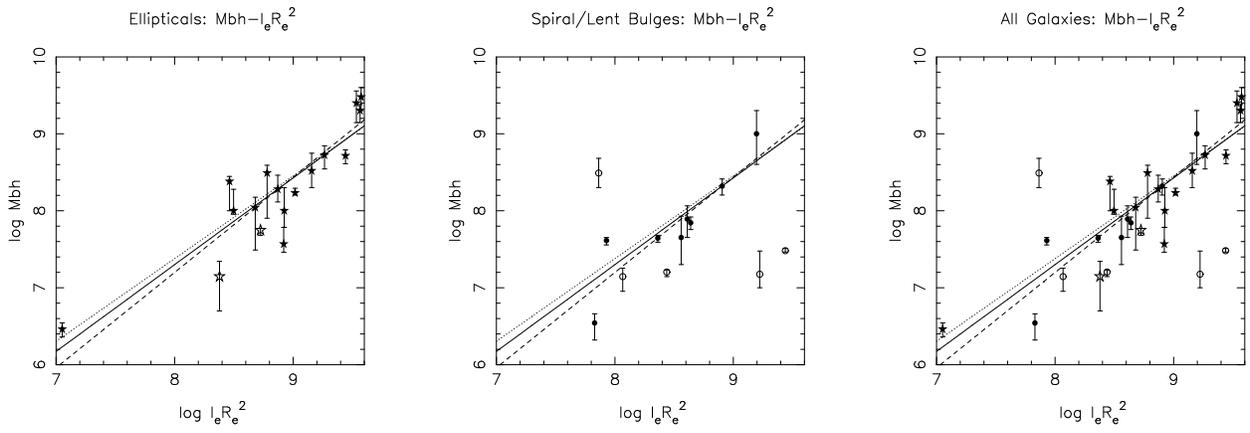}
\caption{Same as figure~\ref{Fig5}, but for $I_eR_e^2$ \label{Fig8}}
\end{figure}

Finally, a commonly proposed predictor of SMBH mass is luminosity. 
\citet{Marconi} determined that
$\log{(M_{\bullet}/M_{\odot})} = (8.21 \pm 0.07) + (1.13 \pm 0.12)(\log{(L_{K,bulge})-10.9})$
with an rms scatter of 0.31 (and similar slope and scatter in B, J and H).
Using $I_eR_e^2$ as a surrogate for luminosity
(Table~\ref{Tab8}, Figure~\ref{Fig8}) it is clear that for ellipticals, $M_{\bullet}(I_eR_e^2)$ is inferior to 
$M_{\bullet}(\sigma)$ as a SMBH mass predictor. For bulges `luminosity' is minimally better, although (see Figure~\ref{Fig8}) there
is still visible scatter about the fit if non-fitting sample galaxies are considered. For all samples
the slope is roughly consistent with 1.0, as in the literature; the combined slope is in fact identical to the \citet{Marconi} K-band
result with almost the same uncertainty. 

\subsection{Gravitational Binding Energy\label{energy}}

The multivariate fitting [$M_{\bullet}(I_e,R_e,\sigma)$] suggests that gravitational binding energy 
is a comparable, or better, predictor of SMBH mass
than bulge velocity dispersion. The more formal fit of $M_{\bullet}(E_g)$, where binding energy ($E_g$) is calculated 
utilizing previously discussed SB
profiles (Section~\ref{Data}) and geometrical assumptions (Section~\ref{DetVals}), with no contribution from dark halos and disks,
supports this conclusion. Throughout this discussion, and the accompanying figures and tables, the term gravitational
binding energy ($E_g$) is defined such that $E_g = -PE$, where PE is the gravitational potential energy.

The most notable characteristic of the $M_{\bullet}(E_g)$ fit is the \textit{stability} of the slope with fitting
algorithm, \textit{weightcent}, and even with galaxy morphology. The best-fit slope is approximately
0.6 (Table~\ref{Tab9}) for the \textit{15E}, \textit{Cap8}, and \textit{23gal} samples, both using
the \textit{lsqfit} and the \textit{robust} fitting algorithms. This slope, 0.6, was also determined
in the previous section using the crude approximation ($\Upsilon^2I_e^2R_e^3$) for gravitational binding 
energy; the method for calculating the binding energy for the galaxy has no significant impact on the derived slope
of the predictive relationship.
The variation in the slope based on the \textit{weightcent} selection, as illustrated by the $a_{max}$
and $a_{min}$ values in Table~\ref{Tab9}, is minimal. 
Even the addition of uncertainty in the $E_g$ value, $\epsilon_x=0.3$ (100\%) (Table~\ref{Tab10}) using the \textit{fitexy} algorithm, results
in a slope which is, at most, 0.03-0.04 higher and within the stated uncertainties for the \textit{lsqfit} 
($\sim$ 0.6) slope. 

The bootstrap algorithm results, implemented as for the $M_{\bullet}(\sigma)$ predictor, further
illustrate the \textit{stability} of the slope. For the \textit{15E} sample,
the mean and the standard deviation of the slope are identical to the original (single-run \textit{lsqfit}) results when
using the bootstrap \textit{lsqfit} algorithm (Table~\ref{Tab11}). 
The remaining samples (\textit{8B, Cap8, 23gal})
show only slightly lower slopes and slightly higher uncertainties in the \textit{lsqfit} bootstrap slopes,
when compared with the original slopes.
The large differences between the original and bootstrap slopes (and uncertainties) for the 8-galaxy 
samples (\textit{8B, Cap8}) which were present in the $M_{\bullet}(\sigma)$ predictor are not present here;
the $E_g$ predictor is more \textit{stable} even for small sample sizes.
The \textit{robust}
bootstrap (Table~\ref{Tab12}) is also generally consistent with the original results, for the elliptical and combined
 samples, further
strengthening the argument for $M_{\bullet} \propto E_g^{0.6}$ as a very \textit{stable} and reliable predictor. 

The spiral/lenticular bulge-only fitting sample (\textit{8B}) slope is consistently
lower than for the elliptical samples for the least-squares fitting; this was also seen for the
$M_{\bullet}(\sigma)$ predictor. This lower slope is given little credence, however,
based on the previously mentioned weaknesses in the bulge sample, the fact that the \textit{lsqfit} slope
of 0.47 is consistent with 0.6 within the stated uncertainties, and the fact that the 
slope is not \textit{stable} with fitting selection. The \textit{robust} fitting algorithm predicts a higher slope (0.66),
while the \textit{robust} bootstrap produces a slope of 0.43, consistent with the
\textit{lsqfit} bootstrap
results. This discrepancy is indicative that the bulge sample is a much \textit{weaker} (less consistent) predictor in comparison with elliptical and combined
galaxy samples.
The preferability of the
$\sim$ 0.6 fitting slope for galaxies, in general, is further illustrated in the top panel of Figure~\ref{Fig2}
and in the correlation between the \textit{8B}-fit-residuals and $E_g$ for the elliptical galaxy sample (Figure~\ref{Fig9}). 

The $M_{\bullet}(E_g)$ predictor is not only at least as \textit{stable} a predictor as $M_{\bullet}(\sigma)$ with morphology,
fitting algorithm, and \textit{weightcent}, but it is also comparable in a statistical sense. 
The \textit{scatter} is statistically equivalent to 
$M_{\bullet}(\sigma)$ for all samples, and
minimally
better 
for the elliptical (\textit{15E, Cap8}) samples.
The comparable nature of these two predictors,
particularly for ellipticals, is visually illustrated in Figure~\ref{Fig2}, in which $M_{\bullet}(E_g)$ is
given along the top and $M_{\bullet}(\sigma)$ along the bottom.
This similarity in predictor-quality is remarkable given that $\sigma$ is based on direct
observations, while $E_g$ is dependent upon a series of geometrical assumptions and the combination of
(often discrepant) data; an error in any of these assumptions or data combinations will increase the \textit{scatter}
in a manner not present for the simple $\sigma$ measurement. 
As illustrated in the previous section, improved $\Upsilon$ values
for all \textit{23gal} sample galaxies will likely reduce the \textit{scatter}.
There is no indication for the necessity of either a second parameter ($R_e$) or for a quadratic
term in the residuals (Figure~\ref{Fig9}), as further discussed in Appendix~\ref{Ap-B}.

The replacement of the SMBH mass for NGC 224 with the higher value of 1.4x$10^8 M_{\odot}$ given in \citet{Bender-M31}, and the replacement of the 140 $km s^{-1}$
velocity dispersion, for NGC 2787, with the higher value of 200 $km s^{-1}$, does not change the conclusions drawn from the fits, as shown in Table~\ref{Tab9}. The slope of
both the \textit{23gal} least-squared (\textit{lsqfit}) fit and the \textit{23gal} \textit{robust} fit are unchanged. 
The slope of both the \textit{8B} \textit{lsqfit} and \textit{robust} fits are
only marginally higher; the least-squared fit is raised to $0.49\pm0.14$ from $0.47\pm0.15$, and the \textit{8B} \textit{robust} fit 
slope is raised to 0.71 from 0.66. The \textit{scatter} and requisite intrinsic
uncertainty in $\log{M_{\bullet}}$ ($\epsilon_{yin}$) is smaller for all but the \textit{8B} \textit{robust} fit (for which the \textit{scatter} is
only slightly higher), which, when combined with the increase in \textit{scatter} for the $M_{\bullet}(\sigma)$
fits using these revised values (see Table~\ref{Tab5}) reinforces the relative success of the gravitational binding energy in accurately predicting SMBH masses.

\subsection{Gravitational Potential\label{potential}}

The predictive power of the gravitational potential, $\Phi(R)$, is also explored. This quantity is
determined as
a by-product of the binding energy calculation (see equation~\ref{E3}).
The gravitational potential is evaluated as a SMBH mass predictor, $M_{\bullet}(\Phi(R))$, at four radii:
20 pc, 100 pc, $R_e$, and $R_e / 8$
(Table~\ref{Tab13}). 

Examination of the fits at fixed radii (i.e. 20 pc and 100 pc) indicates that there is no single predictive
relationship for the black hole mass which is applicable to all radii or to all morphological classes.
The \textit{strength} of the fit decreases as radius increases for the elliptical samples (\textit{15E, Cap8}),
but not for the combined or spiral/lenticular galaxy samples.
For all samples (but most notably for the elliptical 
samples) the slope decreases with increasing radius. 
The variations in slope with \textit{weightcent}, 
as given by $a_{max}$ and $a_{min}$ (Table~\ref{Tab13}), are within the stated uncertainties for each sample, but the range in these values indicates
that the predictor is not
as \textit{stable} as the gravitational binding energy or bulge velocity dispersion. Furthermore, the elliptical sample slopes
are irreconcilably higher than the bulge sample slope for both the 100 pc and the 20 pc radii fits; these
are two separate populations. The combined sample slope is intermediate between the elliptical-only and
the spiral/lenticular bulge-only slopes, but it is a poor predictor; it is worse than velocity 
dispersion at a statistically significant level.

The $M_{\bullet}(\Phi(nR_e))$ predictor, evaluated at $R_e / 8$ and $R_e$, does 
not alleviate the problem of an irreconcilable slope between the spiral/lenticular bulge and the elliptical samples.
The combined sample slope is a compromise between the high ($\sim$2) slope from the elliptical samples and the low ($\sim$1)
slope from the bulge sample. The result of fitting these two disparate populations simultaneously is a predictive fit which is
(statistically) worse than velocity dispersion.

The $M_{\bullet}(\Phi(nR_e))$ predictor does, however, exhibit several features common to all morphological
samples. First, as the radius decreases, the slope of the predictor increases, but only slightly, and to a value within the
original stated uncertainties. Second, as the radius decreases, the \textit{scatter} decreases,
meaning that the \textit{strength} of the predictor increases. Third, the \textit{robust} fitting slopes are higher than the least-squares
fitting slopes, irreconcilably so for the spiral/lenticular (\textit{8B}) samples. Thus, this predictor is less \textit{stable} than the energy or
velocity dispersion predictors.

The $M_{\bullet}(\Phi({R_e/8}))$ predictor for the elliptical samples is minimally better than the (statistically equivalent) 
$M_{\bullet}(\sigma)$ predictor; however, it is a much less \textit{stable} predictor. The statistical equivalence is expected since
$\Phi^2 \sim \sigma^4$; in essence these two predictors are fundamentally
related. The lack of \textit{stability} is illustrated by several characteristics. First, the \textit{robust} slopes for the \textit{15E} and
\textit{Cap8} samples do not agree with each other. Second, there is a noticeable range in slope encompassed by $a_{min}$ and $a_{max}$ 
for the \textit{lsqfit}
results. Third, when uncertainty in $\Phi({R_e/8})$ is included (Table~\ref{Tab10}), there is often no viable $\epsilon_{yin}$ selection 
to produce $\chi_r^2=1.0$, and higher slopes result when a fit is possible. Finally, the bootstrap fitting (Tables~\ref{Tab11} -~\ref{Tab12}) shows 
discrepancies between the mean and the median slope, particularly for the \textit{Cap8} sample. 

The $M_{\bullet}(\Phi({R_e/8}))$ predictor for the spiral/lenticular bulge sample is not an improvement on $M_{\bullet}(\sigma)$. 
First, based
on the \textit{scatter}, it is minimally worse for the \textit{lsqfit} predictor, and statistically significantly worse for the
\textit{robust} predictor. Second, like the elliptical sample there is no slope \textit{stability} between the \textit{robust} predictor
slope and the \textit{lsqfit} slope. Third, there is a discrepancy between the \textit{robust} bootstrap mean and median slope.
Finally, when large
uncertainty in $\epsilon_x$ is added, the slope of the predictor becomes irreconcilably higher.

The large variations in both the original fits and in the bootstrap results with fitting algorithm and \textit{weightcent} selection,
combined with
the high slope for ellipticals ($\geq 2.0$) and the low slope for the bulges (0.98-1.37),
as illustrated in Figure~\ref{Fig10}, are not reconcilable
within the estimated uncertainties, and argue against gravitational potential being a strong or \textit{stable} predictor of black hole
mass for most galaxies. It is, at best, an adequate predictor to use for elliptical galaxies \textbf{only}.

The fit residuals plotted against both $\Phi({R_e/8})$ (Figure~\ref{Fig11}: left) and $R_e / 8$ (Figure~\ref{Fig11}: right) 
show neither a clear need for a quadratic, nor the need for an additional parameter, as confirmed
in Appendix~\ref{Ap-B}.
However, the correlation of the \textit{15E}-fit-residuals with $\Phi({R_e/8})$ for the spiral/lenticular galaxies
and the \textit{8B}-fit-residuals with $\Phi({R_e/8})$ for the elliptical galaxies reconfirms the assertion that these
are different populations with irreconcilable slopes. 
The slight \textit{23gal}-fit-residual correlation with $R_e$ for the elliptical galaxies is a result of the `compromise' 
combined (\textit{23gal}) slope; the elliptical-only-fit-residuals (top panel) do not show this correlation.
It is also notable (Appendix~\ref{Ap-B}) that the $M_{\bullet}[\Phi({R_e/8}),R_e]$ multivariate
fit for spiral/lenticular galaxies suggests a property akin to mass may be slightly preferred to potential alone. 

\subsection{Bulge Mass\label{mass}}

The galaxy bulge mass predictor has been found by previous
authors to correlate (roughly) linearly with SMBH mass.
In previous studies it was found that 
$\log{(M_{\bullet}/M_{\odot})} = (8.28 \pm 0.06) + (0.96 \pm 0.07)(\log{M_{bulge}-10.9)}$ where
$M_{bulge}=3R_e\sigma_e^2G^{-1}$, with an rms scatter of 0.25, \citep{Marconi}, and, for
a 90 galaxy sample of both active and inactive galaxies,
$M_{\bullet} \propto M_{bulge}^{0.95 \pm 0.05}$ \citep{McLure}.
Using more detailed mass-modeling 
(either the Jeans equation or alternative dynamical models in the literature), \citet{Haring} found
$\log{(M_{\bullet}/M_{\odot})} = (8.20 \pm 0.10) + (1.12 \pm 0.06)\log{(M_{bulge}/10^{11}M_{\odot}})$,
with a scatter of 0.3 dex,
 using a bisector linear
regression fit (and an error in $M_{bulge}$ of 0.18 dex). When \textit{fitexy}, the algorithm used here,
is applied, they quote a slope of $1.32 \pm 0.17$.

The galaxy mass explored here as a predictor of SMBH mass is the bulge stellar mass enclosed within a sphere of
radius R, $M(R)$, derived from the SB profile (Section~\ref{SB}) and the previously mentioned geometrical and physical (i.e.
no dark halo or disk contribution) assumptions (Section~\ref{DetVals}). As with the gravitational potential, this enclosed mass
is examined both at fixed radii (1 kpc and 10 kpc) and as a function of the galaxy size ($R_e$ and $10R_e$). 
Of these four radii, $10R_e$ is the best substitute for the bulge mass; there should be very little galaxy mass
beyond a few $R_e$, and so \textit{all} of the mass from the bulge should be enclosed by the extreme limit of $10R_e$.

The predictors based on the mass enclosed by the fixed radii (i.e. 1 kpc and 10 kpc) are 
not an improvement on the bulge velocity dispersion predictor.
For the combined galaxy (\textit{23gal}) sample, the $M(1 kpc)$ predictor is statistically worse than $M_{\bullet}(\sigma)$. The $M(1 kpc)$
fit 
for the remaining samples, and the $M(10 kpc)$ fit for all samples, exhibit \textit{scatter} which is statistically equivalent to, but minimally worse than,
the $M_{\bullet}(\sigma)$ predictor (Table~\ref{Tab14}). 
The comparison of the $M(10 kpc)$ and $M(1 kpc)$ predictors indicates that
the $M(10 kpc)$ predictor is \textit{better} (it has minimally lower \textit{scatter} for all but the \textit{15E} sample
and a more stable slope with morphological selection)
than the $M(1 kpc)$ predictor. 
The $M(10 kpc)$ predictor slopes are \textit{stable} (at 0.9) with both morphological sample and with \textit{weightcent} selection.
Given the consistency of this
slope and the fact that statistically it is equivalent to the $M_{\bullet}(\sigma)$ predictor, it is adequate as
a SMBH mass predictor, but it is not a significant improvement over either velocity dispersion or gravitational
binding energy. 

The $M(R_e)$ and $M(10R_e)$ predictive fits probe a more physically uniform region of the galaxies than the fixed-radius fits,
but do not produce a predictor which is an improvement on velocity dispersion.
The \textit{scatter} is never lower than for 
the velocity dispersion predictor. 
The $M(R_e)$ \textit{15E} and \textit{23gal} sample predictors have statistically significantly worse fits, while the
remaining fits are generally statistically equivalent to $M_{\bullet}(\sigma)$.
The $M(10R_e)$ predictor fits always
have lower \textit{scatter} than for the $M(R_e)$ fit. 
Given that the $M(R_e)$ predictor is \textit{weaker} than the $M(10R_e)$ predictor, it will
not be discussed further, other than to note that the slopes are consistent with 0.8 for all morphologies
and \textit{weightcent} selections. 

The slope of the $M(10R_e)$ predictor is very consistent ($\sim$ 0.8), for the \textit{15E}, \textit{Cap8} and
\textit{23gal} fitting samples, with morphology, fitting algorithm and \textit{weightcent} selection. 
The \textit{robust} slopes are slightly lower, 0.77 in comparison with 0.81-0.83, but
well-within the stated uncertainties. 
The bootstrap fitting (Tables~\ref{Tab11} -~\ref{Tab12}) 
further illustrates the \textit{stability} of the 0.8 slope for these three samples.
In general, the fitted relationships are roughly linear and the slopes are consistent with \citet{McLure} and
\citet{Marconi}, within the stated uncertainties, and slightly lower than \citet{Haring}, likely
due to slight differences in the mass determination, sample selection and fitting. 

Although it exhibits a consistent slope, the $M_{\bullet}(M(10R_e)$ predictor is not as \textit{stable} overall
when compared with predictors such as $M_{\bullet}(E_g)$.
First, the \textit{robust} fits are statistically worse than the velocity dispersion predictor
for both the elliptical (\textit{15E}) and the combined (\textit{23gal}) galaxy samples.
Second, 
the addition of $\epsilon_x=0.3$ (100\%) (Table~\ref{Tab10}) results
in a noticeable slope increase of 0.12 for the \textit{23gal} sample
and an increase of 0.26 for the \textit{8B} sample,
both outside of the original stated uncertainties.
Even for the full-elliptical sample (\textit{15E}), the slope is increased by 0.10, at the limits of the
stated uncertainties.
While these may be extreme limits, it exhibits a lack of the \textit{stability} in slope, in comparison with the
gravitational binding energy prediction.
Third, the spiral/lenticular bulge sample (\textit{8B}) has a slightly
higher \textit{lsqfit} slope, 0.88, well within the stated uncertainties, but shows inconsistencies
by exhibiting a noticeably higher slope with the \textit{robust} fitting algorithm (1.23), and
a noticeably lower slope (0.75) with the \textit{lsqfit} bootstrap fitting. It does center on the
more common 0.8 slope for the \textit{robust} bootstrap fitting. 

The fits for 
$M_{\bullet}$ versus $(M(10R_e))$, shown in Figure~\ref{Fig12}, illustrate the similarity in the slope for all of our galaxy
samples. The residuals of the fits (Figure~\ref{Fig13}), plotted against $(M(10R_e))$ and $10R_e$  show neither a strong  
indication for a quadratic, nor for the addition of another parameter for the bulge and combined samples; this is reconfirmed in Appendix~\ref{Ap-B}.
The addition of $R_e$ does improve the \textit{scatter} in the elliptical samples: the best fit is 
$M_{\bullet} \propto (M({10R_e}))^{1.5}R_e^{-1.2}$, an indication that pure elliptical galaxy bulge mass is not the strongest predictor; rather 
a physical quantity closer to potential or binding energy is suggested.

\section{DISCUSSION \& SUMMARY \label{summary}}

The goal of these calculations was to determine (1) whether the addition of a second (or third) host-galaxy-bulge
parameter
to the bulge velocity dispersion predictor of SMBH mass would result in an improved predictor (smaller residuals) 
and (2) if an alternative
predictor (such as gravitational binding energy or gravitational potential) is equivalent to, or superior to, the well-accepted velocity
dispersion predictor.
These calculations were undertaken for four (sub)samples of galaxies, all taken from the \citet{Tremaine} 31-galaxy
sample: a sample of 15 elliptical galaxies (\textit{15E}), a sample of 8 spiral/lenticular bulges (\textit{8B}),
the 23 galaxy combination of these two samples (\textit{23gal}), and finally, a subsample of 8 elliptical galaxies (\textit{Cap8})
with $\Upsilon$ available from \citet{Cappellari-SAURON}. 
The bulge velocity dispersion predictor, using dispersions from \citet{Tremaine}, was evaluated for each 
of these 4 samples to provide a reference point against which the other predictors were compared. 
The remaining predictors 
were based on SB profiles and galaxy parameters taken from the literature, along with the assumption of an oblate
spheroidal geometry with no disk or dark halo contributions. Thus, the comparisons between the predictors
are based on \textbf{identical samples} and the \textbf{same data set and assumptions}, allowing for a 
direct comparison of the predictor strengths and weaknesses.

In answer to the first question posed, no additional parameter, when combined with $\sigma$, was found to produce
a better predictor than $M_{\bullet}(\sigma)$ for the elliptical and combined galaxy samples;
however, the only galaxy parameters examined in a multivariate fit were $I_e$ and $R_e$.
Multivariate fits of $M_{\bullet}(I_e,\sigma)$ and $M_{\bullet}(R_e,\sigma)$ were not better than $M_{\bullet}(\sigma)$
for the \textit{15E} and \textit{23gal} samples, and had negligible coefficients for $I_e$ and $R_e$.
The combination of all three parameters ($M_{\bullet}(I_e,R_e,\sigma)$) for these two samples
yielded the suggestion that an alternative fit depending only on $M_{\bullet}(I_e,R_e)$
was warranted. The $M_{\bullet}(I_e,R_e)$ predictor shows an indication, for ellipticals and the combined sample,
that $M_{\bullet} \propto [I_e^2R_e^3]^x \propto E^x$, where `E' is the binding energy of the galaxy bulge. 

The $M_{\bullet}(\sigma)$ fit was also examined as function of morphology.
While
the predictive-fit slopes were different (the spiral/lenticular sample fit exhibits a lower slope than the 
elliptical samples), they were statistically reconcilable with each other.
However, the $M_{\bullet}(\sigma)$ residuals for the spiral/lenticular galaxies do correlate with $R_e$, suggesting
the necessity of a multivariate fit, a phenomenon which is not suggested for the elliptical galaxies.
\citet{Marconi} likewise identified a correlation between the $M_{\bullet}(\sigma)$ residuals and $R_e$, but did not
differentiate based on galaxy morphology, and attributed the correlation to all galaxies, rather than to just the
spiral/lenticular galaxies which visibly show the correlation.

In answer to the second question posed, an alternative predictor to $M_{\bullet}(\sigma)$, gravitational
binding energy, was found to be as (statistically) good as the $\sigma$-predictor for all samples.
Considering that it is equivalent, even with the simplifying assumptions and data constraints, this may be
an indication that gravitational binding energy is a fundamental predictor.
Using
\citet{Cappellari-SAURON} mass-to-light ratios for available galaxies, the \textit{scatter} becomes noticeably
lower. This implies that if such mass-to-light ratios, or other improvements in the data,
were available for all galaxies, the \textit{scatter} could be further reduced, potentially making this predictor (statistically)
superior to bulge velocity dispersion.
It may further be hypothesized that since the modeled-$E_g$ predictor, with its many simplifying assumptions
and data imperfections is as good as $M_{\bullet}(\sigma)$, that if the \textit{real}, physical-$E_g$ were known
precisely for each galaxy, as the velocity dispersion is known, the $E_g$ predictor would become (statistically)
superior to velocity dispersion. Given that replacing the surrogate-energy ($\Upsilon^2I_e^2R_e^3$) with the more formally calculated
$E_g$ decreased the \textit{scatter} but did not have an impact on the slope of the $E_g$ predictor, it is anticipated
that replacing the modeled-$E_g$ with the \textit{real}-$E_g$ would not alter the
fundamental
relationship (and slope), but only the predictive
\textit{strength}.

The slope of the predictor, $M_{\bullet} \propto E_g^{0.6}$,
is remarkably \textit{stable}. It shows minimal variation with changes in fitting algorithm (least-squares versus robust),
least-square centering of $M_{\bullet}$, least-square weighting selection (both $M_{\bullet}$ and $E_g$),
or method of calculation of $E_g$ (crude [$\Upsilon^2I_e^2R_e^3$] versus formally calculated). Calculating the
energy as $M(10R_e)\sigma^2$ or $\Upsilon{I_e}{R_e}^2{\sigma}^2$ produces the same slope, 0.6, as well. Even
utilizing the $\Upsilon_{bpC}$-values does not affect the slope. The bootstrap
(with both least-squares and robust algorithms) further reinforces the slope \textit{stability}.
There is, however, some variation with morphology. The
bulges (\textit{8B}) have a lower (0.47) least-squares slope [similar to $M(\sigma)$], 
but this is equally likely to reflect a statistical weakness in the sample as a
true morphologically-based difference in the predictive relationship.

In general, the spiral/lenticular bulge (\textit{8B}) sample predictors are inferior to the \textit{15E}
or \textit{23gal} predictors.
First, the sample size is (statistically) significantly smaller; the \textit{8B} sample is
$\sim$ one-half the size of the \textit{15E} sample and $\sim$ one-third the size of the \textit{23gal} sample.
Second,  for the \textit{8B} sample,  in contrast with the \textit{15E} sample, observational-uncertainty-only
weighed least-squares fits give very different results from the unweighted fits, suggesting errors
in $M_{\bullet}$ (or the associated observed uncertainties). 
Eliminating additional
\textit{8B} sample galaxies to alleviate the problem is not feasible as this would further weaken the sample
in a statistical sense. Third, other than the velocity dispersion predictor, \textbf{all} of the predictors
are dependent on a (literature-based) bulge-disk decomposition with the assumption that the galaxy has only two components.
These added structural and geometrical assumptions for the spiral/lenticular galaxies may have a weakening
effect on the predictors; false assumptions or over-simplifications could make a predictor appear weak for the spiral/lenticular bulge
sample, when in truth, it is the structural simplifications which are at fault, not the fundamental predictor. 
In this sense, the assumption-free, directly observable velocity dispersion may be a more ``reliable'' predictor of SMBH
mass than any of the other predictors.
This is similar to the conclusion reached by \citet{McLure}.

The calculations of the gravitational binding energy and the gravitational potential omitted the
contribution from the spiral/lenticular disk.
Examining Figure~\ref{Fig2} (for
the gravitational binding energy) and Figure~\ref{Fig10} (for the gravitational potential), it is
obvious that while the spiral/lenticular bulge (\textit{8B}) and elliptical (\textit{15E}) populations do not have the same
slope (the bulges prefer a slightly shallower slope),
they are occupying the same region of parameter-space; the bulge-population is not offset from
the elliptical-population, and the two predictive relations cross in the center of the
fitting range. 
The inclusion of the spiral/lenticular disk would increase the galaxy mass and energy and likely result in an offset between the elliptical and spiral/lenticular
galaxy populations; however, the formal calculation of the impact of including the disk-contribution is beyond this scope of this paper.

For the observational host galaxy-SMBH correlations in the literature, including the log-linear velocity dispersion \citep{Tremaine}, 
luminosity \citep{Marconi}, and bulge mass \citep{McLure,Marconi,Haring} 
predictive relationships and the log-quadratic velocity dispersion predictor \citep{Wyithe}, the fits obtained here are 
generally consistent with the published results. The log-linear velocity dispersion predictor determined here for the full galaxy
sample exhibits a slightly lower slope than that in \citet{Tremaine}, but is fully consistent within the stated uncertainties and 
with statements in \citet{Tremaine} regarding the slope as a function of sample selection. 
The luminosity was not investigated here, but the luminosity-surrogate ($I_eR_e^2$) predictor was found to have a roughly linear 
relationship with SMBH mass, and to exhibit the same slope as determined in \citet{Marconi} for the luminosity.
The bulge mass, as estimated here by summing all of the visible stellar bulge light under the assumptions of 
an oblate spheroidal geometry for 
the bulge and a constant mass-to-light ratio, is found to correlate roughly linearly with SMBH mass, as also seen in the
literature by \citet{McLure} and \citet{Marconi}. The bulge masses, determined using the Jeans equation and other dynamical models
from the literature, used by \citet{Haring} led them to find a slightly higher slope than we determine here; however,
we do not believe the difference to be significant.
Finally, we find log-quadratic fits which are consistent with the fits determined by \citet{Wyithe}. However, using our
indicator of statistical superiority (the Snedecor F-test instead of the Bayseian analysis employed in \citet{Wyithe}) 
we do not find these fits to be better than the log-linear fits. 

Our findings are also generally consistent with the theoretical results from \citet{Hopkins}, who probed predictors of SMBH mass,
and the SMBH mass fundamental plane, using major galaxy merger simulations. They find, in their simulations, that 
$M_{\bullet} \propto [M_{*}\sigma^2]^{0.71 \pm 0.03}$, with a smaller (simulated) scatter than for their $M_{\bullet}(\sigma)$ or $M_{\bullet}(M_{*})$ 
relations, where $M_{*}$ is the stellar mass. 
This result is consistent, within the stated uncertainties, with our observationally-based determination 
that $M_{\bullet} \propto E_g^{0.62 \pm 0.06}$ for the elliptical galaxy population (\textit{15E}), and confirms a `tilted' 
relationship between supermassive black hole mass and gravitational binding energy. The \citet{Hopkins} simulations also find the existence of a 
supermassive black hole fundamental plane, $M_{\bullet} \propto \sigma^{2.90 \pm 0.38}R_e^{0.54 \pm 0.11}$, and claim this to be
better than any single-variate predictor of SMBH mass, as determined examining $\sigma$, $M_{*}$, $M_{dyn}$, or $R_e$, at
a greater than 3-sigma level. Their stated scatter for this relationship (0.21), however, is identical to that given for their gravitational
binding energy predictor; we estimate the \textit{best} predictor using a comparison of the scatter, and so under our statistical
methodology their gravitational binding energy predictor would be considered equal to this fundamental plane predictor.
In examining our (weaker) spiral/lenticular 
(\textit{8B}) 
sample, we find 
$M_{\bullet} \propto \sigma^{2.90 \pm 0.52}R_e^{0.94 \pm 0.33}$ to be better than $M_{\bullet}(\sigma)$ at a statistically significant level, 
which is consistent with 
the \citet{Hopkins} fundamental plane, within the stated uncertainties.
However, for our sample containing only 
ellipticals (\textit{15E}), we find $M_{\bullet} \propto \sigma^{3.73 \pm 0.71}R_e^{0.05 \pm 0.24}$; this result has a 
substantially different dependence on $R_e$ than that found by \citet{Hopkins}, and we do not identify this as being an improvement
on the $M_{\bullet}(\sigma)$ predictor.
The reasons for this discrepancy, in ascertaining the \textit{best} predictor, 
with \citet{Hopkins} likely stem, in part, from the difference between the modeled/simulated galaxy parameters and the 
observationally-determined galaxy parameters, which inherently possess additional scatter that may obscure a fundamentally 
multivariate relationship.

The main conclusions of this study can be summarized as follows.
\begin{enumerate}
\item {Gravitational binding energy ($M_{\bullet} \propto E_g^{0.6}$) is as strong a predictor of $M_{\bullet}$ as velocity dispersion.
\begin{enumerate}
\item{The \textit{scatter} is statistically equivalent for both $M_{\bullet}(E_g)$ and $M_{\bullet}(\sigma)$.}
\item{The slope is very \textit{stable} and does not vary with data-weighting selections, fitting algorithm applied 
(least-squares or robust), or with which sample galaxies are selected for inclusion in the fitting (bootstrap). 
The slope resulting from the full data set (spiral/lenticular bulges \textit{and} ellipticals) is the same as that
for elliptical galaxies only.} 
\item{Energy is suggested by multivariate $M_{\bullet}(I_e,R_e,\sigma)$ fitting for ellipticals.}
\end{enumerate}}
\item{The spiral/lenticular bulges and elliptical galaxies may be different populations.
\begin{enumerate}
\item{$M_{\bullet}(\sigma)$ residuals correlate with $R_e$ for spiral/lenticular galaxies, but not for ellipticals. This
is in slight contrast to \citet{Marconi} who identified a correlation but did not differentiate based on morphology.}
\item{Multivariate fitting suggests different physical predictors for bulges (a mass-like quantity) and ellipticals
(an energy-like quantity).}
\item{Spiral/lenticular bulges have lower (but statistically reconcilable) slopes for $M_{\bullet}(E_g)$ and $M_{\bullet}(\sigma)$,
and substantially different and irreconcilable slopes for the gravitational potential predictors. Some of these differences \textit{may}
stem from a statistically weaker spiral/lenticular bulge population sample when compared with the elliptical sample.}
\end{enumerate}}
\item{Bulge mass is an adequate predictor of $M_{\bullet}$, with a roughly linear slope, confirming the results of
\citet{McLure}, \citet{Marconi}, and \citet{Haring}. However, the bulge mass predictor is not as \textit{stable} with variations
in fitting algorithm (least squares versus robust) or with data-weighting selections as either $M_{\bullet}(E_g)$ or $M_{\bullet}(\sigma)$.
Furthermore, the multivariate fit for elliptical galaxies suggests a combination of mass and $R_e$, closer to energy, is preferred over mass.}
\item{Gravitational potential is an adequate predictor of $M_{\bullet}$ for elliptical galaxies. However, it provides no new insight
over the comparable $M_{\bullet}(\sigma)$ predictor and it is an inferior predictor for spiral/lenticular bulges when compared
with $M_{\bullet}(E_g)$ and $M_{\bullet}(\sigma)$. Furthermore, the spiral/lenticular bulge and elliptical populations exhibit 
irreconcilably different slopes, and there is a slight indication from multivariate fitting that a mass-like quantity is preferred for
the spiral/lenticular bulges. The
slope is not very \textit{stable} with variations in fitting algorithm (least squares versus robust) or data-weighting selections for 
any of the fitting samples.} 
\item{`Luminosity' ($I_eR_e^2$) is an adequate predictor of $M_{\bullet}$ with a roughly linear slope, as seen in the
literature \citep{Marconi}. It is statistically equivalent to $M_{\bullet} (\sigma)$ for the spiral/lenticular bulge sample, and for the
combined galaxy sample, but it is statistically worse for the elliptical galaxy sample.}
\item{Improved values of the mass-to-light ratio would improve the \textit{scatter} in the $M_{\bullet}(E_g)$ fit.}
\item{There is no \textbf{strong} evidence (see Appendix~\ref{logquad}) for the need for a log-quadratic fitting function,
preferred by \citet{Wyithe}, 
for any of the parameters examined
(velocity dispersion, gravitational binding energy, gravitational potential, or bulge mass).
Using the criteria specified in this paper to select the \textit{best} form of the predictors (the Snedecor F-test),
the log-quadratic predictors
are statistically equivalent to the log-linear predictors. Therefore, the simpler, log-linear predictors are preferred. However,
the form of the log-quadratic predictors are compatible with those in \citet{Wyithe}, and it is possible that with a sample
that includes more high and low mass galaxies, that the necessity for the log-quadratic predictor would become more apparent
using our selection criteria.}
\end{enumerate}

\acknowledgments

We acknowledge grants HST GO 09107.01 and NASA NAG5-8238 which provided support for this research.
We also thank Dr. K. Gebhardt for providing the luminosity deprojection code
and the deprojected luminosity profiles for galaxies included in \citet{Gebhardt-12}
and Dr. C. Siopis for assistance in utilizing the deprojection code and for supplying the
data from NGC 4258. 
Additionally, we thank the anonymous referee for the many helpful suggestions to improve this paper. 
This research has made use of the NASA/IPAC Extragalactic Database (NED) which is operated by the
Jet Propulsion Laboratory, California Institute of Technology, under contract with the National
Aeronautics and Space Administration.

\appendix

\section{SENSITIVITY TO VARIATIONS IN AXIS RATIO, INCLINATION, AND MASS-TO-LIGHT RATIO \label{Ap-A}}

In modeling the galaxy as an oblate spheroid in order to calculate the gravitational binding energy,
the gravitational potential, and the bulge mass, it is crucial to have accurate estimates of the
input parameters for each galaxy (see Section~\ref{HostGal}). In the literature, however, there is often a wide range of published
values for a given galaxy, particularly for the axis ratio, inclination and mass-to-light ratio. In this section, the effects of
having selected the ``wrong'' value for each of these input parameters, is explored. 
In addition, those galaxies with the largest likelihood of having such an incorrect value selection are identified.

The intrinsic axis ratio (p) is used to project the observed major axis surface brightness (mass) profile to a three-dimensional
description of the galaxy (see Sections~\ref{Data} -~\ref{DetVals}).
The intrinsic axis ratio depends on both the observed inclination ($\theta$) of the galaxy and on the observed axis ratio (q).
The effect on the calculated gravitational binding energy of varying each
of these observed quantities over the full range of physically allowable values,
$0.0\leq q,p \leq 1.0$ and $0.0\degr \leq \theta \leq 90.0\degr$,
is illustrated in Figure~\ref{Fig14} for
three elliptical galaxies: 
NGC 221 (the smallest galaxy), NGC 4486 (the largest galaxy and one of the most inclined) and NGC 6251 (an intermediate galaxy).
In this plot $q / q_{base}$ (bottom panel) and $\theta / \theta_{base}$ (top panel), where the `base' value is the 
value determined in Sections~\ref{Data} -~\ref{DetVals}, is plotted against $E_g / {E_g}_{base}$.
Only one of the observed quantities is varied at a time while the other is left fixed to the value given in Section~\ref{Data}; the
two observed quantities are never simultaneously varied.

The inclination ($\theta$) is difficult to determine for the diskless elliptical galaxies, which are usually taken to 
be at, or near, edge on, while for the spiral/lenticular galaxies the disks provide a better constraint, but can
still be affected by assumptions about the disk thickness or the presence of disk warping or non-round disks.
Those galaxies with
a literature-value 10\% larger or smaller than the \textit{best} value used in Section~\ref{Data} are given in Table~\ref{Tab15}.
(It is possible, throughout this section, that there
exists an unidentified literature source which would cite a still more discrepant value, or a discrepant
value for one of the other sample galaxies.)
Based on Table~\ref{Tab15}, $E_g / {E_g}_{base} = 1.1$ for NGC 221 ($\theta=50\degr$), 
$E_g / {E_g}_{base} =0.98$ for NGC 4486 ($\theta=51\degr$; for $\theta=90\degr$ it would be 0.95), and
for NGC 6251 the variation would be negligible. Inspecting Figure~\ref{Fig14}, it is evident that incorrect inclinations,
even those off by a substantial amount, will have a minimal effect on the calculations.

The observed axis ratio (q) also exhibits variation in the literature-quoted values which may reflect
an axis ratio (ellipticity) that is dependent on the isophotal fitting algorithm, a strong color-dependence of the
shape (and ellipticity) of the galaxy, or a physical structure in the galaxy which renders the axis ratio not constant
with radius (in violation of the simplifying modeling assumptions adopted here).
The galaxies for which there is a bulge q-value in the literature differing from the \textit{best} value by 10\%
or more are given in Table~\ref{Tab16}. 
The plots (Figure~\ref{Fig14}: bottom panel) all exhibit a quadratic relationship between the ratio of the axis ratios
($q / q_{base}$) and the ratio of 
energies ($E_g / {E_g}_{base}$). 
Based on Table~\ref{Tab16}, the range of observed $E_g / {E_g}_{base}$ values would be 0.8-1.3 for NGC 221, 
0.9-1.1 for NGC 4486 (the stated lower limit of q=0.65 is unphysical at
an inclination of 42\degr), and 0.97-1.2 for NGC 6251.
Looking at the full range of $E_g / {E_g}_{base}$ for these representative galaxies, the binding energy could physically
vary by an order of magnitude, but \textit{only} in the case of an axis ratio which is in error by a factor of two, which is 
not seen for these galaxies and which is unlikely for any of the galaxies
based on Table~\ref{Tab16}. For the majority of the galaxies, the variations in the energy, based on q, are at most
20-30\%, which is accounted for in the modeling.

Finally, the calculated gravitational binding energy is strongly dependent on the mass-to-light ratio adopted;
$E \propto M^2 \propto \Upsilon^2$, so for an error in $\Upsilon$ such that $\Upsilon=x\Upsilon_{true}$, $E=x^2E_{true}$.
The values quoted in the literature show variations resulting from different estimates of the bulge luminosities,
different methods for dynamically estimating the bulge mass, and possibly from a radial variation in the mass-to-light ratio.
The least-well-determined
$\Upsilon$ values, in the literature, tend to be for the spiral/lenticular bulges. It is possible to obtain an estimate
of $M_{bulge}$ from the rotation curve for these galaxies to produce an alternative $\Upsilon$ estimate,
as was done for three of the galaxies (NGC 2787, 4459 and 4596), but for the majority
of the galaxies, the bulge-only light profile, which is believed to be a source of \textit{scatter} in and of itself, will result in
an $\Upsilon$ which is, at best, a minimal improvement over the virial/dynamical mass determinations from the literature.
In Table~\ref{Tab17}, those galaxies for which there is a mass-to-light ratio in the literature $\geq$ 10\% different from the $\Upsilon_{bpC}$ value for
the mass-to-light ratio given in Table~\ref{Tab2} are listed, and the most extreme mass-to-light ratios located in the literature are
enumerated in order to probe the full range of possible variations which might exist in the computed energy.
The \citet{Cappellari-SAURON} mass-to-light ratios are excluded from consideration in the construction of this table, although they are taken
to be superior and are used for all galaxies where available (see Section~\ref{sigma-other}).
Table~\ref{Tab17} shows that for a few of the galaxies, if the extreme limits to the mass-to-light ratio are adopted, there could be
a substantial variation in the computed gravitational binding energy. However, in including large uncertainties in the energy
during the course of the fitting, it was found that this would not have a substantial effect on the derived $M_{\bullet}(E_g)$ predictor
(see Section~\ref{energy}). 

\section{LOG-QUADRATIC AND MULTIVARIABLE FITTING RESULTS \label{Ap-B}}

\subsection{Log-Quadratic Fits\label{logquad}}

\citep{Wyithe} suggests that both $M_{\bullet}(\sigma)$ and $M_{\bullet}(M_{bulge})$ are not
log-linear functions, but are, in fact, log-quadratic expressions such that 
$\log{M_{\bullet}} = (8.05 \pm 0.06) + (4.2 \pm 0.37)\log{(\sigma_n)} + (1.6 \pm 1.3)[\log{(\sigma_n)}]^2$
(intrinsic scatter $0.275 \pm 0.05$) where $\sigma_n \equiv \sigma / 200 km s^{-1}$  and
$\log{M_{\bullet}} = (8.05 \pm 0.1) + (1.15 \pm 0.18)\log{(M_{bn})} + 
(0.12 \pm 0.14)[\log{(M_{bn})}]^2$ 
(intrinsic scatter $0.41 \pm 0.07$) where $M_{bn} \equiv M_{bulge} / 10^{11}M_{\odot}$. This argument is based in part upon a rigorous statistical analysis
of the data, and in part on the positive log-linear-fit residuals for \textit{both} the smallest and the largest
galaxies in the samples. 

For the sample of galaxies used here, the $M_{\bullet}(\sigma)$ residuals for fitting-sample galaxies, 
when plotted against $\sigma$ (Figure~\ref{Fig3}),
show no indication for a pattern of positive residuals with high- \textit{and} low-$\sigma$ galaxies in either the elliptical
galaxy or in the spiral/lenticular galaxy subsamples. Likewise the fitting-sample galaxy
$M_{\bullet}(E_g)$ residuals vs $E_g$ (Figure~\ref{Fig9}),
the $M_{\bullet}(M(10R_e))$ residuals vs $M(10R_e)$ (Figure~\ref{Fig13}) and the \textit{23gal}-fit
$M_{\bullet}(\Phi({R_e/8}))$ residuals vs $\Phi({R_e/8})$ (Figure~\ref{Fig11}) 
show no evidence for a pattern of larger positive residuals for the high \textit{and} low mass galaxies. (The \textit{8B}-sample-fit
$M_{\bullet}(\Phi({R_e/8}))$ residuals vs $\Phi({R_e/8})$ for the elliptical galaxies
and the \textit{15E}-sample-fit
$M_{\bullet}(\Phi({R_e/8}))$ residuals vs $\Phi({R_e/8})$ for the spiral/lenticular galaxies exhibit 
inadequate fits; this is evidenced by linear correlations, but there is no indication of a quadratic). 
However, for all of these predictors, the three largest elliptical galaxies exhibit positive residuals,
when compared with other sample-galaxies. 

When log-quadratic fits are performed for these four predictors (X= $\sigma$, $E_g$, $\Phi({R_e/8})$, and
$M(10R_e)$), using the multivariate fitting algorithm \textit{lsqfit}, with normalized fitting parameters
such that $X_n = {X /  X_0}$ (see Table~\ref{Tab3}), there is no strong evidence for the necessity
of a log-quadratic expression.
The log-quadratic fits 
are tabulated in Table~\ref{Tab18}, where for the function 
\begin{equation}\label{EB1}
\log{M_{\bullet}} = a_1[\log{X_n}]^2 + a_2\log{X_n} + d
\end{equation}
the best fit coefficients ($a_i$, d) are given for each of the morphologically-separated fitting samples
(\textit{15E}, \textit{Cap8}, \textit{8B}, and \textit{23gal}), along with the requisite intrinsic uncertainty,
the \textit{scatter}, the maximum and minimum $a_1$ coefficient obtained by varying the \textit{weightcent} selections,
and the F-test ratio and its statistical significance (see Section~\ref{Predictors}), both when compared with the log-linear $M_{\bullet}(\sigma)$ predictor
and when compared with the log-linear $M_{\bullet}(X)$ predictor.
For \textit{every} predictor and \textit{every} fitting sample the log-linear and log-quadratic fits are statistically equivalent,
using our statistical indicator, the Snedecor F-test.

For the $M_{\bullet}(\sigma)$ predictor, the coefficient of the quadratic term ($a_1$) varies strongly with morphological
fitting sample and exhibits uncertainties larger than the coefficient itself. For the combined galaxy sample
the best fit is $\log{M_{\bullet}} = (8.13 \pm 0.08) + (3.90 + 0.37)\log{\sigma_n} + (0.97 \pm 1.22)[\log{\sigma_n}]^2$
(intrinsic uncertainty 0.25). These coefficients are consistent with the \citet{Wyithe} result. 

For the $M_{\bullet}(E_g)$ predictor, the coefficient of the quadratic term for all of the fitting samples is
minimal. For the combined sample the best fit is
$\log{M_{\bullet}} = (8.06 \pm 0.09) + (0.61 \pm 0.06)\log{{E_g}_n} + (0.05 \pm 0.04)[\log{{E_g}_n}]^2$
(intrinsic uncertainty 0.27). 
The dominant term remains $M_{\bullet} \propto E_g^{0.6}$, as
previously seen in the log-linear calculations. 

For the $M_{\bullet}(\Phi({R_e/8}))$ predictor,
the coefficient of the quadratic term ($a_1$) shows large variations with morphological sample. 
It varies from $-0.21 \pm 0.4$
for the elliptical (\textit{15E}) sample to $0.25 \pm 1.06$ for the spiral/lenticular (\textit{8B}) sample to
$0.43 \pm 0.48$ for the combined galaxy (\textit{23gal}) sample
The combined galaxy sample fit is 
$\log{M_{\bullet}} = (8.10 \pm 0.12) + (1.71 \pm 0.24)\log{\Phi({R_e/8})_n} + 
(0.43 \pm 0.48)[\log{\Phi({R_e/8})_n}]^2$ (intrinsic uncertainty 0.36). As
with the log-linear expression, this is not a strong predictor for galaxies as a whole because of
the strongly differing elliptical and spiral/lenticular galaxy populations. 

For the $M_{\bullet}(M(10R_e))$ predictor, the coefficient of the quadratic term ($a_1$) is always consistent
with zero. The quadratic and linear ($a_2$) terms are larger for the spiral/lenticular sample than for the
elliptical and combined galaxy samples (0.18 versus 0.08 for $a_1$ and 1.0 versus 0.87-0.89 for $a_2$). 
The combined galaxy sample fit is given by
$\log{M_{\bullet}} = (8.02 \pm 0.09) + (0.87 \pm 0.09)\log{M(10R_e)_n)} + (0.08 \pm 0.08)[\log{M(10R_e)_n)}]^2$ 
(intrinsic uncertainty 0.29). These coefficients are barely consistent, within the limits of the uncertainties,
with the \citet{Wyithe} result, although this may be, in part, because of the slightly different normalization
of the mass. (Unlike in a log-linear fit, selecting a different normalization for the log-quadratic fits
will result in different  coefficients.)

Although none of the predictive fits for any of the fitting samples showed a \textbf{strong} indication 
for the necessity of a log-quadratic, as opposed to a log-linear, fit, neither do any of the predictors refute
that a quadratic does fit the data, and that the fits are compatible with the results quoted by \citet{Wyithe}. 
For all of the predictive fits, the log-quadratic fits were
statistically equivalent to the log-linear fits, using the statistical indicator adopted here, the Snedecor F-test. 
This is a different statistical indicator than the Bayseian techniques employed by \citet{Wyithe}.
The $M_{\bullet}(X)$ residuals versus X illustrated positive residuals for the three largest elliptical
galaxies, but matching positive residuals are not seen for the small number of low mass galaxies included in the our fitting sample.
Two of the smallest dispersion galaxies in \citet{Tremaine}, NGC 4742 and the Milky Way, are not used in our fitting samples,
but are included by \citet{Wyithe}.
It
is possible that with a revised sample including more high and low mass galaxies that the need for a log-quadratic
expression would become more evident using the statistical criteria employed in this paper. 
However, given the current sample, the simpler log-linear expressions are preferred
over the log-quadratic expressions for all predictors.

\subsection{Multivariate Fits \label{multivar}}

\begin{figure}
\includegraphics[scale=0.60]{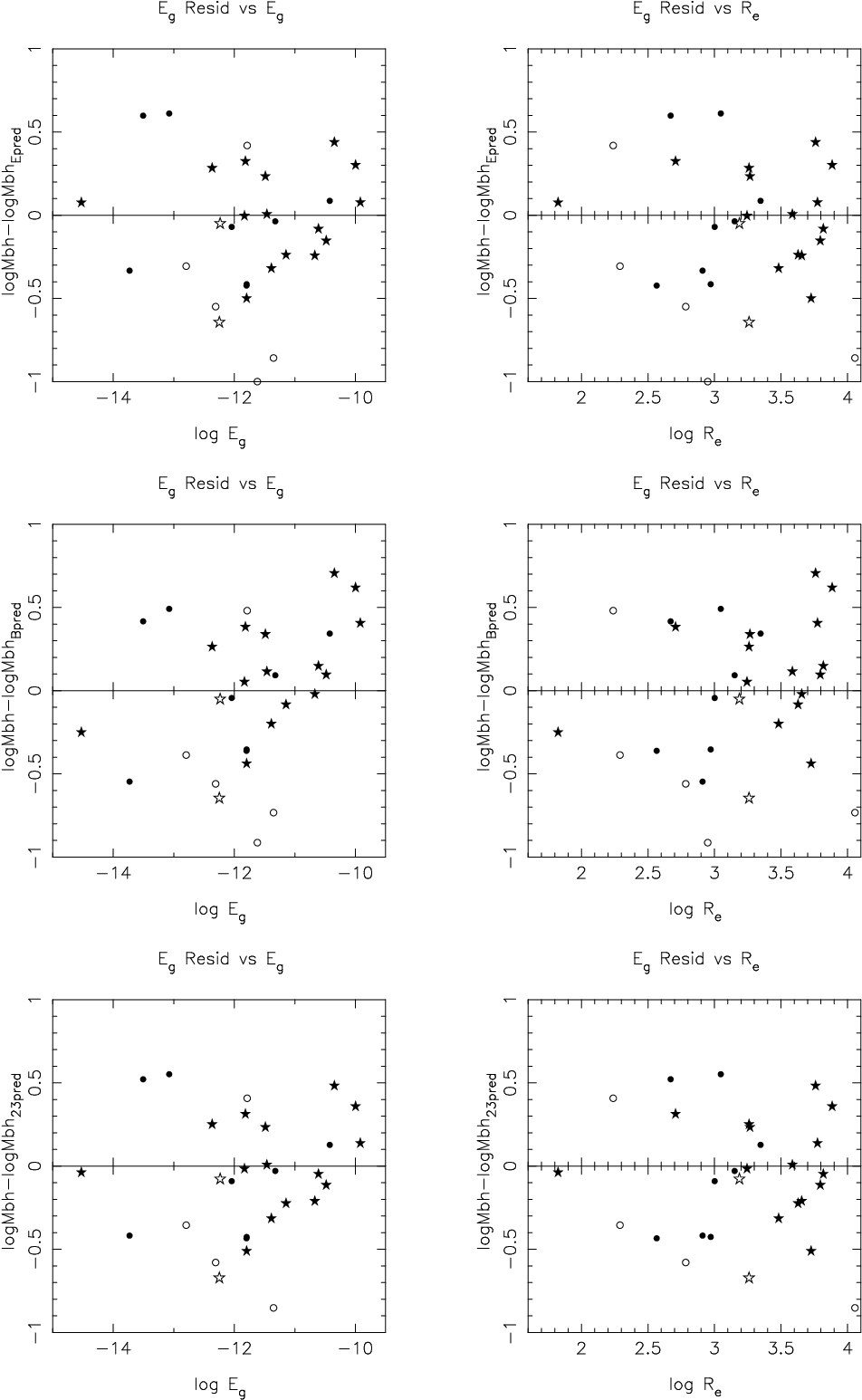}
\caption{Residuals from the gravitational binding energy fits ($M_{\bullet obs}-M_{\bullet pred}$).
The top row shows the \textit{15E} elliptical fit, the middle row the \textit{8B} bulge fit and the bottom row the \textit{23gal} fit.
The left column plots this residual as a function of $E_g$ and the right column plots the residual as a
function of bulge effective radius.
The symbols are as in Figure~\ref{Fig2}.
The galaxies with $|residuals| \geq 0.6$ 
in the top panel (\textit{15E} fit) are 
NGC 1068 (open circle), NGC 2778 (open star), NGC 4258 (open circle), and 
NGC 4596 (\textit{8B}, filled circle), in the middle panel (\textit{8B} fit) are
NGC 1068, NGC 2778, NGC 4258, NGC 4486 (\textit{15E, Cap8}, filled star)
and IC 1459(\textit{15E}, filled star),
and in the bottom panel (\textit{23gal} fit) are
NGC 1068, NGC 2778, and NGC 4258.
\label{Fig9}}
\end{figure}

\begin{figure}
\plotone{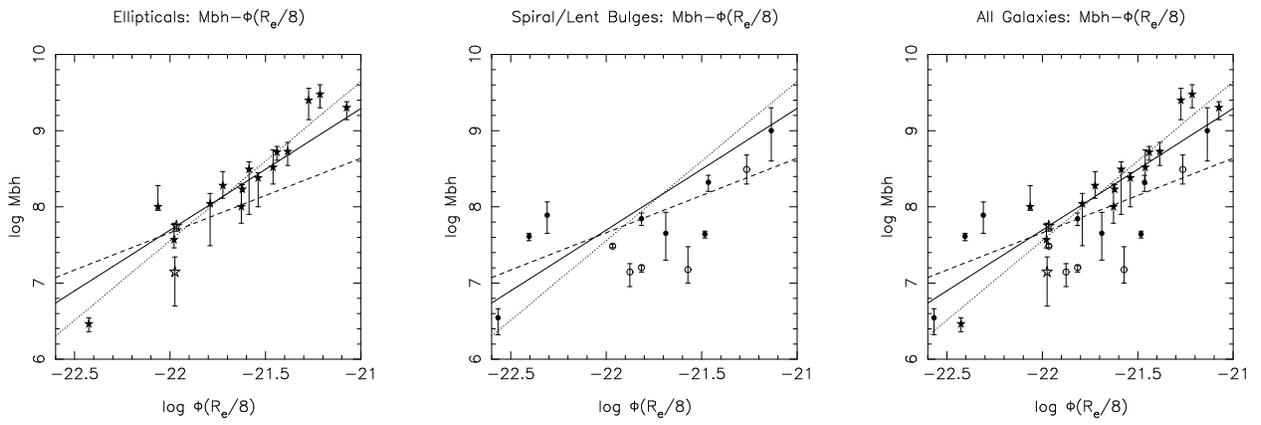}
\caption{Same as Figure~\ref{Fig5}, but for Gravitational Potential at $R_e/8$. \label{Fig10}}
\end{figure}

\begin{figure}
\includegraphics[scale=0.60]{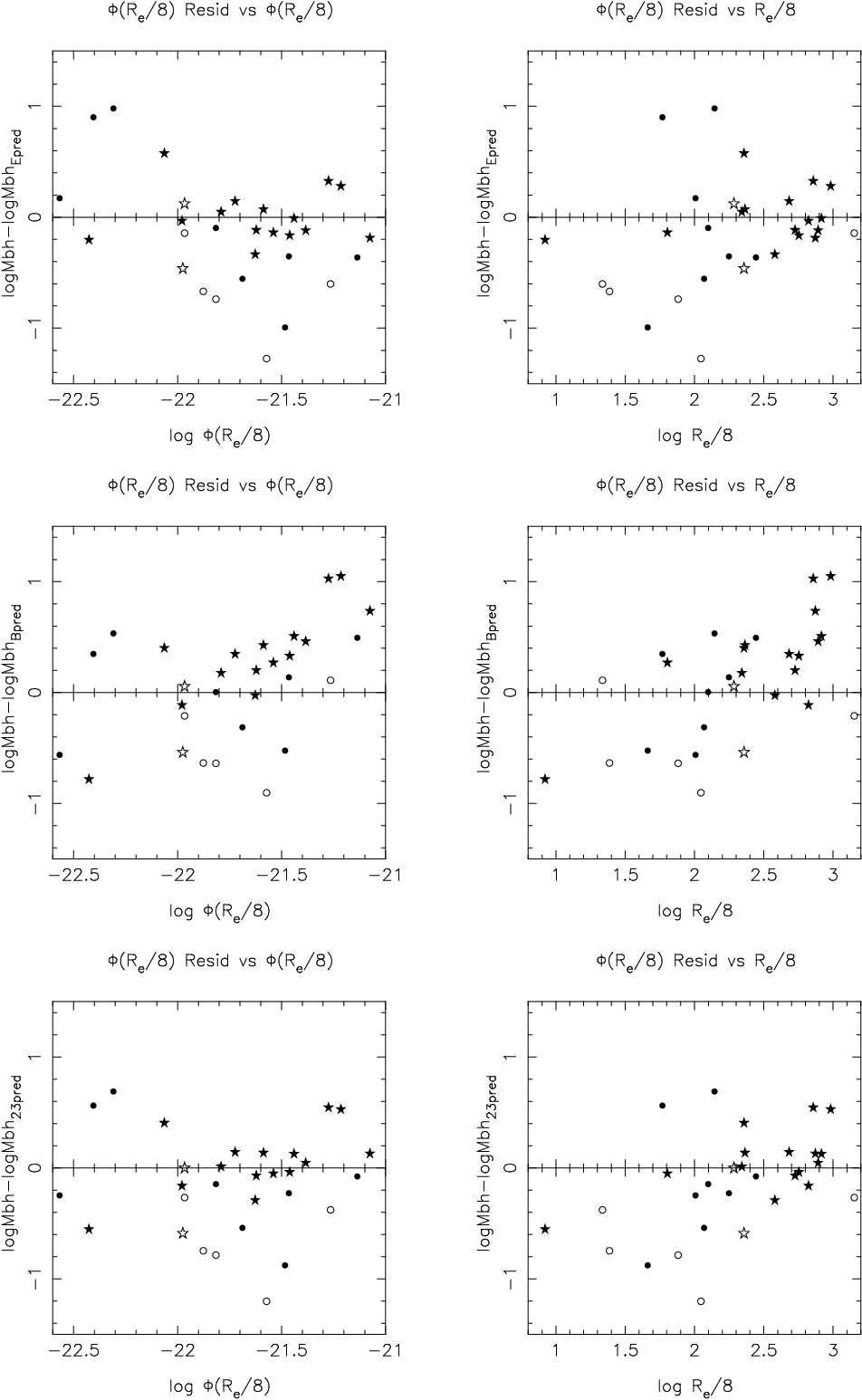}
\caption{Same as Figure~\ref{Fig9}, but for Gravitational Potential at $R_e/8$.
The galaxies with $|residuals| \geq 0.6$ 
in all panels are
NGC 1068 (open circle), NGC 3384 (open circle) and NGC 4742 (open circle).
Additionally, in the top panel (\textit{15E} fit): NGC 1023 (\textit{8B}, filled circle),
NGC 2787 (\textit{8B}, filled circle), NGC 4342 (open circle) and NGC 4596 (\textit{8B}, filled circle),
in the middle panel (\textit{8B} fit): NGC 221 (\textit{15E,Cap8}, filled star), 
NGC 4486 (\textit{15E,Cap8}, filled star), NGC 4649 (\textit{15E}, filled star), 
and IC 1459 (\textit{15E}, filled star) and in the bottom panel (\textit{23gal} fit):
NGC 1023 (\textit{8B}, filled circle) and NGC 4596 (\textit{8B}, filled circle)
are outliers. 
It is notable that for the bulge (\textit{8B}) fit the high-mass and low-mass elliptical galaxies
are all outliers, while for the elliptical fit (\textit{15E}) fit 3 out of the 8 fitted bulges are outliers; this
further illustrates the discrepancy between these two sample fits.
 \label{Fig11}}
\end{figure}

\begin{figure}
\plotone{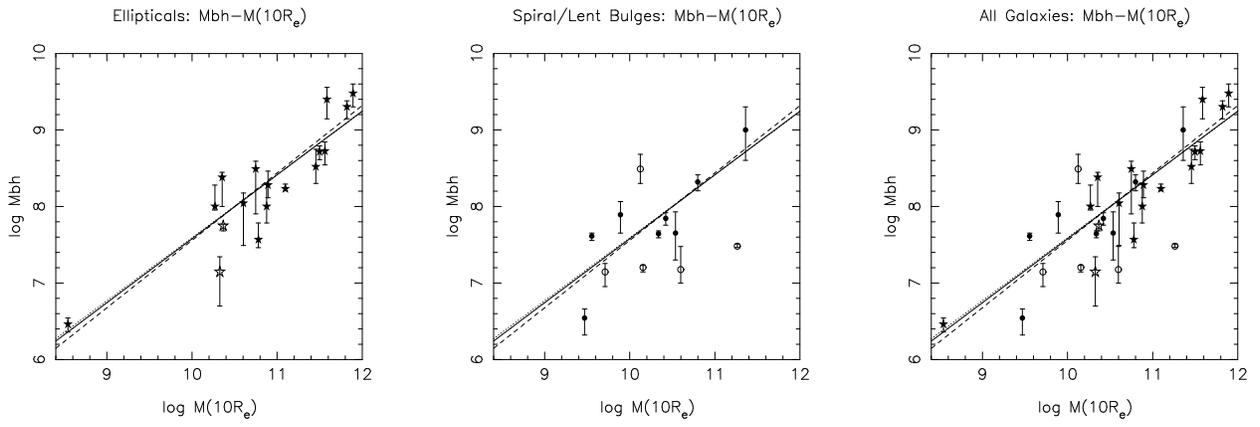}
\caption{Same as Figure~\ref{Fig5}, but for $M(10R_e)$.
\label{Fig12}}
\end{figure}

\begin{figure}
\includegraphics[scale=0.60]{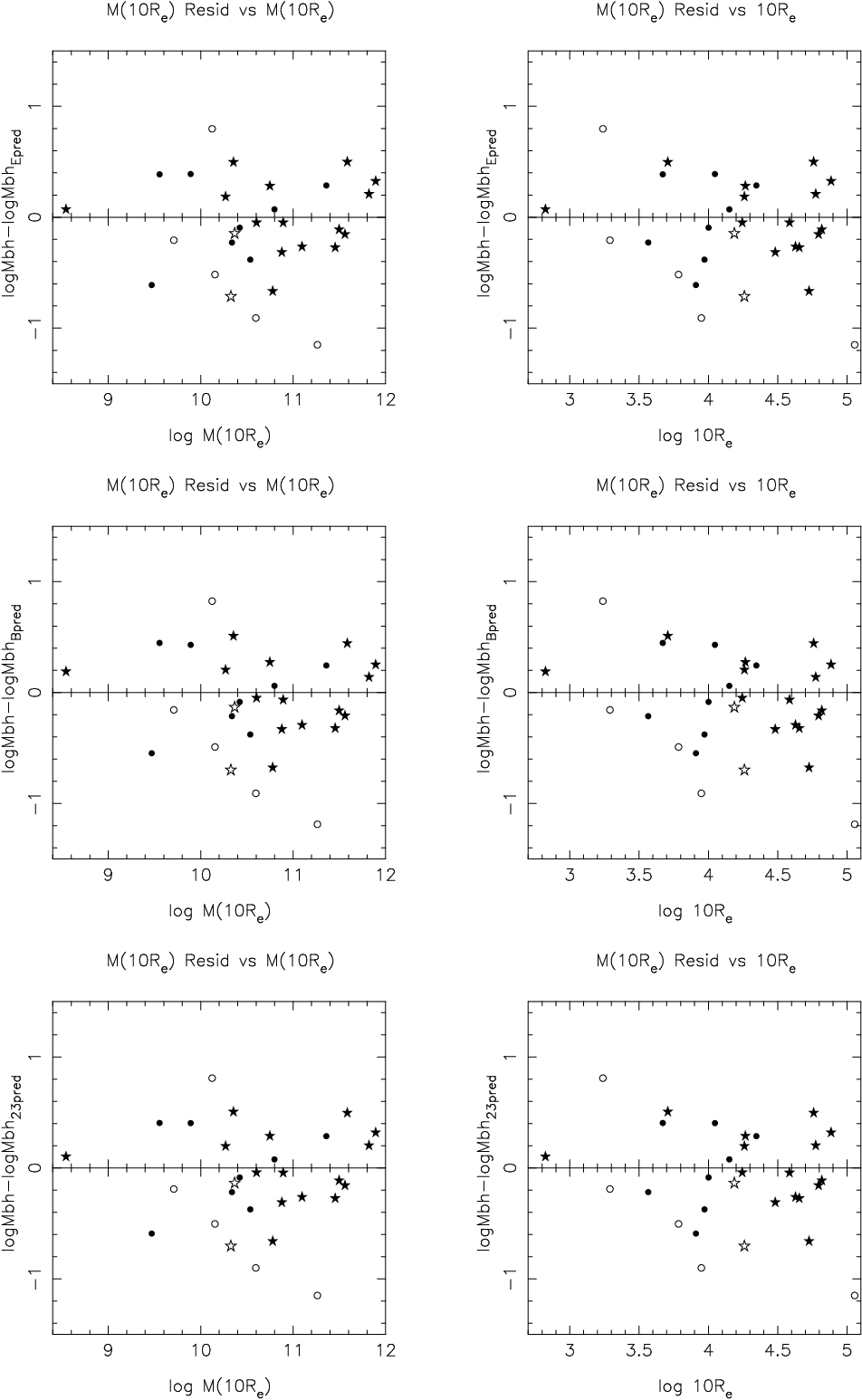}
\caption{Same as Figure~\ref{Fig9}, but for $M(10R_e)$.
The galaxies with $|residuals| \geq 0.6$
in all panels are
NGC 821 (\textit{15E,Cap8}, filled star), NGC 1068 (open circle), NGC 2778 (open star),
NGC 4342 (open circle) and NGC 4258 (open circle). Additionally, in the top panel (\textit{15E} fit)
NGC 7457 (\textit{8B}, filled circle) is an outlier. 
\label{Fig13}}
\end{figure}

\begin{figure}
\plotone{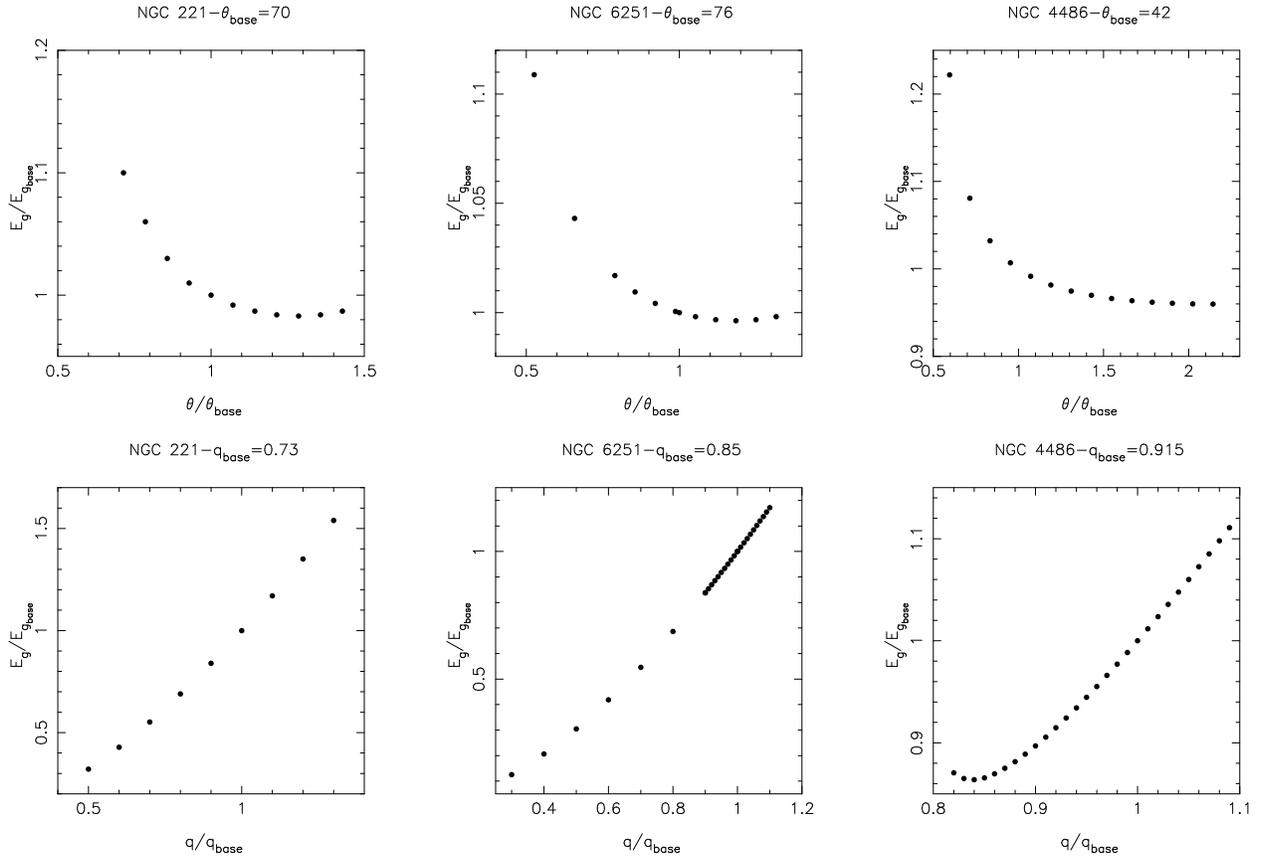}
\caption{
Results on the calculated gravitational binding energy when the inclination angle (top) and observed axis ratio (bottom) are varied
over the range of physically allowed values for three galaxies: NGC 221 (left),
NGC 6251 (middle) and NGC 4486 (right).
\label{Fig14}}
\end{figure}

In addition to considering the log-quadratic expression, it is important to explore the $M_{\bullet}(X,R_e)$ multivariate
predictors for $X = E_g, \Phi({R_e/8})$, and $M(10R_e)$. When this multivariate expression was explored for
$X = \sigma$, it was found that the addition of $R_e$ was an improvement for the spiral/lenticular galaxies (\textit{8B}),
but not for the elliptical galaxy (\textit{15E}) sample. Examining the residuals for the $M_{\bullet}(X)$ residuals versus $R_e$
for the gravitational binding energy (Figure~\ref{Fig9}), the gravitational potential (Figure~\ref{Fig11}) and
the bulge mass (Figure~\ref{Fig13}), there is no evidence for a residual-$R_e$ correlation for the binding energy or
bulge mass predictors. For the $\Phi({R_e/8})$ predictor, there is a hint of a correlation for elliptical
galaxies (\textit{15E}, filled stars) when using the \textit{8B}-fit and the \textit{23gal} fit, but not when using the better-fitting
\textit{15E}-fit. This is an indication of the lower slopes producing poor predictions of $M_{\bullet}$ for elliptical galaxies,
rather than a fundamental correlation.

The tabulated multivariable $M_{\bullet}(X,R_e)$ fits for the gravitational binding energy, gravitational potential and
bulge mass are given in Tables ~\ref{Tab19}, ~\ref{Tab20}, and ~\ref{Tab21} respectively. These tables enumerate the fits
expressed as $\log{(M_{\bullet})} = a_1\log{(X_n)} + a_2\log{({R_e}_n)} + d$, such that for each of the four morphologically based
samples  
(\textit{15E}, \textit{Cap8}, \textit{8B}, and \textit{23gal}), the best-fit coefficients ($a_i$, d) are given
along with the requisite intrinsic uncertainty,
the \textit{scatter}, the maximum and minimum $a_1$ coefficient obtained by varying the \textit{weightcent} selections,
and the F-test ratio and its statistical significance (see Section~\ref{Predictors}), both when compared with the log-linear $M_{\bullet}(\sigma)$ predictor
and when compared with the log-linear $M_{\bullet}(X)$ predictor.

For the $M_{\bullet}(E_g,R_e)$ predictor, the single variable, log-linear $M_{\bullet}(E_g)$ predictor still appears to be the
most reliable. For the elliptical samples, the $M_{\bullet}(E_g,R_e)$ predictor is not statistically significantly better than
the $M_{\bullet}(E_g)$ predictor, with a predicted fit of $M_{\bullet} \propto E_g^{0.88 \pm 0.12}R_e^{-0.62 \pm 0.26}$ (\textit{15E}).
The spiral/lenticular sample (\textit{8B}) suggests a relation of 
$M_{\bullet} \propto E_g^{0.41 \pm 0.17}R_e^{0.59 \pm 0.76}$, which is incompatible with the elliptical fit.
The combined sample, however, still suggests that 
 $M_{\bullet} \propto E_g^{0.6}$ ($M_{\bullet} \propto E_g^{0.57 \pm 0.10}R_e^{0.02 \pm 0.24}$) with little contribution
from $R_e$. Given the non-statistically-significant improvement for the elliptical galaxies when $R_e$ is added and the discrepant
spiral/lenticular and elliptical predictors with the inclusion of $R_e$, the single-variable $M_{\bullet}(E_g)$ predictor will be taken to be superior
to $M_{\bullet}(E_g,R_e)$ for galaxies in general.

The $M_{\bullet}(\Phi({R_e/8}),R_e)$ and $M_{\bullet}(\Phi({R_e}),R_e)$ predictors show a slight, 
but not statistically significant, decrease in the residuals,
when compared with $M_{\bullet}(\Phi({R_e/8}))$ and $M_{\bullet}(\Phi({R_e}))$ respectively.
However, although the \textit{scatter} is slightly
lower, the fits are still irreconcilably different for the elliptical ($M_{\bullet} \propto [\Phi(R_e/8)]^{1.81 \pm 0.26}R_e^{0.22 \pm 0.16}$)
and the spiral/lenticular ($M_{\bullet} \propto [\Phi(R_e/8)]^{0.82 \pm 0.35}R_e^{0.93 \pm 0.72}$) populations. The
combined sample fit of 
$M_{\bullet} \propto [\Phi(R_e/8)]^{1.15 \pm 0.22}R_e^{0.55 \pm 0.17}$ 
fits neither population particularly well. Thus, since there was no substantial improvement to the predictors, nor were the discrepant
spiral/lenticular and elliptical predictors brought into closer functional agreement, the $M_{\bullet}(\Phi(R_e/8))$ predictor
will be preferred over $M_{\bullet}(\Phi({R_e/8}),R_e)$.
However, the spiral/lenticular fit implies 
$M_{\bullet} \propto [\Phi({R_e/8})]^{0.82}R_e^{0.93} \sim M^{0.82}R_e^{0.11} \sim M^{0.82}$. This, in turn, implies that a quantity
akin to mass is a strong predictor of $M_{\bullet}$ for the spiral/lenticular galaxy population.

Finally, the $M_{\bullet}(M(10R_e),R_e)$ predictor shows a significant improvement (lower residuals) over the $M_{\bullet}(M(10R_e))$ predictor
for the elliptical samples. The multivariate fit of $M_{\bullet} \propto [M(10R_e)]^{1.54 \pm 0.23}R_e^{-1.17 \pm 0.35}$ for the \textit{15E}
elliptical sample suggests that a physical quantity similar to energy ($M_{\bullet} \propto M^{1.54}R_e^{-1.17} \approx M^2R_e^{-1} \propto E$)
is a better predictor than pure mass for the elliptical galaxies. For the spiral/lenticular sample, 
$M_{\bullet} \propto [M(10R_e)]^{0.82 \pm 0.33}R_e^{0.26 \pm 0.80}$, the contribution from $R_e$ is minimal and the \textit{scatter}
is (non-statistically-significantly) worse, suggesting that mass alone may be an an adequate predictor of $M_{\bullet}$. 
The ``pure'' mass-predictor for spiral/lenticular galaxies was also suggested by the 
$M_{\bullet}(\Phi({R_e}),R_e)$ predictor.

\clearpage
\begin{deluxetable}{lclclcc}
\tabletypesize{\scriptsize} 
\tablecaption{Surface Brightness Profile Sources \label{Tab1}}
\tablewidth{0pt}
\tablehead{
\colhead{Galaxy}
 & \colhead{Ref} & \colhead{Form} & \colhead{Range} & \colhead{H/G} &  \colhead{Color(Ref)} & \colhead{$R_{corr}$(Ref)}
}
\startdata
NGC 221&1&D&0.02-1.1&H&\nodata&\nodata \\
''&2&D&0.85-116&G&ri=P(2)\tablenotemark{a}&MN:$\epsilon$=P(2) \\
NGC 224&1&D&0.01-10.5&H&\nodata&\nodata\\
''&3&D&40.78-176.9&G&vg=-0.03-0.37gr;gr=P(3)&\nodata\\
''&4&Fd&130-5600&G&\nodata&\nodata\\
NGC 821&5,6&D&0.2-327.9&H,G&\nodata&\nodata\\ 
NGC 1023&7&D&0.2-8.7&H&\nodata&\nodata\\
''&4&Fd&6-284&G&\nodata&\nodata\\
\textit{NGC 1068}&\textit{8}&\textit{D}&\textit{0.7-11}&\textit{G}&\nodata&\nodata\\
''&\textit{4}&\textit{Fd}&\textit{12-280}&\textit{G}&\nodata&\nodata\\
\textit{NGC 2778}&\textit{5,6}&\textit{D}&\textit{0.02-348.5}&\textit{H,G}&\nodata&\nodata\\
NGC 2787&9&F&\phn-18&H&bv=1.06(10)&MN:$\epsilon$=0.23(9) \\
''&4&Fd&19-126&G&bv=1.06(10)&\nodata\\
NGC 3115&7&D&0.2-3.9&H&\nodata&\nodata\\
''&4&Fd&5-360&G&\nodata&\nodata\\
NGC 3245&11&D&0.01-3&H&\nodata\tablenotemark{a}&MN:$\epsilon$=P(11)\\
''&4&Fd&4-110&G&br=1.39,bv=0.87(12)&\nodata\\
NGC 3377&5,6&D&0.02-367.1&H,G&\nodata&\nodata\\
NGC 3379&13&D&0.02-14.6&H&\nodata&\nodata\\
''&14&D&0.4-4.8&G&vr=0.89(18)&\nodata\\
''&15&D&6-170&G&bv=0.96(10)&\nodata\\
''&16&D&8-255&G&bv=0.96(10)&\nodata\\
''&17&D&8.3-159&G&bv=0.96(10)&\nodata\\
''&18&D&8.3-191&G&bg=0.65bv(18),bv=0.96(10)&\nodata\\
''&19&D&8.3-347&G&bv=0.96(10)&MN:c/a=P(19) \\
\textit{NGC 3384}&\textit{20}&\textit{D}&\textit{0.02-14.6}&\textit{H}&\nodata&\nodata\\
''&\textit{4}&\textit{Fd}&\textit{4-184}&\textit{G}&\nodata&\nodata\\
NGC 3608&5,6&D&0.02-321.6&H,G&\nodata&\nodata\\
\textit{NGC 4258}&\textit{21,22}&\textit{D/Fd}&\textit{0.04-149.4}&\textit{H,G}&\nodata&\nodata\\
NGC 4261&14&D&1.3-4.8&G&vr=0.91bv(14),bv=0.99(10)&\nodata\\
''&23&F&0.1-10.25&H&\nodata&\nodata\\
''&15&D&10.6-140&G&vr=0.62(24),bv=1.0(10)&\nodata\\
NGC 4291&5,6&D&0.02-327.9&H,G&\nodata&\nodata\\
\textit{NGC 4342}&\textit{25}&\textit{Fd}&\textit{0.01-5.51}&\textit{H}&\textit{vi=1.3(25,26)}&\textit{MI:$\epsilon$=0.275 (25b)} \\
NGC 4459&27&D&0.3-9.4&G&\nodata&\nodata\\
''&19&D&1.4-10.6&G&\nodata&MN:c/a=P(19) \\
''&4&Fd&10-128&G&bv=0.98(10,28)&\nodata\\
NGC 4473&5,6&D&0.05-355.5&H,G&\nodata&\nodata\\
NGC 4486&29&D&0.02-19.5&H&vi=1.73(29)&\nodata\\
''&27&D&1-983&G&bv=0.96(10)&\nodata\\
''&19&D&1.4-84.7&G&bv=0.96(10)&MN:c/a=P(19) \\
''&17&D&1.6-138&G&bv=0.96(10)&\nodata\\
''&15&D&4.6-163&G&bv=0.96(10)&\nodata\\
\textit{NGC 4564}&\textit{30}&\textit{F}&\textit{0.2-18}&\textit{H}&\textit{bv=0.89,vi=1.14,br=1.44(31)}&\nodata\\
''&\textit{32}&\textit{F}&\textit{15-119}&\textit{G}&\textit{bv=0.89,vi=1.14(31)}&\nodata\\
NGC 4596&4&Fd&3-146&G&bv=0.96(10,28)&\nodata\\
NGC 4649&5,6&D&0.02-381.9&H,G&\nodata&\nodata\\
NGC 4697&5,6&D&0.02-382.2&H,G&\nodata&\nodata\\
\textit{NGC 4742}&\textit{7}&\textit{D}&\textit{0.2-2.4}&\textit{H}&\textit{bv=0.81(10)}&\nodata\\
''&\textit{33}&\textit{Fd}&\textit{2-80}&\textit{G}&\nodata&\nodata\\
NGC 5845&5,6&D&0.02-342.3&H,G&\nodata&\nodata\\
NGC 6251&34&D&0.03-1.1&H&$ub_{FOC} \cong 1.8ub_J+0.4$;ub=P(34),bv=1.0(10b)&MN:$\epsilon$=P,0.112(35)\\
''&35&D&1.5-6.1&G&vr=0.91bv(14);bv=1.0(10b)&MN:$\epsilon$=P(35)\\
''&35&D&2.2-18.4&G&$br_g=0.34+1.57bv$(36);bv=1.0(10b)&MN:$\epsilon$=P(35)\\
NGC 7052&14&D&0.4-2.2&G&vr=0.91bv(14);bv=1.0,vi=1.4(37)&\nodata\\
''&37&D&2.5-15.8&H&\nodata&\nodata \\
''&38&D&6.7-31&G&$br_g=0.34+1.57bv$(36);bv=1.0,vi=1.4(37)&MN:$\epsilon$=P(38)\\
''&39&D&28-119&G&vi=1.4(37)&\nodata\\
NGC 7457&20&D&0.03-5.5&H&\nodata&\nodata\\
''&40&Fd&6-40&G&br=1.287,rk=2.403(41),bv=0.89(10)&MN:$\epsilon=0.35$(40)\\
IC 1459&42&D&0.05-10.6&H&vi=A(43)&MN:$\epsilon$=P,0.263(42) \\
''&44&D&1.5-77&G&br=P(44);bv=0.98(10);vi=A(43)&MN:$\epsilon$=P(44) \\
\enddata
\tablenotetext{a} {For conversion to V for SB profile comparisons the following conversions were used: N221 (vi=1.133 \citep{Tonry}) and
NGC 3245 (br=1.39,bv=0.87 \citep{MP})}
\tablecomments{Description of columns: (1) galaxy name; (2) SB profile reference; (3) the form of the data: F (fit such as
Nuker-law, de Vaucouleurs or Sersic profile), Fd (fit for the bulge portion of a bulge-disk decomposition),
 or D (original data points); (4) radial range (in \arcsec) along major axis;
(5) ground (G) or HST (H) based data; (6) color correction(s) applied to
original SB profile such that a color correction of X-Y will be given as `xy'. For those cases where a single color correction
was used for multiple radii, that is given. For those galaxies where the color correction was a function of radius, the value is
listed only as P here, and for those where there were several values within specified aperture radii the value is listed as A,
 with the source of the correction; (7) radial correction(s) applied to correct the original SB profile to the major
axis listed such that MN implies a correction from the mean axis and MI implies a correction from the minor axis, with the ellipticity ($\epsilon$) or
axial ratio (q) used in the conversion given such that if a single value was used it is stated, or if the ellipticity as a function of radius was
used, this is given as P, with the source of the correction. The galaxies not belonging to the \textit{23gal} sample, i.e. those galaxies not used in the primary fitting
functions, are denoted in italics.}
\tablerefs{(1) \citet{Lauer-M31-2}, (2) \citet{Peletier-32}, (3) \citet{Hoessel} - ``positions with respect to the
 center have been multiplied by a factor of 1.035'' as given by \citet{Walterbos}, (4) \citet{Baggett}, (5) Full profile obtained directly
from K. Gebhardt, (6) \citet{Gebhardt-12}, (7) \citet{LauerHST}, (8) \citet{Schild}, (9) \citet{Peng}, (10) \citet{rc3},
(10b) typical value in \citet{rc3} for other sample galaxies,
(11) taken from Figure 2 of \citet{Barth}, (12) \citet{MP}, (13) \citet{G3379}, (14) \citet{Lauer}, (15) \citet{Peletier}, (16) \citet{BursteinD},
(17) \citet{Davis}, (18) \citet{Kormendy}, (19) \citet{Michard}, (20) \citet{Lauer5}, (21) Full profile obtained directly from
C. Siopis, (22) \citet{Siopis}, (23) \citet{Ferrarese} with 0.5 mag added to match other data, (24) \citet{F4261}, (25) \citet{Scorza},
(25b) \citet{Scorza} Figure 1, (26) \citet{N4342}, (27) \citet{Caon1}, (28) \citet{Sarzi}, (29) \citet{Lauer-M87}, (30) \citet{Rest},
(31) \citet{MichardC}, (32) \citet{CaonCO}, (33) \citet{Vigroux}, (34) \citet{Crane} Figure 23 assuming
$(U-B)_{FOC} \cong 1.8(U-B)_J+0.4$ may be applied to all radii and that $U_{FOC} \sim U_J$,
(35) \citet{Young}, (36) \citet{deJuan,Kent}, (37) \citet{v7052} Figure 3,
(38) \citet{deJuan}, (39) \citet{GonzalezS} Figure 6, (40) \citet{Khor} with 0.79 mag added to match other data, (41) \citet{PB},
(42) \citet{Forbes}, (43) \citet{Carollo}, (44) \citet{Franx}
}
\end{deluxetable}

\clearpage

\begin{deluxetable}{lccccccccccccccl}
\tabletypesize{\scriptsize}
\rotate
\tablecaption{Host Galaxy Parameters Used in Modeling \label{Tab2}}
\tablewidth{0pt}
\tablehead{
\colhead{Galaxy} & \colhead{Type\tablenotemark{a,b}} & \colhead{$M_{\bullet}$(low,high)\tablenotemark{c}} &
\colhead{Ref} & \colhead{D\tablenotemark{d}} & \colhead{$\Upsilon_{bpC}$} & \colhead{Ref} &  \colhead{$\Upsilon_{Cap}\tablenotemark{e}$}
& \colhead{Band} & \colhead{q} & \colhead{Ref} & \colhead{$\theta$} & \colhead{Ref} &
\colhead{$A_e$} & \colhead{Ref} & \colhead{Sample} \\
\colhead{1} & \colhead{2} & \colhead{3} &
\colhead{4} & \colhead{5} & \colhead{6} & \colhead{7} &  \colhead{8}
& \colhead{9} & \colhead{10} & \colhead{11} & \colhead{12} & \colhead{13} &
\colhead{14} & \colhead{15} & \colhead{16}
}
\startdata
NGC 221&cE2-3&2.9E6(2.3,3.5)&1&0.81&1.6&1&1.37&I&0.73&1&70&1&0.078&2&15E,Cap8\\
NGC 224&SA(s)b&4.5E7(2.0,8.5)\tablenotemark{f}&3,4,5,6&0.76&5.0&6&\nodata&V&0.825&5,7&77&3,5,8&1.04&9&8B\\
NGC 821&E6&3.7E7(2.9,6.1)&10&24.1&7.6&10&4.01&V&0.62&11&90&10&6.76&10&15E,Cap8\\
NGC 1023&SB(rs)0-&4.4E7(3.9,4.8)&12&11.4&4.97&12&\nodata&V&0.72&12&90&12&0.434&9&8B\\
\textit{NGC 1068}&\textit{(R)SA(rs)b}&\textit{1.5E7(1.0,3.0)}&\textit{6,13}&\textit{15.0}&\textit{1.21}&\textit{14}&\nodata&\textit{V}&\textit{0.875}&\textit{15}&\textit{38.3}&\textit{16}&\textit{1.13}&\textit{9}&\nodata\\
\textit{NGC 2778}&\textit{E2}&\textit{1.4E7(0.5,2.2)}&\textit{10}&\textit{22.9}&\textit{8.0}&\textit{10}&\nodata&\textit{V}&\textit{0.78}&\textit{11}&\textit{90}&\textit{10}&\textit{2.06}&\textit{10}&\nodata\\
NGC 2787&SB(r)0+&4.1E7(3.6,4.5)&17&7.5&3.21&18&\nodata&B&0.77&19&50&17&0.630&9&8B\\
NGC 3115&S0-&1.0E9(0.4,2.0)&6,20,21&9.7&6.9&6&\nodata&V&0.6&21&86&21,22&2.87&9&8B\\
NGC 3245&SA(r)0&2.1E8(1.6,2.6)&23&20.9&3.74&23&\nodata&R&0.85&23b&63&23&1.58&9&8B\\
NGC 3377&E5-6&1.0E8(0.9,1.9)&10&11.2&2.9&10&2.69&V&0.50&11&90&10&2.57&10&15E,Cap8\\
NGC 3379&E1&1.0E8(0.61,2.0)&24&10.6&4.76&24&4.35&V&0.90&25&90&24&3.20&26&15E,Cap8\\
\textit{NGC 3384}&\textit{SB(s)0-}&\textit{1.6E7(1.4,1.7)}&\textit{10}&\textit{11.6}&\textit{2.5}&\textit{10}&\nodata&\textit{V}&\textit{0.60}&\textit{11}&\textit{90}&\textit{10}&\textit{0.786}&\textit{9}&\nodata\\
NGC 3608&E2&1.9E8(1.3,2.9)&10&22.9&3.7&10&4.66&V&0.82&11&90&10&4.25&10&15E,Cap8\\
\textit{NGC 4258}&\textit{SAB(s)bc}&\textit{3.04E7(2.91,3.24)}&\textit{27}&\textit{7.28}&\textit{3.3}&\textit{27}&\nodata&\textit{V}&\textit{0.65}&\textit{27}&\textit{72}&\textit{27}&\textit{14.65}&\textit{27}&\nodata\\
NGC 4261&E2-3&5.2E8(4.1,6.2)&28&31.6&4.75&28&\nodata&V&0.75&28,29&64&28&8.00&26&15E\\
NGC 4291&E3&3.1E8(0.8,3.9)&10&26.2&5.5&10&\nodata&V&0.76&11&90&10&2.12&10&15E\\
\textit{NGC 4342}&\textit{S0-}&\textit{3.1E8(2.0,4.8)}&\textit{30}&\textit{15.3}&\textit{6.2}&\textit{30}&\nodata&\textit{I}&\textit{0.75}&\textit{31}&\textit{90}&\textit{30}&\textit{0.200}&\textit{2}&\nodata\\
NGC 4459&SA(r)0+&7.0E7(5.7,8.3)&17&16.1&6.3&32&4.22&B&0.85&33&41&17&1.30&9&8B\\
NGC 4473&E5&1.1E8(0.31,1.5)&6&15.7&6.0&10&3.66&V&0.62&11&72&10&2.34&10&15E,Cap8\\
NGC 4486&E+0-1&3.0E9(2.0,4.0)&6,34,35&16.1&4.0&6&8.27&V&0.915&36&42&35&8.59&37&15E,Cap8\\
\textit{NGC 4564}&\textit{E6}&\textit{5.6E7(4.8,5.9)}&\textit{10}&\textit{15.0}&\textit{2.0}&\textit{10}&\nodata&\textit{I}&\textit{0.70}&\textit{11}&\textit{90}&\textit{10}&\textit{1.84}&\textit{10}&\nodata\\
NGC 4596&SB(r)0+&7.8E7(4.5,11.6)&17&16.8&1.2&38&\nodata&B&0.85&39&42&17&1.42&9&8B\\
NGC 4649&E2&2.0E9(1.4,2.4)&10&16.8&8.5&10&\nodata&V&0.88&11&90&10&6.34&10&15E\\
NGC 4697&E6&1.7E8(1.6,1.9)&10&11.7&4.7&10&\nodata&V&0.6&11&90&10&5.49&10&15E\\
\textit{NGC 4742}&\textit{S0}&\textit{1.4E7(0.9,1.8)}&\textit{6,40}&\textit{15.5}&\textit{2.67}&\textit{26}&\nodata&\textit{B}&\textit{0.6}&\textit{41}&\textit{58}&\textit{42}&\textit{0.339}&\textit{43}&\nodata\\
NGC 5845&E3&2.4E8(1.0,2.8)&10&25.9&5.5&10&4.54&V&0.71&11&90&10&.605&10&15E,Cap8\\
NGC 6251&E2&5.3E8(3.5,7.0)&44&93.0&9.7&44&\nodata&V&0.85&45&76&44&6.81&2(\&45)&15E\\
NGC 7052&E4&3.3E8(2.0,5.6)&46&58.7&6.3&6,46&\nodata&I&0.55&47&70&46&6.69&48&15E\\
NGC 7457&SA(rs)0-&3.5E6(2.1,4.6)&10&13.2&3.2&10&2.13&V&0.7&11&90&10&0.972&2&8B\\
IC 1459&E3&2.5E9(1.4,3.6)&49&29.2&3.2&49&\nodata&I&0.725&50&90&49&6.75&49&15E\\
\enddata
\tablenotetext{a} {Galaxy types are from NED for all galaxies except for NGC 2778, NGC 4742, NGC 5845, NGC 6251 and NGC 7052.
The type given
here for those galaxies is from \citet{Gebhardt-12,Tremaine} for NGC 2778 and NGC 5845, from \citet{Tremaine,FF6251} for NGC 6251, and from \citet{Tremaine,v7052} for NGC 7052.
NGC 4742 is classed as S0 in \citet{LauerHST} but as E4 in NED,\citet{Tremaine}.}
\tablenotetext{b} {Galaxy classification in the literature differs from NED as follows: NGC 821 - classified as E4 \citep{Gebhardt-12,Tremaine},
NGC 3115 classed as an E \citep{LauerHST}, NGC 4291 classed as E2 \citep{Gebhardt-12,Tremaine},NGC 4342 noted as E7 \citep{Scorza} ,
NGC 4564 classed as E3 \citep{Gebhardt-12,Tremaine}, NGC 4649 classed as E1 \citep{Gebhardt-12,Tremaine} and NGC 4697 classed as E4 \citep{Gebhardt-12,Tremaine}.}
\tablenotetext{c} {The values in \citet{Tremaine} for the black hole masses (and limits) differ from those here for the following galaxies. The value in \citet{Tremaine}:
NGC 221 (2.5x$10^6$ (2.0,3.0)), NGC 1023 (upper limit of 4.9), NGC 3379 (5.0x$10^7$ and 1.6x$10^8$ as limits),
NGC 4258 (3.9x$10^7$(3.8,4.0) from \citet{Herrnstein}), NGC 4342 (3.0x$10^8$ (2.0,4.7)), NGC 4596 (upper limit of 12.0), IC 1459 (limits (2.1,3.0)).}
\tablenotetext{d} {The distance for each of the galaxies is taken directly from \citet{Tremaine}. (\citet{Tremaine} used
SBF distances \citep{Tonry}, when available, and, when not available, from the redshift assuming a Hubble constant
of 80 $km s^{-1} Mpc^{-1}$.)}
\tablenotetext{e} {The color terms used for conversion to the default wavelength-band are as follows with V-I(vi) colors from \citet{Tonry} and B-V(bv) color from
\citet{Sarzi}: NGC 821 vi=1.196, NGC 3377 vi=1.114, NGC 3379 vi=1.193, NGC 3608 vi=1.156, NGC 4459 vi=1.187, bv=0.98, NGC 4473 vi=1.158, NGC 4486 vi=1.244,
NGC 5845 vi=1.124, \& NGC 7457 vi=1.104. The solar colors were taken from \citet{Allen} such that $bv_{\odot}$=0.650, $ub_{\odot}$=0.195,
$uv_{\odot}$=0.845, $vr_{\odot}$=0.54, $vi_{\odot}$=0.88, $vk_{\odot}$=1.49.}
\tablenotetext{f}{\citet{Bender-M31} give a higher SMBH mass of 1.4x$10^8 M_{\odot}$ (1.1,2.3), based on HST STIS spectroscopy and a revised model for the nucleus of NGC 224. The
implications of using this alternate mass are discussed in Section~\ref{Predictors}.}
\tablecomments{Description of columns: (1) galaxy name; (2) galaxy type from NED; (3) black hole mass (in  $M_{\odot}$) from (4) reference, rescaled
to (5) distance (in Mpc); (6) best-pre-\citet{Cappellari-SAURON} mass-to-light ratio, $\Upsilon_{bpC}$, (in $M_{\odot}/L_{\odot}$), from (7) reference, and
(8) \citet{Cappellari-SAURON} mass-to-light
ratio, $\Upsilon_{Cap}$, (in $M_{\odot}/L_{\odot}$),
both in (9) wavelength-band, at the distance given in col. 5; (10) observed axis ratio from (11) reference; (12) inclination (in \degr) from (13) reference; (14) bulge effective
radius along the major axis (in kpc) from (15) reference; (16) sample
membership denoted as 8B, 15E and/or Cap8. All members of these sub-samples also belong to the \textit{23gal} sample. 
The galaxies not belonging to the \textit{23gal} sample, i.e. those galaxies not used in the primary fitting
functions, are denoted in italics.}
\tablerefs{(1) \citet{N221}, (2) crude $r^{1/4}$ law fit to composite SB profile,
 (3) \citet{Bacon-M31}, (4) \citet{Kormendy-M31}, (5) \citet{Tremaine-M31}, (6) \citet{Tremaine},
(7) \citet{Kent2}, (8) \citet{Walterbos}, (9) \citet{Baggett}, (10) \citet{Gebhardt-12},
(11) private communication with K. Gebhardt, (12) \citet{N1023}, (13) \citet{Greenhill},
(14) \citet{Takamiya} Figure 4 inner regions, (15) average of \citet{rc3} (0.85) and \citet{marcum} (0.90),
(16) \citet{Sanchez}, (17) \citet{Sarzi}, (18)
Calculated using (B) $L_{bulge}=3.25x10^9 L_{\odot}$ \citep{Sarzi} and \citet{BinneyT} eq. 4-80b
($M \sim {{<v^2>r_h}/{(0.4G)}}$) with $\sigma_e=184.9 {km s^{-1}}$ \citep{Sarzi} to obtain
$\Upsilon_B=3.38$. A second calculation wherein $M={{v_{max}^2R_e/G}}$
($v_{max}=v \sin{\theta}$) and using $\theta=50\degr$ \citep{Sarzi} and $v_{max} = 150 {km s^{-1}}$
from Figure 4 of \citet{Sarzi}, resulted in $\Upsilon_B=3.03$. These were averaged to 3.21.,
(19) \citet{Peng}, (20) \citet{N3115}, (21) \citet{N3115old}, (22) \citet{Capaccioli},
(23) \citet{Barth}, (23b) Based on Figure 2 of \citet{Barth}, (24) \citet{G3379}, (25) central value of \citet{Faber,Bender88,Bacon,deV,BursteinD,Peletier,vdm90,vdm91}
values which are 0.89-0.91, (26) \citet{Bacon}, (27) \citet{Siopis}, (28) \citet{F4261},
(29) \citet{Ferrarese}, (30) \citet{N4342}, (31) \citet{Caon,Scorza},
(32) \citet{FaberJ}, (33) \citet{Michard,Caon1} SB profiles,
(34) \citet{M4486}, (35) \citet{H4486}, (36) central value of \citet{Faber,Peletier,Bacon},
(37) \citet{Peletier}, (38)
Calculated, as in (18), using $L_{bulge}=9.61x10^9 L_{\odot}$, $\sigma_e=136.3 km s^{-1}$,
$v_{max}=80 {km s^{-1}}$, and $\theta=42\degr$ \citep{Sarzi}, and obtaining
$\Upsilon_B=1.47$ in the first calculation and $\Upsilon_B=0.91$ for the second, with an average
value of 1.19.,
(39) \citet{Wozniak} SB profile, (40) \citet{Tremaine} (which references the unavailable M.E. Kaiser et al. 2002, in preparation)
(41) based on \citet{LauerHST} SB profile,
(42) using method in \citet{Yasuda} based on \citet{Hubble}
($i=cos^{-1}\sqrt{{{(q_{25}^2-0.2^2)}/{(1-.2^2)}}}$) which here would result in 58\degr  with the
\citet{rc3} value of $q_{25}=0.56$, (43) \citet{Vigroux}, (44) \citet{FF6251},
(45) \citet{Bender88}, (46) \citet{v7052}, (47) \citet{Faber}, (48) \citet{GonzalezS},
(49) \citet{IC1459}, (50) average of \citet{Faber} ESO, FW and \citet{Franx,VK}.}
\end{deluxetable}

\clearpage

\begin{deluxetable}{lccccccccccc}
\rotate
\tabletypesize{\scriptsize}
\tablecaption{Galaxy Parameters Used in the Fits\label{Tab3}}
\tablewidth{0pt}
\tablehead{
\colhead{Galaxy} & \colhead{$\sigma$} & \colhead{$I_{eV}$} &\colhead{$M(10R_e)$} & \colhead{$M(R_e)$} & \colhead{$M(10kpc)$}&
 \colhead{$M(kpc)$}
& \colhead{$\Phi(R_e)$} & \colhead{$\Phi({R_e/8})$}& \colhead{$\Phi(20 pc)$} & \colhead{$\Phi(100pc)$} & \colhead{$E_g$}\\
\colhead{1} & \colhead{2} & \colhead{3} &
\colhead{4} & \colhead{5} & \colhead{6} & \colhead{7} &  \colhead{8}
& \colhead{9} & \colhead{10} & \colhead{11} & \colhead{12}
}
\startdata
NGC 0221&75&2.52E+03&3.48E+08&1.37E+08&3.60E+08&3.55E+08&1.53E-23&3.75E-23&2.72E-23&1.14E-23&2.96E-15\\
NGC 0224&160&4.12E+02&3.44E+10&1.24E+10&3.46E+10&1.31E+10&1.03E-22&2.05E-22&2.38E-22&2.09E-22&1.60E-12\\
NGC 0821&209&2.95E+01&6.02E+10&3.05E+10&4.13E+10&9.25E+09&3.26E-23&1.05E-22&2.39E-22&1.87E-22&1.60E-12\\
NGC 1023&205&1.70E+03&2.18E+10&7.43E+09&2.32E+10&1.49E+10&1.57E-22&3.29E-22&3.74E-22&2.74E-22&1.59E-12\\
\textit{NGC 1068}&\textit{151}&\textit{2.11E+03}&\textit{3.96E+10}&\textit{1.50E+10}&\textit{4.02E+10}&\textit{1.65E+10}&\textit{1.18E-22}&\textit{2.68E-22}&\textit{3.26E-22}&\textit{2.74E-22}&\textit{2.40E-12}\\
\textit{NGC 2778}&\textit{175}&\textit{7.22E+01}&\textit{2.12E+10}&\textit{1.06E+10}&\textit{2.00E+10}&\textit{6.86E+09}&\textit{3.75E-23}&\textit{1.06E-22}&\textit{2.12E-22}&\textit{1.41E-22}&\textit{5.67E-13}\\
NGC 2787&140\tablenotemark{a}&3.87E+02&3.59E+09&1.05E+09&3.88E+09&1.96E+09&1.82E-23&3.94E-23&4.90E-23&3.41E-23&3.11E-14\\
NGC 3115&230&3.18E+02&2.29E+11&9.55E+10&1.95E+11&5.02E+10&2.75E-22&7.32E-22&1.16E-21&9.36E-22&3.79E-11\\
NGC 3245&205&3.99E+02&6.29E+10&2.73E+10&5.99E+10&2.12E+10&1.31E-22&3.43E-22&5.23E-22&4.00E-22&4.75E-12\\
NGC 3377&145&9.49E+01&1.86E+10&9.20E+09&1.74E+10&5.94E+09&2.79E-23&8.64E-23&1.87E-22&1.18E-22&4.30E-13\\
NGC 3379&206&9.12E+01&7.54E+10&3.77E+10&6.45E+10&1.73E+10&8.11E-23&2.36E-22&3.79E-22&3.32E-22&4.10E-12\\
\textit{NGC 3384}&\textit{143}&\textit{7.39E+02}&\textit{1.43E+10}&\textit{5.76E+09}&\textit{1.50E+10}&\textit{7.92E+09}&\textit{6.21E-23}&\textit{1.53E-22}&\textit{1.87E-22}&\textit{1.43E-22}&\textit{4.93E-13}\\
NGC 3608&182&5.00E+01&7.80E+10&3.82E+10&5.93E+10&1.50E+10&6.26E-23&1.89E-22&3.62E-22&2.99E-22&3.45E-12\\
\textit{NGC 4258}&\textit{105}&\textit{2.13E+01}&\textit{1.83E+11}&\textit{7.24E+10}&\textit{6.61E+10}&\textit{6.82E+09}&\textit{4.19E-23}&\textit{1.08E-22}&\textit{2.30E-22}&\textit{1.90E-22}&\textit{4.44E-12}\\
NGC 4261&315&6.39E+01&3.15E+11&1.29E+11&1.63E+11&2.65E+10&1.28E-22&3.63E-22&5.41E-22&5.19E-22&2.46E-11\\
NGC 4291&242&1.76E+02&5.61E+10&2.61E+10&5.14E+10&1.72E+10&9.00E-23&2.59E-22&3.84E-22&3.26E-22&3.26E-12\\
\textit{NGC 4342}&\textit{225}&\textit{2.43E+03}&\textit{1.33E+10}&\textit{5.75E+09}&\textit{1.41E+10}&\textit{1.22E+10}&\textit{2.23E-22}&\textit{5.44E-22}&\textit{5.55E-22}&\textit{3.09E-22}&\textit{1.63E-12}\\
NGC 4459&186&4.33E+02&2.64E+10&9.51E+09&2.63E+10&9.48E+09&6.77E-23&1.53E-22&1.83E-22&1.60E-22&9.04E-13\\
NGC 4473&190&1.55E+02&4.02E+10&1.69E+10&3.65E+10&1.13E+10&5.98E-23&1.62E-22&2.14E-22&1.90E-22&1.47E-12\\
NGC 4486&375&6.36E+01&7.81E+11&2.55E+11&3.04E+11&3.16E+10&2.62E-22&6.10E-22&7.95E-22&7.77E-22&1.01E-10\\
\textit{NGC 4564}&\textit{162}&\textit{2.24E+02}&\textit{2.33E+10}&\textit{9.70E+09}&\textit{2.25E+10}&\textit{6.61E+09}&\textit{4.43E-23}&\textit{1.08E-22}&\textit{1.72E-22}&\textit{1.28E-22}&\textit{5.85E-13}\\
NGC 4596&152&3.28E+02&7.78E+09&3.21E+09&7.67E+09&2.97E+09&1.87E-23&4.91E-23&6.08E-23&5.31E-23&8.41E-14\\
NGC 4649&385&1.04E+02&6.62E+11&2.81E+11&3.95E+11&5.37E+10&3.37E-22&8.42E-22&1.11E-21&1.08E-21&1.21E-10\\
NGC 4697&177&5.75E+01&1.25E+11&6.19E+10&9.26E+10&1.71E+10&8.36E-23&2.39E-22&3.95E-22&3.42E-22&7.17E-12\\
\textit{NGC 4742}&\textit{90}&\textit{3.06E+03}&\textit{5.13E+09}&\textit{1.60E+09}&\textit{5.73E+09}&\textit{4.32E+09}&\textit{5.51E-23}&\textit{1.33E-22}&\textit{1.41E-22}&\textit{8.00E-23}&\textit{1.60E-13}\\
NGC 5845&234&1.12E+03&2.26E+10&9.51E+09&2.33E+10&1.48E+10&1.28E-22&2.89E-22&3.32E-22&2.63E-22&1.53E-12\\
NGC 6251&290&4.73E+01&3.64E+11&1.52E+11&2.07E+11&2.32E+10&1.72E-22&4.13E-22&6.69E-22&5.73E-22&3.31E-11\\
NGC 7052&266&7.00E+01&2.85E+11&8.84E+10&1.61E+11&2.17E+10&1.43E-22&3.46E-22&4.50E-22&4.36E-22&2.14E-11\\
NGC 7457&67&1.02E+02&2.95E+09&1.37E+09&2.99E+09&1.57E+09&1.07E-23&2.71E-23&3.49E-23&2.72E-23&1.86E-14\\
IC 1459&340&1.03E+02&3.85E+11&1.64E+11&2.23E+11&4.16E+10&1.87E-22&5.32E-22&8.91E-22&8.21E-22&4.50E-11\\
\tableline
\tableline
Normalization&200.&200.&4.0E{10}&1.6E{10}&3.6E{10}&1.1E{10}&
8.0E{-23}&2.0E{-22}&3.0E{-22}&2.4E{-22}&2.0E{-12}\\
\enddata
\tablenotetext{a}{The value of 140 $km s^{-1}$ for $\sigma$, the rms stellar velocity dispersion within a slit aperture of length $2R_e$, is taken from \citet{Tremaine}. It is noted that sources
in the literature quote a much higher \textit{central} value, e.g., \citet{Barth-2787} claim 202 $km s^{-1}$, \citet{Neistein} claim 205 $km s^{-1}$, \citet{Sarzi} claim
210 $km s^{-1}$, and \citet{Erwin} claim 257 $km s^{-1}$. This discrepancy between the $\sigma$ value in \citet{Tremaine} and the central value may stem from the central
peak in the velocity dispersion profile, as illustrated in \citet{Bertola}, \citet{Erwin} and \citet{Neistein}. The implications of using a higher velocity dispersion, 200 $km s^{-1}$, for NGC
2787 are discussed in Section~\ref{Predictors}.}
\tablecomments{The values of the bulge parameters used in searching for an alternative predictor of
$M_{\bullet}$ to the $M_{\bullet}(\sigma)$ relationship are presented here for the (1) full sample of
30 galaxies; not all 30 galaxies are used in the fitting as discussed in the following sections.
The (2) effective stellar bulge velocity dispersion in $km s^{-1}$ is taken from \citet{Tremaine} for all galaxies
except NGC 4258 \citep{Siopis}. The remaining parameters are in units of pc, $L_{\odot}$, $M_{\odot}$
and seconds, or appropriate combinations thereof, with the remaining columns as follows:
 (3) V-band $I_e$;
modeled mass enclosed by sphere of radius (4) $10R_e$, (5) $R_e$, (6) 10 kpc, (7) 1kpc; modeled gravitational potential
at (8) $R_e$, (9) ${R_e/8}$, (10) 20 pc, (11) 100 pc; and (12) modeled gravitational binding energy.
The bottom row contains the normalization applied to the data prior to computing the $M_{\bullet}$-predictor
fits. This normalization value is based on the mean or median (in log-space), or a compromise between the
two values. The remaining normalizations not included in the table are:
$R_e$ (1600), $\Upsilon_{best}^2I_e^2R_e^3$ (3.2x$10^{15}$), $I_e^2R_e^3$ (1.8x$10^{14}$),
$I_eR_e^2$ (5.5x$10^8$), and $\Upsilon_{bpC}^2I_e^2R_e^3$ (3.98x$10^{15}$). 
The galaxies not belonging to the \textit{23gal} sample, i.e. those galaxies not used in the primary fitting
functions, are denoted in italics.}
\end{deluxetable}

\clearpage

\begin{deluxetable}{lll}
\tabletypesize{\scriptsize}
\tablecaption{Quick Reference Table of Analysis Abbreviations\label{Tab4}}
\tablewidth{0pt}
\tablehead{
\colhead{Abbrv.} & \colhead{Cat.} & \colhead{Definition} 
}
\startdata
basic eq&E&$Y = \sum_i{a_iX_i} + d$; $Y=\log{M_{\bullet}}$; $X_i=\log{x_i}$; $x_i$=host gal. param. \\
\textit{lsqfit}*&A&multivariate least-squared fitting algorithm; utilizes $\epsilon_y$ \\
\textit{fitexy}&A&linear least squares fitting algorithm; utilizes $\epsilon_y$ and $\epsilon_x$ \citep{Press} \\
\textit{robust}&A&unweighted robust (absolute deviation minimization) fitting algorithm (\citet{Press} \textit{medfit})\\
$\epsilon_y$&WC&uncertainty in Y \\
$\epsilon_x$&WC&uncertainty in $X_1$ (default = 0) \\
\textit{Avgerr}*&WC&$\epsilon_y$ is symmetrized by averaging upper and lower Y-uncertainties\\
\textit{Recent}&WC&Y recentered between maximum and minimum Y (based on Y-uncertainties) \\
$\epsilon_{yin}$&WC&``intrinsic'' uncertainty in Y selected such that $\chi_r^2 = 1$ for fit \\
$\epsilon_{yo}$&WC&``observed'' uncertainty in Y \\
\textit{OBS}&WC&$\epsilon_y = \epsilon_{yo}$ \\
\textit{INT}&WC&$\epsilon_y = \epsilon_{yin}$ \\
\textit{OBS+INT}*&WC&$\epsilon_y = \sqrt{\epsilon_{yo}^2+\epsilon_{yin}^2}$ \\
\textit{weightcent}&WC&Selection of Y-centering (\textit{Avgerr/Recent}) and Y-weighting (\textit{OBS,OBS+INT,INT})\\
N&E&number of host galaxy `data points' in fit \\
DOF&E&degrees of freedom in fit \\
$N_{\chi}$&E&number of `new' data sets in bootstrap with $\chi_r^2 = 1.0$ \\
\textit{23gal}&S&final galaxy sample of 23 elliptical, spiral \& lenticular bulges \\
\textit{15E}&S&15 elliptical galaxies in \textit{23gal} sample \\
\textit{8B}&S&8 spiral/lenticular bulges in \textit{23gal} sample \\
\textit{Cap8}&S&8 galaxy subset of \textit{15E} sample with $Y_{Cap}$ available \\
$\sigma_{fy}$&Q&$\sigma_{fy}^2=DOF^{-1}{\sum_i{\left({Y_i-(\sum_j{a_jX_{ji}} + d)}\right)}^2} $ \\
\textit{NSS}&Q&difference between fits not statistically significant ($<$ 75\%) \\
\textit{75SS}&Q&difference between fits significant at 75-90\% level \\
\textit{90SS}&Q&difference between fits significant at 90-95\% level \\
\textit{95SS}&Q&difference between fits significant at 95-99\% level \\
\textit{99SS}&Q&difference between fits significant at $\geq$ 99\% level \\
`minimal improvement'&Q&\textit{scatter} lower than reference (e.g. $M_{\bullet}(\sigma)$), but not at 75\% probability level \\
\textit{strength}&Q&a \textit{stronger} fit has lower scatter \\
\textit{stable}&Q&a \textit{stable} fit shows little variation with fitting algorithm, \textit{weightcent}, and/or morphology \\
\textit{best} fit&Q& the most \textit{stable} and \textit{strongest} fit\\
\enddata
\tablecomments{This table presents a reference to the abbreviations and codes (column 1) used throughout Sections~\ref{Predictors}-~\ref{summary}. 
The second column lists the general category of the abbreviated
quantity: A=fitting algorithm, E= equation or variables therein, Q = quality of fit, S = fitting sample
and WC=weighting/Y-centering. The final column gives a brief definition for the abbreviation.
Starred parameters denote the default selections.}
\end{deluxetable}

\clearpage

\begin{deluxetable}{llccccccc}
\tabletypesize{\scriptsize}
\tablecaption{Fitting Parameters: $\log{M_{\bullet}} = a\log{\sigma \over \sigma_0}+d$ \label{Tab5}}
\tablewidth{0pt}
\tablehead{
\colhead{Sample} & \colhead{Algorithm} & \colhead{a} & \colhead{d} &
\colhead{$\epsilon_{yin}$} & \colhead{$\epsilon_x$} & \colhead{$\sigma_{fy}$} &
\colhead{$a_{max}$} & \colhead{$a_{min}$}
}
\startdata
15E&lsqfit&3.86$\pm$0.38&8.18$\pm$0.08&0.23&0.000&0.28&3.89$\pm$0.41(AI)&3.63$\pm$0.36(RO)\\
23gal\tablenotemark{a}&lsqfit&3.75$\pm$0.32&8.16$\pm$0.06&0.24&0.000&0.30&3.79$\pm$0.34(AI)&3.71$\pm$0.32(RB)\\
8B\tablenotemark{a}&lsqfit&3.27$\pm$0.77&8.09$\pm$0.14&0.30&0.000&0.38&3.64$\pm$0.80(RI)&3.27$\pm$0.77(AB)\\
Cap8&lsqfit&3.92$\pm$0.65&8.16$\pm$0.14&0.32&0.000&0.36&4.00$\pm$0.49(AO)&3.79$\pm$0.73(RI)\\
\tableline
15E&fitexy&3.90$\pm$0.39&8.18$\pm$0.08&0.22&0.021&0.28&3.94$\pm$0.42(AI)&3.80$\pm$0.41(RB)\\
23gal&fitexy&3.79$\pm$0.32&8.16$\pm$0.06&0.23&0.021&0.30&4.01$\pm$0.18(AO)&3.75$\pm$0.33(RB)\\
8B&fitexy&3.31$\pm$0.79&8.09$\pm$0.14&0.29&0.021&0.38&3.69$\pm$0.81(RI)&3.31$\pm$0.79(AB)\\
Cap8&fitexy&3.96$\pm$0.66&8.16$\pm$0.14&0.31&0.021&0.36&4.07$\pm$0.28(AO)&3.82$\pm$0.74(RI)\\
\tableline
15E&robust&3.99&8.16&0.00&0.00&0.28&\nodata&3.82(R)\\
23gal\tablenotemark{a}&robust&3.70&8.19&0.00&0.00&0.30&\nodata&3.46(R)\\
8B\tablenotemark{a}&robust&3.34&8.13&0.00&0.00&0.37&3.48(R)&\nodata\\
Cap8&robust&3.91&8.13&0.00&0.00&0.36&\nodata&3.62(R)\\
\enddata
\tablenotetext{a}{If the higher $M_{\bullet}$ value of 1.4x$10^8 M_{\odot}$ for NGC 224 \citep{Bender-M31} and the higher velocity dispersion
of 200 $km s^{-1}$ for NGC 2787 (see Table~\ref{Tab3}) are used, then the fits are altered as follows:\\
23gal(lsqfit): a=$3.71\pm0.37$; d=$8.15\pm0.07$; $\epsilon_{yin}=0.29$; $\sigma_{fy}=0.33$; $a_{max}=3.73\pm0.38$(AI); $a_{min}=3.66\pm0.37$(RB)\\
8B(lsqfit): a=$3.10\pm0.96$; d=$8.07\pm0.17$; $\epsilon_{yin}=0.40$; $\sigma_{fy}=0.45$; $a_{max}=3.39\pm0.98$(RI); $a_{min}=3.10\pm0.96$(AB)\\
23gal(robust): a=3.49; d=8.20; $\sigma_{fy}=0.33$; $a_{min}=3.46$(R)\\
8B(robust): a=3.66; d=8.28; $\sigma_{fy}=0.47$; $a_{max}=3.74$(R)}
\tablecomments{The columns are as follows: (1) galaxy sample, (2) fitting algorithm, (3) slope (a), (4) zero-point (d),
(5) intrinsic uncertainty in $Y=\log{M_{\bullet}}$ ($\epsilon_{yin}$) selected such that $\chi_r^2 = 1.0$, 
(6) uncertainty in $X=\log{\sigma}$ ($\epsilon_x$), (7) \textit{scatter} of points around
fit relationship ($\sigma_{fy}$), (8) maximum slope for sample with (\textit{weightcent} method) and (9) minimum
slope for sample with (\textit{weightcent} method). The \textit{weightcent} method designations are
A:Avgerr or  R:Recent and  I=INT, O=OBS, or B =OBS+INT. The \textit{OBS} fit for the \textit{8B}
and \textit{23gal} samples are omitted, for reasons previously discussed.}
\end{deluxetable}

\clearpage

\begin{deluxetable}{lcccccccc}
\tabletypesize{\scriptsize}
\tablecaption{Bootstrap Statistics: $M_{\bullet}(\sigma)$ \label{Tab6}}
\tablewidth{0pt}
\tablehead{
\colhead{Parameter}& \colhead{lsqfit} & \colhead{lsqfit} & \colhead{lsqfit} & \colhead{lsqfit} &
\colhead{robust} & \colhead{robust} & \colhead{robust} & \colhead{robust}
}
\startdata
sample&15E&23gal&8B&Cap8&15E&23gal&8B&Cap8 \\
$N_{\chi}$&991&1000&952&959&1000&1000&1000&1000\\
\tableline
a&&&&&&&\\
\tableline
Mean&3.77&3.81&4.04&3.80&3.73&3.79&4.20&3.69\\
Median&3.84&3.82&3.75&4.03&3.91&3.85&3.66&3.91\\
Adev&0.27&0.20&1.01&0.54&0.46&0.29&1.73&1.22\\
Sdev&0.38&0.27&1.45&0.98&0.65&0.38&3.90&8.12\\
Skew&-0.47&-0.60&1.43&-3.04&-0.79&-0.51&12.74&-8.30\\
Kurt&4.43&3.04&6.41&13.24&1.80&1.89&247.91&453.99\\
\tableline
d&&&&&&&\\
\tableline
Mean&8.22&8.18&8.18&8.20&8.20&8.18&8.21&8.12\\
Median&8.22&8.18&8.20&8.22&8.16&8.16&8.28&8.26\\
Adev&0.06&0.05&0.12&0.11&0.09&0.06&0.21&0.24\\
Sdev&0.08&0.06&0.15&0.13&0.11&0.08&0.44&1.74\\
Skew&-0.54&-0.17&-0.40&-0.79&0.54&0.59&12.27&-19.22\\
Kurt&1.44&0.24&0.13&0.94&0.21&0.60&224.60&388.43\\
\tableline
$\sigma_{fy}$&&&&&&&\\
\tableline
Mean&0.31&0.35&0.40&0.39&0.25&0.28&0.33&0.37\\
Median&0.29&0.35&0.38&0.36&0.24&0.27&0.30&0.30\\
Adev&0.06&0.05&0.10&0.09&0.04&0.04&0.10&0.15\\
Sdev&0.07&0.06&0.12&0.12&0.06&0.05&0.27&1.05\\
Skew&0.95&0.33&0.53&1.29&0.48&0.09&16.77&18.61\\
Kurt&0.76&-0.02&0.43&3.23&0.11&-0.03&339.89&357.76\\
\tableline
$\epsilon_{yin}$&&&&&&&\\
\tableline
Mean&0.25&0.30&0.32&0.32&\nodata&\nodata&\nodata&\nodata\\
Median&0.24&0.30&0.30&0.30&\nodata&\nodata&\nodata&\nodata\\
Adev&0.07&0.06&0.11&0.12&\nodata&\nodata&\nodata&\nodata\\
Sdev&0.09&0.07&0.14&0.15&\nodata&\nodata&\nodata&\nodata\\
Skew&0.65&0.06&0.44&0.77&\nodata&\nodata&\nodata&\nodata\\
Kurt&0.16&-0.10&0.09&1.24&\nodata&\nodata&\nodata&\nodata\\
\enddata
\tablecomments{
The results from a bootstrap (1000 runs) for each galaxy sample for the $M_{\bullet}(\sigma)$ predictor using the
default \textit{weightcent} \textit{lsqfit} and \textit{robust} fitting algorithms are given.
The mean, median, average deviation (Adev), standard deviation (Sdev), skew and kurtosis (Kurt) are given
for the slope (a), zero-point (d), calculated \textit{scatter} ($\sigma_{fy}$) and intrinsic uncertainty in $\log{M_{\bullet}}$ ($\epsilon_{yin}$) needed
to obtain a $\chi_r^2=1.0$ for the least-squares fitting (counting only unique galaxies in the DOF). }
\end{deluxetable}

\clearpage

\begin{deluxetable}{llccccccc}
\tabletypesize{\scriptsize}
\tablecaption{Fitting Parameters: $\log{M_{\bullet}} = a_1\log{I_e \over {I_e}_0}+a_2\log{R_e \over {R_e}_0}+
a_3\log{\sigma \over \sigma_0}+d$ \label{Tab7}}
\tablewidth{0pt}
\tablehead{
\colhead{Sample} & \colhead{$a_1$} &  \colhead{$a_2$} &
 \colhead{$a_3$} &\colhead{d} &
\colhead{$\epsilon_{yin}$} & \colhead{$\sigma_{fy}$} &
\colhead{$F_{\sigma_y}$} & \colhead{Sig} 
}
\startdata
15E&0.00$\pm$0.00&0.05$\pm$0.24&3.73$\pm$0.71&8.18$\pm$0.08&0.25&0.29&0.92&NSS\\
23gal&0.00$\pm$0.00&0.28$\pm$0.16&3.16$\pm$0.46&8.15$\pm$0.06&0.23&0.30&1.01&NSS\\
8B&0.00$\pm$0.00&0.94$\pm$0.33&2.90$\pm$0.52&8.32$\pm$0.14&0.16&0.24&2.42&75SS\\
\tableline
15E&0.03$\pm$0.18&0.00$\pm$0.00&3.90$\pm$0.50&8.19$\pm$0.09&0.25&0.29&0.93&NSS\\
23gal&-0.16$\pm$0.12&0.00$\pm$0.00&3.56$\pm$0.35&8.15$\pm$0.06&0.24&0.30&0.96&NSS\\
8B&-1.02$\pm$0.24&0.00$\pm$0.00&4.68$\pm$0.61&8.54$\pm$0.15&0.10&0.22&3.06&75SS\\
\tableline
15E&-0.71$\pm$0.36&0.00$\pm$0.00&0.00$\pm$0.00&8.20$\pm$0.20&0.68&0.70&0.16&99SS\\
23gal&-0.62$\pm$0.29&0.00$\pm$0.00&0.00$\pm$0.00&8.13$\pm$0.15&0.70&0.72&0.17&99SS\\
8B&0.68$\pm$0.77&0.00$\pm$0.00&0.00$\pm$0.00&7.59$\pm$0.33&0.66&0.71&0.28&90SS\\
\tableline
15E&0.00$\pm$0.00&1.07$\pm$0.25&0.00$\pm$0.00&8.13$\pm$0.15&0.49&0.53&0.29&95SS\\
23gal&0.00$\pm$0.00&1.13$\pm$0.21&0.00$\pm$0.00&8.10$\pm$0.11&0.49&0.53&0.32&99SS\\
8B&0.00$\pm$0.00&1.56$\pm$0.90&0.00$\pm$0.00&8.20$\pm$0.31&0.56&0.59&0.41&75SS\\
\tableline
15E&2.48$\pm$0.88&3.31$\pm$1.16&-0.25$\pm$1.48&8.23$\pm$0.07&0.17&0.24&1.38&NSS\\
23gal&0.77$\pm$0.51&1.32$\pm$0.70&1.91$\pm$0.93&8.13$\pm$0.06&0.22&0.28&1.17&NSS\\
8B&-1.33$\pm$0.68&-0.32$\pm$0.67&5.32$\pm$1.37&8.60$\pm$0.20&0.12&0.24&2.47&75SS\\
\tableline
15E&2.34$\pm$0.32&3.12$\pm$0.29&0.00$\pm$0.00&8.23$\pm$0.07&0.16&0.23&1.50&75SS\\
23gal&1.72$\pm$0.26&2.68$\pm$0.26&0.00$\pm$0.00&8.11$\pm$0.06&0.24&0.29&1.05&NSS\\
8B&1.47$\pm$0.42&2.39$\pm$0.60&0.00$\pm$0.00&7.98$\pm$0.20&0.30&0.34&1.21&NSS\\
\enddata
\tablecomments{Exploration of multivariate fitting parameter combinations.
The columns are as follows: (1) galaxy sample, (2) $a_1$, (3) $a_2$, (4) $a_3$, (5) d,
(6) $\epsilon_{yin}$ selected such that $\chi_r^2=1.0$, (7) \textit{scatter} of points around
fit relationship ($\sigma_{fy}$), (8) F-test ratio based on comparison of
$M_{\bullet}(\sigma)$ and $M_{\bullet}(fit)$ and (9) the F-test probabilistic comparison
that the fits are different: NSS (not significantly different), 75SS (different at 75\% level),
90SS (different at 90\% level), 95SS (different at 95\% level) and 99SS (different at 99\% level).
All fits are determined using the \textit{lsqfit} algorithm with default \textit{weightcent} selections.}
\end{deluxetable}

\clearpage

\begin{deluxetable}{llccccccc}
\tabletypesize{\scriptsize}
\tablecaption{Fitting Parameters: $\log{M_{\bullet}} = a_1\log{X \over X_0}+a_2\log{R_e \over {R_e}_0}+
d$ \label{Tab8}}
\tablewidth{0pt}
\tablehead{
\colhead{X} & \colhead{Sample} & \colhead{$a_1$} &  \colhead{$a_2$} &
\colhead{d} &
\colhead{$\epsilon_{yin}$} & \colhead{$\sigma_{fy}$} &
\colhead{$F_{\sigma_y}$} & \colhead{Sig}
}
\startdata
$I_e^2R_e^3$&15E&0.92$\pm$0.07&0.00$\pm$0.00&8.22$\pm$0.06&0.18&0.25&1.28&NSS\\
$I_e^2R_e^3$&23gal&0.91$\pm$0.08&0.00$\pm$0.00&8.14$\pm$0.06&0.24&0.28&1.10&NSS\\
$I_e^2R_e^3$&8B&0.76$\pm$0.16&0.00$\pm$0.00&7.96$\pm$0.12&0.27&0.32&1.39&NSS\\
$I_e^2R_e^3$&Cap8&0.95$\pm$0.13&0.00$\pm$0.00&8.25$\pm$0.12&0.27&0.32&1.26&NSS\\
\tableline
$\Upsilon_{bpC}^2I_e^2R_e^3$&15E&0.64$\pm$0.10&0.00$\pm$0.00&8.23$\pm$0.10&0.35&0.39&0.52&75SS\\
$\Upsilon_{bpC}^2I_e^2R_e^3$&Cap8&0.65$\pm$0.17&0.00$\pm$0.00&8.24$\pm$0.19&0.48&0.51&0.51&75SS\\
\tableline
$\Upsilon_{best}^2I_e^2R_e^3$&15E&0.63$\pm$0.06&0.00$\pm$0.00&8.18$\pm$0.07&0.22&0.29&0.97&NSS\\
$\Upsilon_{best}^2I_e^2R_e^3$&23gal&0.60$\pm$0.06&0.00$\pm$0.00&8.16$\pm$0.07&0.29&0.34&0.78&NSS\\
$\Upsilon_{best}^2I_e^2R_e^3$&8B&0.47$\pm$0.15&0.00$\pm$0.00&8.03$\pm$0.17&0.39&0.44&0.73&NSS\\
$\Upsilon_{best}^2I_e^2R_e^3$&Cap8&0.65$\pm$0.09&0.00$\pm$0.00&8.23$\pm$0.12&0.26&0.32&1.31&NSS\\
\tableline
$I_eR_e^2$&15E&1.07$\pm$0.14&0.00$\pm$0.00&8.17$\pm$0.09&0.30&0.35&0.65&75SS\\
$I_eR_e^2$&23gal&1.13$\pm$0.11&0.00$\pm$0.00&8.13$\pm$0.07&0.28&0.34&0.79&NSS\\
$I_eR_e^2$&8B&1.24$\pm$0.28&0.00$\pm$0.00&8.12$\pm$0.14&0.28&0.33&1.35&NSS\\
$I_eR_e^2$&Cap8&1.02$\pm$0.22&0.00$\pm$0.00&8.15$\pm$0.16&0.41&0.45&0.65&NSS\\
\tableline
$\Upsilon_{best}^2I_e^2R_e^3$&15E&0.73$\pm$0.13&-0.24$\pm$0.26&8.21$\pm$0.08&0.22&0.29&0.95&NSS\\
$\Upsilon_{best}^2I_e^2R_e^3$&23gal&0.53$\pm$0.10&0.19$\pm$0.22&8.15$\pm$0.07&0.29&0.34&0.76&NSS\\
$\Upsilon_{best}^2I_e^2R_e^3$&8B&0.41$\pm$0.16&0.74$\pm$0.71&8.20$\pm$0.24&0.39&0.43&0.76&NSS\\
$\Upsilon_{best}^2I_e^2R_e^3$&Cap8&0.79$\pm$0.19&-0.29$\pm$0.34&8.27$\pm$0.13&0.27&0.32&1.26&NSS\\
\enddata
\tablecomments{The columns are as follows: (1) fit parameter, (2) galaxy sample, (3) $a_1$, (4) $a_2$,  (5) d,
(6) $\epsilon_{yin}$, (7) $\sigma_{fy}$, (8) the F-test ratio, and its (9) significance as in Table~\ref{Tab7}.
All fits are determined using the \textit{lsqfit} algorithm with default \textit{weightcent}.}
\end{deluxetable}

\clearpage

\begin{deluxetable}{llcccccccc}
\tabletypesize{\scriptsize}
\tablecaption{Fitting Parameters: $\log{M_{\bullet}} = a\log{E_g \over {E_g}_0}+d$ \label{Tab9}}
\tablewidth{0pt}
\tablehead{
\colhead{Sample} & \colhead{Algorithm} & \colhead{a} & \colhead{d} &
\colhead{$\epsilon_{yin}$} &
\colhead{$\sigma_{fy}$} &
\colhead{$a_{max}$} & \colhead{$a_{min}$} &
\colhead{$F_{\sigma_y}$} & \colhead{Sig}
}
\startdata
15E&lsqfit&0.62$\pm$0.06&8.13$\pm$0.07&0.21&0.28&0.62$\pm$0.07(AI)&0.59$\pm$0.04(RO)&1.01&NSS\\
23gal\tablenotemark{a}&lsqfit&0.58$\pm$0.05&8.13$\pm$0.07&0.28&0.32&0.59$\pm$0.06(AI)&0.57$\pm$0.05(RB)&0.84&NSS\\
8B\tablenotemark{a}&lsqfit&0.47$\pm$0.15&8.05$\pm$0.17&0.38&0.43&0.50$\pm$0.14(AI)&0.47$\pm$0.15(AB)&0.78&NSS\\
Cap8&lsqfit&0.62$\pm$0.09&8.16$\pm$0.12&0.27&0.32&0.62$\pm$0.09(AB)&0.61$\pm$0.09(RI)&1.24&NSS\\
\tableline
15E&robust&0.59&8.14&0.0&0.28&\nodata&0.59(R)&1.0&NSS\\
23gal\tablenotemark{a}&robust&0.59&8.12&0.0&0.33&\nodata&0.58(R)&0.83&NSS\\
8B\tablenotemark{a}&robust&0.66&8.07&0.0&0.48&\nodata&0.65(R)&0.59&NSS\\
Cap8&robust&0.59&8.14&0.0&0.33&0.60(R)&\nodata&1.20&NSS\\
\enddata
\tablenotetext{a}{If the higher $M_{\bullet}$ value of 1.4x$10^8 M_{\odot}$ for NGC 224 \citep{Bender-M31} and the higher velocity dispersion
of 200 $km s^{-1}$ for NGC 2787 (see Table~\ref{Tab3}) are used, then the fits are altered as follows:\\
23gal(lsqfit): a=$0.58\pm0.05$; d=$8.15\pm0.07$; $\epsilon_{yin}=0.27$; $\sigma_{fy}=0.31$; $a_{max}=0.58\pm0.06$(AI); $a_{min}=0.57\pm0.05$(RB)\\
8B(lsqfit): a=$0.49\pm0.14$; d=$8.11\pm0.16$; $\epsilon_{yin}=0.37$; $\sigma_{fy}=0.40$; $a_{max}=0.53\pm0.13$(RI); $a_{min}=0.49\pm0.14$(AB)\\
23gal(robust): a=0.59; d=8.14; $\sigma_{fy}=0.31$; $a_{min}=0.59$(R)\\
8B(robust): a=0.71; d=8.09; $\sigma_{fy}=0.49$; $a_{min}=0.69$(R)}
\tablecomments{The columns are as follows: (1) galaxy sample, (2) fitting algorithm, (3) slope (a), (4) zero-point (d),
(5) $\epsilon_{yin}$, (6) $\sigma_{fy}$, (7)
maximum slope for sample with (\textit{weightcent}), (8) minimum
slope for sample with (\textit{weightcent}), (9) F-test ratio and its (10) significance specified as in Table~\ref{Tab7}.}
\end{deluxetable}

\clearpage

\begin{deluxetable}{lcccccc}
\tabletypesize{\scriptsize}
\tablecaption{Fitting Parameters: $\log{M_{\bullet}} \propto a\log{X \over X_0}$\label{Tab10}}
\tablewidth{0pt}
\tablehead{
\colhead{Sample}
&\colhead{} & \colhead{$\epsilon_x=0.1$} & \colhead{}
&\colhead{} & \colhead{$\epsilon_x=0.3$} & \colhead{}\\
\colhead{} &
\colhead{a} & \colhead{$\epsilon_{yin}$}
& \colhead{$\sigma_{fy}$} &
\colhead{a} & \colhead{$\epsilon_{yin}$}
& \colhead{$\sigma_{fy}$}
}
\startdata
\tableline
X=$E_g$&&&&&&\\
\tableline
15E&0.62(0.62,0.60)&0.20&0.28&0.65(0.66,0.64)&0.10&0.28\\
23gal&0.58(0.59,0.58)&0.27&0.32&0.61(0.62,0.61)&0.21&0.33\\
8B&0.48(0.50,0.48)&0.38&0.43&0.51(0.53,0.51)&0.35&0.43\\
Cap8&0.63(0.63,0.62)&0.26&0.32&0.65(0.66,0.64)&0.19&0.33\\
\tableline
X=$M({10R_e})$&&&&&&\\
\tableline
15E&0.83(0.84,0.83)&0.27&0.34&0.92(0.94,0.91)&0.11&0.35\\
23gal&0.85(0.86,0.84)&0.27&0.34&0.95(0.97,0.95)&0.11&0.35\\
8B&0.90(0.94,0.90)&0.34&0.40&1.14(1.14,1.14)&0.18&0.42\\
Cap8&0.82(0.83,0.81)&0.34&0.40&0.88(0.91,0.87)&0.24&0.40\\
\tableline
X=$\Phi({R_e/8})$&&&&&&\\
\tableline
15E&\nodata(2.23,2.16)&\nodata&\nodata&\nodata(2.27,2.27)&\nodata&\nodata\\
23gal&1.69(1.69,1.66)&.32&0.39&\nodata(\nodata)&\nodata&\nodata\\
8B&1.01(1.09,1.01)&0.42&0.47&1.45(1.49,1.45)&0.25&0.52 \\
Cap8&2.35(2.37,2.20)&0.11&0.32&\nodata(2.45,2.42)&\nodata&\nodata\\
\enddata
\tablecomments{The columns are as follows: (1) galaxy sample; (2) the slope(max,min), (3) requisite $\epsilon_{yin}$ for
$\chi_r^2=1.0$, (4) \textit{scatter} ($\sigma_{fy}$) for $\epsilon_x=0.1$; the (5) slope(max,min),
(6) $\epsilon_{yin}$, (7) $\sigma_{fy}$ for $\epsilon_x=0.3$.}
\end{deluxetable}

\clearpage

\begin{deluxetable}{lcccccccccccc}
\tabletypesize{\scriptsize}
\tablecaption{Bootstrap Statistics: $E_g$, $\Phi({R_e/8})$, and $M({10R_e})$ with \textit{lsqfit} \label{Tab11}}
\tablewidth{0pt}
\tablehead{
\colhead{Parameter}& \colhead{$E_g$} & \colhead{$E_g$} & \colhead{$E_g$} & \colhead{$E_g$} &
\colhead{$\Phi$} & \colhead{$\Phi$} & \colhead{$\Phi$} & \colhead{$\Phi$} &
\colhead{M} & \colhead{M} & \colhead{M} & \colhead{M}  
}
\startdata
sample&15E&23gal&8B&Cap8&15E&23gal&8B&Cap8&15E&23gal&8B&Cap8\\
$N_{\chi}$&998&1000&984&927&979&1000&983&839&1000&1000&958&958\\
\tableline
a&&&&&&&&&&&\\
\tableline
Mean&0.62&0.57&0.42&0.60&2.03&1.55&0.88&2.13&0.83&0.82&0.75&0.76\\
Median&0.61&0.57&0.42&0.63&2.07&1.54&0.90&2.21&0.82&0.82&0.74&0.83\\
Adev&0.04&0.05&0.12&0.08&0.20&0.20&0.30&0.25&0.07&0.06&0.21&0.16\\
Sdev&0.06&0.07&0.15&0.15&0.25&0.25&0.38&0.35&0.10&0.08&0.26&0.31\\
Skew&1.12&0.00&0.19&-2.43&-0.47&0.13&0.25&-1.77&0.65&0.15&-0.03&-3.61\\
Kurt&4.45&1.42&0.45&12.14&-0.03&0.05&2.40&5.21&6.77&1.04&0.19&15.51\\
\tableline
d&&&&&&&&&&&\\
\tableline
Mean&8.15&8.15&8.08&8.17&8.21&8.19&8.01&8.28&8.12&8.10&8.09&8.15\\
Median&8.16&8.15&8.09&8.19&8.20&8.20&8.02&8.29&8.12&8.11&8.10&8.16\\
Adev&0.06&0.06&0.10&0.09&0.06&0.07&0.11&0.08&0.07&0.05&0.08&0.10\\
Sdev&0.08&0.07&0.12&0.11&0.08&0.09&0.14&0.10&0.10&0.07&0.10&0.13\\
Skew&-1.03&-0.01&-0.34&-0.88&0.31&-0.06&-0.68&0.19&-1.14&-0.10&-0.59&-0.63\\
Kurt&2.88&0.08&0.08&0.84&-0.45&-0.03&2.39&0.24&4.43&-0.12&1.11&0.88\\
\tableline
$\sigma_{fy}$&&&&&&&&&&&\\
\tableline
Mean&0.32&0.38&0.47&0.38&0.30&0.49&0.51&0.38&0.39&0.39&0.44&0.42\\
Median&0.32&0.38&0.46&0.37&0.29&0.48&0.51&0.38&0.38&0.39&0.43&0.41\\
Adev&0.04&0.04&0.10&0.07&0.05&0.07&0.10&0.08&0.05&0.04&0.09&0.10\\
Sdev&0.05&0.05&0.13&0.09&0.06&0.08&0.13&0.10&0.07&0.05&0.12&0.13\\
Skew&0.56&0.29&0.53&0.88&0.46&0.29&0.36&1.03&0.35&0.34&0.58&0.89\\
Kurt&0.68&0.19&1.50&3.14&0.51&0.29&1.13&3.16&0.32&0.10&0.50&1.19\\
\tableline
$\epsilon_{yin}$&&&&&&&&&&&\\
\tableline
Mean&0.26&0.34&0.41&0.30&0.23&0.45&0.45&0.32&0.33&0.34&0.38&0.36\\
Median&0.26&0.33&0.40&0.31&0.23&0.45&0.45&0.32&0.33&0.33&0.37&0.35\\
Adev&0.05&0.05&0.11&0.09&0.07&0.07&0.11&0.10&0.06&0.05&0.11&0.13\\
Sdev&0.06&0.06&0.15&0.12&0.08&0.09&0.15&0.13&0.08&0.06&0.14&0.16\\
Skew&0.23&0.17&0.41&0.41&-0.03&0.10&0.21&0.18&0.18&0.24&0.35&0.51\\
Kurt&0.38&0.14&0.94&1.31&0.00&0.28&0.77&1.26&0.14&0.10&0.20&0.29\\
\enddata
\tablecomments{
For \textit{lsqfit} (default \textit{weightcent}) the results are given from 1000 bootstrap
`runs' for each galaxy sample when fitting SMBH mass against the gravitational binding energy, gravitational potential at $R_e/8$ and
the bulge mass, $M({10R_e})$. The mean, median, average deviation (Adev), standard deviation (Sdev), skew and kurtosis (Kurt) are given
for the slope (a), zero-point (d), \textit{scatter} ($\sigma_{fy}$) and intrinsic uncertainty ($\epsilon_{yin}$) needed
to obtain a $\chi_r^2=1.0$ for the least-squares fitting (counting only unique galaxies in the DOF). }
\end{deluxetable}

\clearpage

\begin{deluxetable}{lcccccccccccc}
\tabletypesize{\scriptsize}
\tablecaption{Bootstrap Statistics: $E_g$, $\Phi({R_e/8})$, and $M({10R_e})$ with \textit{robust} \label{Tab12}}
\tablewidth{0pt}
\tablehead{
\colhead{Parameter}& \colhead{$E_g$} & \colhead{$E_g$} & \colhead{$E_g$} & \colhead{$E_g$} &
\colhead{$\Phi$} & \colhead{$\Phi$} & \colhead{$\Phi$} & \colhead{$\Phi$} &
\colhead{M} & \colhead{M} & \colhead{M} & \colhead{M}
}
\startdata
sample&15E&23gal&8B&Cap8&15E&23gal&8B&Cap8&15E&23gal&8B&Cap8\\
\tableline
a&&&&&&&&&&&\\
\tableline
Mean&0.60&0.57&0.43&0.68&2.03&1.63&0.79&2.15&0.81&0.80&0.82&0.80\\
Median&0.59&0.58&0.42&0.62&2.13&1.76&0.94&2.48&0.77&0.77&0.77&0.89\\
Adev&0.07&0.08&0.25&0.24&0.30&0.33&0.67&0.55&0.12&0.12&0.46&0.17\\
Sdev&0.10&0.11&0.35&3.06&0.37&0.39&2.59&1.85&0.29&0.16&1.43&0.38\\
Skew&0.85&-0.14&4.26&29.85&-0.48&-0.55&-17.84&-17.92&15.56&0.27&8.72&1.67\\
Kurt&5.16&1.31&54.38&928.65&0.64&-0.64&353.75&538.13&380.59&0.33&373.24&59.03\\
\tableline
d&&&&&&&&&&&\\
\tableline
Mean&8.16&8.14&8.09&8.18&8.18&8.18&7.94&8.29&8.11&8.11&8.10&8.18\\
Median&8.14&8.12&8.07&8.21&8.20&8.15&8.03&8.27&8.05&8.07&8.07&8.20\\
Adev&0.10&0.08&0.20&0.23&0.09&0.09&0.30&0.12&0.14&0.11&0.24&0.15\\
Sdev&0.14&0.11&0.45&1.46&0.11&0.11&1.77&0.86&0.27&0.12&1.18&0.18\\
Skew&-0.95&-0.16&13.75&-9.20&0.28&0.50&-19.13&24.36&-14.74&0.19&0.98&-0.86\\
Kurt&2.72&2.11&260.07&332.33&0.51&-0.63&388.99&657.84&351.03&-0.22&362.95&12.06\\
\tableline
$\sigma_{fy}$&&&&&&&&&&&\\
\tableline
Mean&0.26&0.31&0.40&0.34&0.24&0.39&0.46&0.32&0.32&0.32&0.40&0.32\\
Median&0.26&0.31&0.40&0.29&0.24&0.39&0.43&0.30&0.32&0.32&0.38&0.32\\
Adev&0.04&0.03&0.11&0.13&0.05&0.05&0.15&0.11&0.05&0.03&0.15&0.09\\
Sdev&0.05&0.04&0.20&0.90&0.06&0.07&0.86&0.53&0.11&0.04&0.76&0.11\\
Skew&0.49&0.69&14.34&17.83&0.17&0.03&22.45&23.68&18.86&0.31&20.08&0.37\\
Kurt&0.58&3.34&346.76&320.34&-0.18&-0.08&547.67&632.84&496.01&0.25&424.37&0.30\\
\enddata
\tablecomments{
For \textit{robust} the results are given from 1000 bootstrap
`runs' for each galaxy sample when fitting SMBH mass against the gravitational binding energy, gravitational potential at $R_e/8$ and
the bulge mass, $M({10R_e})$. The mean, median, average deviation (Adev), standard deviation (Sdev), skew and kurtosis (Kurt) are given
for the slope (a), zero-point (d), and \textit{scatter} ($\sigma_{fy}$). }
\end{deluxetable}

\clearpage

\begin{deluxetable}{lllcccccccc}
\tabletypesize{\scriptsize}
\tablecaption{Fitting Parameters: $\log{M_{\bullet}} = a\log{\Phi(X) \over {\Phi(X_0)}}+d$ \label{Tab13}}
\tablewidth{0pt}
\tablehead{
\colhead{$\Phi(X)$} & \colhead{Sample} & \colhead{Algorithm} & \colhead{a} & \colhead{d} &
\colhead{$\epsilon_{yin}$} & \colhead{$\sigma_{fy}$} &
\colhead{$a_{max}$} & \colhead{$a_{min}$} &
\colhead{$F_{\sigma_y}$} & \colhead{Sig}
}
\startdata
$\Phi({20pc})$&15E&lsqfit&1.85$\pm$0.16&8.22$\pm$0.07&0.19&0.26&1.88$\pm$0.18(AI)&1.74$\pm$0.13(RO)&1.22&NSS\\
$\Phi({20pc})$&23gal&lsqfit&1.49$\pm$0.17&8.22$\pm$0.08&0.33&0.37&1.50$\pm$0.17(AI)&1.48$\pm$0.17(RB)&0.66&75SS\\
$\Phi({20pc})$&8B&lsqfit&0.97$\pm$0.32&8.02$\pm$0.18&0.41&0.45&1.04$\pm$0.31(RI)&0.97$\pm$0.32(AB)&0.70&NSS\\
$\Phi({20pc})$&Cap8&lsqfit&1.84$\pm$0.27&8.23$\pm$0.13&0.29&0.34&1.84$\pm$0.27(AB)&1.80$\pm$0.19(RO)&1.14&NSS\\
\tableline
$\Phi({100pc})$&15E&lsqfit&1.48$\pm$0.14&8.24$\pm$0.07&0.22&0.28&1.51$\pm$0.16(AI)&1.38$\pm$0.11(RO)&1.02&NSS\\
$\Phi({100pc})$&23gal&lsqfit&1.34$\pm$0.14&8.21$\pm$0.08&0.31&0.35&1.36$\pm$0.15(AI)&1.33$\pm$0.14(RB)&0.72&75SS\\
$\Phi({100pc})$&8B&lsqfit&0.96$\pm$0.32&8.02$\pm$0.18&0.41&0.45&1.03$\pm$0.31(RI)&0.96$\pm$0.32(AB)&0.71&NSS\\
$\Phi({100pc})$&Cap8&lsqfit&1.44$\pm$0.24&8.24$\pm$0.14&0.32&0.36&1.44$\pm$0.24(AB)&1.41$\pm$0.25(RI)&0.99&NSS\\
\tableline
$\Phi({R_e})$&15E&lsqfit&1.95$\pm$0.20&8.26$\pm$0.08&0.24&0.29&2.09$\pm$0.17(AO)&1.84$\pm$0.23(RI)&0.92&NSS\\
$\Phi({R_e})$&23gal&lsqfit&1.53$\pm$0.21&8.19$\pm$0.09&0.40&0.43&1.53$\pm$0.21(AB)&1.50$\pm$0.22(RI)&0.49&90SS\\
$\Phi({R_e})$&8B&lsqfit&0.94$\pm$0.37&7.92$\pm$0.18&0.46&0.50&1.00$\pm$0.37(RI)&0.94$\pm$0.37(AB)&0.56&75SS\\
$\Phi({R_e})$&Cap8&lsqfit&2.03$\pm$0.34&8.31$\pm$0.14&0.32&0.37&2.30$\pm$0.30(AO)&1.87$\pm$0.41(RI)&0.97&NSS\\
\tableline
$\Phi({R_e/8})$&15E&lsqfit&2.09$\pm$0.17&8.19$\pm$0.06&0.18&0.25&2.14$\pm$0.14(AO)&2.00$\pm$0.21(RI)&1.29&NSS\\
$\Phi({R_e/8})$&23gal&lsqfit&1.60$\pm$0.20&8.18$\pm$0.08&0.36&0.39&1.60$\pm$0.20(AI)&1.57$\pm$0.20(RB)&0.58&75SS\\
$\Phi({R_e/8})$&8B&lsqfit&0.98$\pm$0.35&7.96$\pm$0.18&0.43&0.48&1.05$\pm$0.35(RI)&0.98$\pm$0.35(AB)&0.63&NSS\\
$\Phi({R_e/8})$&Cap8&lsqfit&2.22$\pm$0.30&8.26$\pm$0.12&0.25&0.32&2.37$\pm$0.23(AO)&2.06$\pm$0.38(RI)&1.31&NSS\\
\tableline
$\Phi({R_e/8})$&15E&robust&2.14&8.17&0.00&0.25&2.13(R)&\nodata&1.25&NSS \\
$\Phi({R_e/8})$&23gal&robust&1.79&8.10&0.00&0.41&1.84(R)&\nodata&0.54&90SS \\
$\Phi({R_e/8})$&8B&robust&1.37&8.00&0.00&0.51&\nodata&1.36(R)&0.53&75SS \\
$\Phi({R_e/8})$&Cap8&robust&2.47&8.26&0.00&0.34&\nodata&2.47(R)&1.12&NSS \\
\enddata
\tablecomments{Gravitational Potential at radii `X', as specified. See Table~\ref{Tab9} for a description of columns 2-11.}
\end{deluxetable}

\clearpage

\begin{deluxetable}{lllcccccccc}
\tabletypesize{\scriptsize}
\tablecaption{Fitting Parameters: $\log{M_{\bullet}} = a\log{M(X) \over {M(X_0)}}+d$ \label{Tab14}}
\tablewidth{0pt}
\tablehead{
\colhead{X} & \colhead{Sample} & \colhead{Algorithm} & \colhead{a} & \colhead{d} &
\colhead{$\epsilon_{yin}$} & \colhead{$\sigma_{fy}$} &
\colhead{$a_{max}$} & \colhead{$a_{min}$} &
\colhead{$F_{\sigma_y}$} & \colhead{Sig}
}
\startdata
$M({R_e})$&15E&lsqfit&0.82$\pm$0.11&8.06$\pm$0.10&0.31&0.36&0.83$\pm$0.10(RI)&0.77$\pm$0.08(RO)&0.63&75SS\\
$M({R_e})$&23gal&lsqfit&0.80$\pm$0.12&8.12$\pm$0.09&0.41&0.44&0.81$\pm$0.12(AI)&0.80$\pm$0.11(RB)&0.45&95SS\\
$M({R_e})$&8B&lsqfit&0.83$\pm$0.25&8.09$\pm$0.18&0.38&0.42&0.88$\pm$0.24(AI)&0.83$\pm$0.26(RB)&0.81&NSS\\
$M({R_e})$&Cap8&lsqfit&0.79$\pm$0.16&8.05$\pm$0.16&0.39&0.43&0.80$\pm$0.10(RO)&0.79$\pm$0.17(AI)&0.69&NSS\\
\tableline
$M({10R_e})$&15E&lsqfit&0.82$\pm$0.10&8.09$\pm$0.09&0.28&0.34&0.82$\pm$0.10(RI)&0.79$\pm$0.07(RO)&0.70&NSS\\
$M({10R_e})$&23gal&lsqfit&0.83$\pm$0.08&8.08$\pm$0.07&0.29&0.34&0.85$\pm$0.09(AI)&0.83$\pm$0.08(RB)&0.76&NSS\\
$M({10R_e})$&8B&lsqfit&0.88$\pm$0.25&8.09$\pm$0.17&0.35&0.40&0.93$\pm$0.23(AI)&0.88$\pm$0.25(AB)&0.91&NSS\\
$M({10R_e})$&Cap8&lsqfit&0.81$\pm$0.15&8.09$\pm$0.14&0.35&0.40&0.81$\pm$0.09(RO)&0.80$\pm$0.15(RI)&0.83&NSS\\
\tableline
\tableline
$M({10R_e})$&15E&robust&0.77&8.04&0.00&0.35&\nodata&0.76(R)&0.64&75SS \\
$M({10R_e})$&23gal&robust&0.77&8.05&0.00&0.35&0.78(R)&\nodata&0.73&75SS \\
$M({10R_e})$&8B&robust&1.23&8.07&0.00&0.47&\nodata&1.19(R)&0.62&NSS \\
$M({10R_e})$&Cap8&robust&0.77&8.05&0.00&0.40&0.78(R)&\nodata&0.81&NSS \\
\tableline
\tableline
$M({10kpc})$&15E&lsqfit&0.92$\pm$0.11&8.15$\pm$0.09&0.29&0.34&0.93$\pm$0.12(AI)&0.85$\pm$0.08(RO)&0.69&NSS\\
$M({10kpc})$&23gal&lsqfit&0.94$\pm$0.10&8.12$\pm$0.07&0.30&0.35&0.96$\pm$0.10(AI)&0.93$\pm$0.09(RB)&0.74&NSS\\
$M({10kpc})$&8B&lsqfit&0.91$\pm$0.26&8.05$\pm$0.16&0.36&0.40&0.96$\pm$0.25(AI)&0.91$\pm$0.26(AB)&0.88&NSS\\
$M({10kpc})$&Cap8&lsqfit&0.90$\pm$0.17&8.15$\pm$0.15&0.36&0.41&0.90$\pm$0.17(AB)&0.89$\pm$0.17(RI)&0.78&NSS\\
\tableline
$M({kpc})$&15E&lsqfit&1.35$\pm$0.15&8.24$\pm$0.08&0.27&0.32&1.37$\pm$0.17(AI)&1.23$\pm$0.11(RO)&0.77&NSS\\
$M({kpc})$&23gal&lsqfit&1.29$\pm$0.15&8.16$\pm$0.08&0.34&0.38&1.31$\pm$0.16(AI)&1.27$\pm$0.15(RB)&0.62&75SS\\
$M({kpc})$&8B&lsqfit&0.97$\pm$0.33&7.94$\pm$0.17&0.41&0.46&1.04$\pm$0.32(RI)&0.97$\pm$0.33(AB)&0.69&NSS\\
$M({kpc})$&Cap8&lsqfit&1.26$\pm$0.25&8.20$\pm$0.16&0.38&0.42&1.26$\pm$0.25(AB)&1.23$\pm$0.27(RI)&0.74&NSS\\
\tableline
\enddata
\tablecomments{Mass enclosed by spherical radius X, as specified in column 1. See Table~\ref{Tab9} for a description of columns 2-11.}
\end{deluxetable}

\begin{deluxetable}{lccccccc}
\tabletypesize{\scriptsize}
\tablecaption{Maximum and Minimum Limits for Galaxy Parameters: Inclination \label{Tab15}}
\tablewidth{0pt}
\tablehead{
\colhead{Galaxy}
& \colhead{$\theta$} & \colhead{Ref} & \colhead{$\theta_{MAX}$}
& \colhead{Ref} & \colhead{$\theta_{MIN}$}& \colhead{Ref} & \colhead{MAX/MIN $\theta$}
}
\startdata
NGC 221&70&1&\nodata&\nodata&50&2&\nodata /28.6\% \\
NGC 224&77&3,4,5&78&6&55&5\tablenotemark{a}&1.3\% /28.6\% \\
NGC 1023&90&7&\nodata&\nodata&60.4&8&\nodata /32.9\% \\
NGC 1068&38.3&8&46&9&25&10&20.1\% /34.7\% \\
NGC 3379&90&2,11&\nodata&\nodata&60\tablenotemark{b}&12&\nodata /33.3\% \\
NGC 3384&90&13&\nodata&\nodata&64&14&\nodata /28.9\% \\
NGC 4261&64&15&\nodata&\nodata&30&12&\nodata /53.1\% \\
NGC 4342&90&16&\nodata&\nodata&78\tablenotemark{c}&17&\nodata /13.3\% \\
NGC 4459&41&18&47&2,18b&\nodata&\nodata&14.6\% /\nodata \\
NGC 4486&42&19&51\tablenotemark{d}&20&\nodata&\nodata&21.4\% /\nodata \\
NGC 6251&76&21&\nodata&\nodata&68&22&\nodata /10.5\% \\
NGC 7457&90&13&\nodata&\nodata&64&2&\nodata /28.9\% \\
IC 1459&90&23&\nodata&\nodata&60&24&\nodata /33.3\% \\
\enddata
\tablenotetext{a} {Nuclear Disk}
\tablenotetext{b} {\citet{Cap91} gives a value of 31.5\degr based on estimate of disk.}
\tablenotetext{c} {\citet{Scorza} gives 83\degr  for the nuclear disk.}
\tablenotetext{d} {\citet{Cappellari-SAURON} gives a value of 90\degr.}
\tablecomments{Description of columns: (1) galaxy name;
(2) inclination (in \degr), from (3); (4) maximum inclination (in \degr), from (5); (6) minimum inclination (in \degr), from (7);
and (8) percentage variation
$[100({{\theta_{MAX}-\theta_{OBS}})/{(\theta_{OBS})}}]$/$[100({{\theta_{OBS}-\theta_{MIN}})/{(\theta_{OBS})}}]$.}
\tablerefs{(1) \citet{N221}, (2) \citet{Cappellari-SAURON}, (3) \citet{Walterbos}, (4) \citet{Tremaine-M31}, (5) \citet{Bacon-M31},
(6) \citet{marcum}, (7) \citet{N1023}, (8) \citet{Sanchez}, (9) \citet{Takamiya}, (10) \citet{Greenhill},
(11) \citet{G3379}, (12) \citet{vdm90}, (13) \citet{Gebhardt-12}, (14) \citet{Busarello}, (15) \citet{F4261}, (16) \citet{N4342}, (17) \citet{Scorza}, (18) \citet{Sarzi},
(18b) \citet{Sarzi} modeled gas disk, (19) \citet{H4486}, (20) \citet{M4486}, (21) \citet{FF6251}, (22) \citet{CraneV},
(23) \citet{IC1459}, (24) \citet{VK}.}
\end{deluxetable}

\clearpage

\begin{deluxetable}{lccccccc}
\tabletypesize{\scriptsize}
\tablecaption{Maximum and Minimum Limits for Galaxy Parameters: Axis Ratio \label{Tab16}}
\tablewidth{0pt}
\tablehead{
\colhead{Galaxy}
& \colhead{q} & \colhead{Ref} & \colhead{$q_{MAX}$} & \colhead{Ref}
& \colhead{$q_{MIN}$} & \colhead{Ref} & \colhead{MAX/MIN q}
}
\startdata
NGC 221&0.73&1&0.87&2&0.63&3&19.2\% /13.7\%\\
NGC 1023&0.72&4&0.799&5&0.649&6&11.0\% /9.9\% \\
NGC 2778&0.78&7&0.889&6&0.72&8&14.0\% /7.7\% \\
NGC 3115&0.60&9&\nodata&\nodata&0.452&5&\nodata /24.7\% \\
NGC 3377&0.50&7&0.64&6&0.491&5&28.0\% /1.8\%\\
NGC 3379&0.90&10&1.0&3&0.85&2&11.1\% /5.6\% \\
NGC 3384&0.60&7&0.82&5&0.585&6&36.7\% /2.5\%\\
NGC 4261&0.75&11&0.89&12&0.74&13&18.7\% /1.3\% \\
NGC 4291&0.76&7&0.85&8&0.74&6&11.8\% /2.6\%\\
NGC 4473&0.62&7&\nodata&\nodata&0.531&6&\nodata /14.4\%\\
NGC 4564&0.70&7&\nodata&\nodata&0.42&12&\nodata /40.0\%\\
NGC 4486&0.915&14&1.0&3,13&0.65&15&9.3\% /29.0\% \\
NGC 6251&0.85&16&0.96&17&0.83&12&12.9\% /2.4\% \\
NGC 7052&0.55&2&0.89&18&0.44&18&61.8\% /20.0\% \\
NGC 7457&0.70&7&0.936&6&0.629&5&33.7\% /10.1\%\\
\enddata
\tablecomments{Description of columns: (1) galaxy name; (2) observed axis ratio (q), from (3); (4) maximum axis ratio, from (5);
(6) minimum axis ratio, from (7); and (8) percentage variation given as $[100({{q_{MAX}-q_{OBS}})/{(q_{OBS})}}]$/$[100({{q_{OBS}-q_{MIN}})/{(q_{OBS})}}]$.}
\tablerefs{(1) \citet{N221}, (2) \citet{Faber} FW value, (3) \citet{Faber} ESO value
(4) \citet{N1023}, (5) \citet{Lauer5} average ellipticity $r > r_b$, (6) \citet{Lauer5} average ellipticity $r \leq r_b$,
(7) private communication with K. Gebhardt, (8) \citet{Bacon}, (9) \citet{N3115old},
(10) central value of \citet{Faber,Bender88,Bacon,deV,BursteinD,Peletier,vdm90,vdm91},
(11) \citet{Ferrarese,F4261}, (12) \citet{rc3}, (13) \citet{Bing}, (14) central value of \citet{Faber,Peletier,Bacon},
(15) \citet{Caon1} SB profile, (16) \citet{Bender88}, (17) 2MASS (from NED), (18) SB profile of \citet{deJuan}}
\end{deluxetable}

\clearpage

\begin{deluxetable}{lcccccccccccc}
\rotate
\tabletypesize{\scriptsize}
\tablecaption{Maximum and Minimum Limits for Galaxy Parameters: Mass-to-Light Ratio \label{Tab17}}
\tablewidth{0pt}
\tablehead{
\colhead{Galaxy}
 & \colhead{$\Upsilon_{bpC}$} & \colhead{Ref} & \colhead{$\Upsilon_{max}$}
& \colhead{Ref} & \colhead{Color} & \colhead{Ref} & \colhead{$\Upsilon_{min}$}& \colhead{Ref} & \colhead{Color} & \colhead{Ref} & \colhead{Band} & \colhead{MAX/MIN $\Upsilon$}
}
\startdata
NGC 221&1.6&1&2.0&2&0.95(bv),1.133(vi)&3(bv),4(vi)&1.4&5&1.133(vi)&4&I&25.0\% /12.5\% \\
NGC 224&5.0&6&5.8&7&\nodata&\nodata&4.1&8&\nodata&\nodata&V&16\% /18.0\% \\
NGC 821&7.6&9&12.5&2&0.99(bv)&3&4.8&10&0.61(vr)&11&V&64.5\% /36.8\% \\
NGC 1023&5.0&12&5.8&10&\nodata&\nodata&\nodata&\nodata&\nodata&\nodata&V&16.0\% /\nodata \\
NGC 1068&1.2&13a&3.0&13&\nodata&\nodata&0.2&10&0.76(vr)&14&V&150\% /83.3\% \\
NGC 2778&8.0&9&\nodata&\nodata&\nodata&\nodata&6.4&6&\nodata&\nodata&V&\nodata /20.0\% \\
NGC 2787&3.2&15&15.4&16&\nodata&\nodata&\nodata&\nodata&\nodata&\nodata&B&381.3\% /\nodata \\
NGC 3115&6.9&6&7.0&10&\nodata&\nodata&2.7&17&\nodata&\nodata&V&1.4\% / 60.9\% \\
NGC 3377&2.9&9&9.4&2&0.86(bv)&3&2.5&10&\nodata&\nodata&V&224.1\% /13.8\% \\
NGC 3379&4.8&18&7.5&19&0.89(vr)&20&4.3&21&0.89(vr)&20&V&56.3\% /10.4\% \\
NGC 3384&2.5&9&2.8&6,10&\nodata&\nodata&\nodata&\nodata&\nodata&\nodata&V&12.0\% /\nodata \\
NGC 3608&3.7&9&6.9&2&0.94(bv)&3&\nodata&\nodata&\nodata&\nodata&V&86.5\% /\nodata \\
NGC 4261&4.8&22&14.3&2&0.99(bv)&3&\nodata&\nodata&\nodata&\nodata&V&197.9\% /\nodata \\
NGC 4291&5.5&9&12.7&2&0.96(bv)&3&4.4&6&\nodata&\nodata&V&130.9\% /20.0\% \\
NGC 4459&6.3&23&9.3&16&\nodata&\nodata&3.6&24&\nodata&\nodata&B&47.6\% /42.9\% \\
NGC 4473&6.0&9&7.0&2&0.96(bv)&3&5.2&10&\nodata&\nodata&V&16.7\% / 13.3\% \\
NGC 4486&4.0&6&11.0&23&0.96(bv)&3&\nodata&\nodata&\nodata&\nodata&V&175.0\% /\nodata \\
NGC 4564&2.0&9&10.8&2&0.93(bv),1.161(vi)&3(bv),4(vi)&1.9&6&\nodata&\nodata&I&440.0\% /5.0\% \\
NGC 4596&1.2&25&8.9&16&\nodata&\nodata&\nodata&\nodata&\nodata&\nodata&B&641.7\% /\nodata \\
NGC 4697&4.7&9,10&7.6&2&0.91(bv)&3&\nodata&\nodata&\nodata&\nodata&V&61.7\% /\nodata \\
NGC 4742&2.7&2&\nodata&\nodata&\nodata&\nodata&1.1&10&1.28(br)&26&B&\nodata /59.3\% \\
NGC 5845&5.5&9,10&6.5&2&1.02(bv)&3&4.8&6&\nodata&\nodata&V&18.2\% /12.7\% \\
NGC 6251&9.7&27&\nodata&\nodata&\nodata&\nodata&7.3&10&0.61(vr)&11&V&\nodata /24.7\% \\
NGC 7052&6.3&6,28&\nodata&\nodata&\nodata&\nodata&2.5&10&0.70(ri)&11&I&\nodata /60.3\% \\
\enddata
\tablecomments{Description of columns: (1) galaxy name;
(2) $\Upsilon_{bpC}$ in $M_{\odot} / L_{\odot}$, from (3) reference; (4) $\Upsilon_{MAX}$ in $M_{\odot} / L_{\odot}$, 
from (5) reference, with (6) color correction, from (7) reference; (8) $\Upsilon_{MIN}$ in $M_{\odot} / L_{\odot}$, from (9) reference, 
with (10) color correction,
from (11) reference; all in (12) wavelength-band; with (13) percentage variation
$[100({{\Upsilon_{MAX}-\Upsilon_{OBS}})/{(\Upsilon_{OBS})}}]$/$[100({{\Upsilon_{OBS}-\Upsilon_{MIN}})/{(\Upsilon_{OBS})}}]$.
\citet{Cappellari-SAURON} $\Upsilon$-values ($\Upsilon_{Cap}$) are excluded from consideration here, although, when 
available, they are superior to those values given in column 2 (see Sections~\ref{HostGal} and~\ref{sigma}).
For those galaxies which are excluded, the previously discussed value (Section~\ref{HostGal}) was accepted as the correct
value. All values have been rescaled to a common distance and wavelength-band. The color corrections are coded such that
color X-Y will be given as `xy'.
In the cases where it was necessary to convert $\Upsilon$ to the default waveband, the 
solar colors were taken from \citet{Allen} such that $bv_{\odot}=0.650$, $ub_{\odot}=0.195$,
$uv_{\odot}=0.845$, $vr_{\odot}=0.54$, $vi_{\odot}=0.88$, $vk_{\odot}=1.49$, and the color terms for the
individual galaxies were taken from the literature. }
\tablerefs{(1) \citet{N221}, (2) \citet{Bacon}, (3) \citet{rc3}, (4) \citet{Tonry}, (5) \citet{Lauer-old32}, (6) \citet{Tremaine},
(7)  \citet{Kormendy-M31}, (8) \citet{Tremaine-M31}, (9) \citet{Gebhardt-12}, (10) \citet{Haring}, (11) \citet{Fukugita} for avg. elliptical,
(12) \citet{N1023}, (13) \citet{Takamiya}, (13a)  \citet{Takamiya} Figure 4, (14) \citet{Schild} SB profile,
(15) 
calculated using (B) $L_{bulge}=3.25x10^9 L_{\odot}$ \citep{Sarzi} and \citet{BinneyT} eq. 4-80b
($M \sim {{<v^2>r_h}/{(0.4G)}}$) with $\sigma_e=184.9 {km s^{-1}}$ \citep{Sarzi} to obtain
$\Upsilon_B=3.38$. A second calculation wherein $M={{v_{max}^2R_e/G}}$
($v_{max}=v \sin{\theta}$) and using $\theta=50\degr$ \citep{Sarzi} and $v_{max} = 150 {km s^{-1}}$
from Figure 4 of \citet{Sarzi}, resulted in $\Upsilon_B=3.03$. These were averaged to 3.21.,
(16) using $M_{bulge}$ and $L_{bulge}$ from \citet{Sarzi}, (17) \citet{N3115,N3115old}, (18) \citet{G3379}, (19) \citet{vdm90},
(20) \citet{Kormendy}, (21) \citet{vdm91}, (22) \citet{F4261}, (23) \citet{FaberJ}, (24) 
calculated, as in (15), using $L_{bulge}=7.16x10^9 L_{\odot}$, $\sigma_e=166.6 km s^{-1}$,
$v_{max}=150 {km s^{-1}}$, and $\theta=41\degr$ \citep{Sarzi}, and obtaining
$\Upsilon_B=3.04$ in the first calculation and $\Upsilon_B=4.1$ for the second, with an average
value of 3.57.,
(25) 
calculated, as in (15), using $L_{bulge}=9.61x10^9 L_{\odot}$, $\sigma_e=136.3 km s^{-1}$, 
$v_{max}=80 {km s^{-1}}$, and $\theta=42\degr$ \citep{Sarzi}, and obtaining
$\Upsilon_B=1.47$ in the first calculation and $\Upsilon_B=0.91$ for the second, with an average
value of 1.19., (26) \citet{Poulain},
(27) \citet{FF6251}, (28) \citet{v7052}.}
\end{deluxetable}

\clearpage

\begin{deluxetable}{lccccccccccc}
\rotate
\tabletypesize{\scriptsize}
\tablecaption{Fitting Parameters: $\log{M_{\bullet}} = a_1[\log{X \over X_0}]^2
+a_2\log{X \over X_0}+d$ \label{Tab18}}
\tablewidth{0pt}
\tablehead{
\colhead{Sample} & \colhead{$a_1$} & \colhead{$a_2$} & \colhead{d} &
\colhead{$\epsilon_{yin}$} & \colhead{$\sigma_{fy}$} &
\colhead{$a_{max}$} & \colhead{$a_{min}$} &
\colhead{$F_{\sigma_y}$} & \colhead{$sig_{\sigma}$} & \colhead{$F_{X}$} & \colhead{$sig_X$}
}
\startdata
\tableline
X=$\sigma$&&&&&&&&&&\\
\tableline
15E&0.33 $\pm$ 1.62&3.89 $\pm$ 0.43&8.17 $\pm$ 0.11&0.25&0.29&0.86 $\pm$ 1.77(RI)&-2.16 $\pm$ 1.27(RO)&0.92&NSS&0.92&NSS\\
23gal&0.97 $\pm$ 1.22&3.90 $\pm$ 0.37&8.13 $\pm$ 0.08&0.25&0.30&1.00 $\pm$ 1.26(AI)&0.78 $\pm$ 1.24(RB)&0.99&NSS&0.99&NSS\\
8B&1.68 $\pm$ 5.39&4.04 $\pm$ 2.51&8.11 $\pm$ 0.17&0.33&0.40&3.89 $\pm$ 5.26(AI)&1.33 $\pm$ 5.24(RB)&0.92&NSS&0.92&NSS\\
Cap8&1.07 $\pm$ 2.75&4.12 $\pm$ 0.89&8.12 $\pm$ 0.17&0.35&0.39&1.13 $\pm$ 2.77(AI)&0.43 $\pm$ 2.64(RO)&0.86&NSS&0.86&NSS\\
\tableline
X=$E_g$&&&&&&&&&&\\
\tableline
15E&0.04 $\pm$ 0.04&0.65 $\pm$ 0.07&8.05 $\pm$ 0.10&0.21&0.29&0.06 $\pm$ 0.03(AO)&0.03 $\pm$ 0.04(AI)&0.96&NSS&0.93&NSS\\
23gal&0.05 $\pm$ 0.04&0.61 $\pm$ 0.06&8.06 $\pm$ 0.09&0.27&0.32&0.05 $\pm$ 0.04(RI)&0.04 $\pm$ 0.04(RB)&0.86&NSS&1.0&NSS\\
8B&0.08 $\pm$ 0.16&0.57 $\pm$ 0.24&8.00 $\pm$ 0.22&0.42&0.45&0.09 $\pm$ 0.15(AI)&0.07 $\pm$ 0.16(RB)&0.71&NSS&0.91&NSS\\
Cap8&0.04 $\pm$ 0.05&0.69 $\pm$ 0.12&8.10 $\pm$ 0.14&0.28&0.34&0.04 $\pm$ 0.05(AO)&0.04 $\pm$ 0.05(RO)&1.13&NSS&0.89&NSS\\
\tableline
X=$\Phi({R_e/8})$&&&&&&&&&&\\
\tableline
15E&-0.21 $\pm$ 0.40&2.06 $\pm$ 0.18&8.22 $\pm$ 0.09&0.19&0.26&0.01 $\pm$ 0.50(RI)&-0.21 $\pm$ 0.40(AB)&1.16&NSS&0.92&NSS\\
23gal&0.43 $\pm$ 0.48&1.71 $\pm$ 0.24&8.10 $\pm$ 0.12&0.36&0.39&0.43 $\pm$ 0.48(AB)&0.39 $\pm$ 0.49(RB)&0.58&75SS&1.0&NSS\\
8B&0.25 $\pm$ 1.06&1.09 $\pm$ 0.59&7.91 $\pm$ 0.27&0.48&0.51&0.40 $\pm$ 0.99(AI)&0.19 $\pm$ 1.07(RB)&0.55&NSS&0.89&NSS\\
Cap8&-0.10 $\pm$ 0.80&2.18 $\pm$ 0.39&8.27 $\pm$ 0.16&0.29&0.35&0.03 $\pm$ 0.98(RI)&-0.48 $\pm$ 0.79(RO)&1.09&NSS&0.84&NSS\\
\tableline
X=$M({10R_e})$&&&&&&&&&&\\
\tableline
15E&0.08 $\pm$ 0.09&0.88 $\pm$ 0.11&8.00 $\pm$ 0.13&0.28&0.35&0.15 $\pm$ 0.07(AO)&0.07 $\pm$ 0.09(AI)&0.66&75SS&0.94&NSS\\
23gal&0.08 $\pm$ 0.08&0.87 $\pm$ 0.09&8.02 $\pm$ 0.09&0.29&0.34&0.08 $\pm$ 0.08(RI)&0.07 $\pm$ 0.08(AI)&0.76&NSS&1.0&NSS\\
8B&0.18 $\pm$ 0.44&1.00 $\pm$ 0.38&8.05 $\pm$ 0.20&0.38&0.42&0.19 $\pm$ 0.41(AI)&0.15 $\pm$ 0.45(RB)&0.80&NSS&0.91&NSS\\
Cap8&0.08 $\pm$ 0.13&0.89 $\pm$ 0.21&8.03 $\pm$ 0.18&0.37&0.42&0.09 $\pm$ 0.11(AO)&0.07 $\pm$ 0.13(AI)&0.73&NSS&0.91&NSS\\
\enddata
\tablecomments{The log-quadratic fits with the columns as follows: (1) galaxy sample, (2) $a_1$, (3) $a_2$,  (4) d,
(5) $\epsilon_{yin}$, (6) $\sigma_{fy}$,
(7) maximum $a_1$ (\textit{weightcent}) and (8) minimum $a_1$ (\textit{weightcent}),
(9) F-test ratio and (10) significance from comparison with the log-linear velocity dispersion predictor and
(11) F-test ratio and (12) significance from comparison with the log-linear fit for X. 
All fits determined using the \textit{lsqfit} algorithm with default \textit{weightcent}.}
\end{deluxetable}

\clearpage

\begin{deluxetable}{lccccccccc}
\tabletypesize{\scriptsize}
\tablecaption{Fitting Parameters: $\log{M_{\bullet}} = a_1\log{E_g \over {E_g}_0}+a_2\log{R_e \over {R_e}_0}+d$ \label{Tab19}}
\tablewidth{0pt}
\tablehead{
\colhead{Sample} & \colhead{$a_1$}&\colhead{$a_2$}  & \colhead{d} &
\colhead{$\epsilon_{yin}$} & \colhead{$\sigma_{fy}$} &
\colhead{$F_{\sigma_y}$} & \colhead{$sig_{\sigma}$} & \colhead{$F_{1}$} & \colhead{$sig_1$}
}
\startdata
15E&0.88$\pm$0.12&-0.62$\pm$0.26&8.16$\pm$0.06&0.16&0.24&1.36&NSS&1.36&NSS\\
23gal&0.57$\pm$0.10&0.02$\pm$0.24&8.13$\pm$0.07&0.28&0.33&0.80&NSS&0.94&NSS\\
8B&0.41$\pm$0.17&0.59$\pm$0.76&8.18$\pm$0.24&0.40&0.44&0.74&NSS&0.96&NSS\\
Cap8&0.92$\pm$0.17&-0.62$\pm$0.33&8.24$\pm$0.11&0.20&0.27&1.73&NSS&1.40&NSS\\
\enddata
\tablecomments{The multivariate ($E_g,R_e$) predictive fits with the columns as follows: (1) galaxy sample, (2) $a_1$, (3) $a_2$,  (4) d,
(5) $\epsilon_{yin}$, (6) $\sigma_{fy}$,
(7) F-test ratio and (8) significance from comparison with the log-linear velocity dispersion predictor and
(9) F-test ratio and (10) significance from comparison with the log-linear fit (for $E_g$).
All fits are determined using the \textit{lsqfit} algorithm with default \textit{weightcent}.}
\end{deluxetable}

\clearpage
\begin{deluxetable}{llllccccccc}
\tabletypesize{\scriptsize}
\tablecaption{Fitting Parameters: $\log{M_{\bullet}} = a_1\log{X \over X_0}+a_2\log{nR_e \over {nR_e}_0}+d$ \label{Tab20}}
\tablewidth{0pt}
\tablehead{
\colhead{X} &
\colhead{Sample} & \colhead{$a_1$}&\colhead{$a_2$}  & \colhead{d} &
\colhead{$\epsilon_{yin}$} &
\colhead{$\sigma_{fy}$} &
\colhead{$F_{\sigma_y}$} & \colhead{$sig_{\sigma}$} & \colhead{$F_{1}$} & \colhead{$sig_1$}
}
\startdata
$\Phi({R_e})$&15E&1.55$\pm$0.23&0.38$\pm$0.15&8.19$\pm$0.07&0.18&0.26&1.18&NSS&1.24&NSS\\
$\Phi({R_e})$&Cap8&1.62$\pm$0.34&0.42$\pm$0.19&8.25$\pm$0.12&0.23&0.30&1.45&NSS&1.52&NSS\\
\tableline
\tableline
$\Phi({R_e/8})$&15E&1.81$\pm$0.26&0.22$\pm$0.16&8.15$\pm$0.07&0.17&0.25&1.33&NSS&1.0&NSS\\
$\Phi({R_e/8})$&23gal&1.15$\pm$0.22&0.55$\pm$0.17&8.14$\pm$0.07&0.29&0.34&0.79&NSS&1.32&NSS\\
$\Phi({R_e/8})$&8B&0.82$\pm$0.35&0.93$\pm$0.72&8.18$\pm$0.24&0.41&0.44&0.73&NSS&1.19&NSS\\
$\Phi({R_e/8})$&Cap8&1.90$\pm$0.41&0.25$\pm$0.22&8.22$\pm$0.12&0.24&0.31&1.34&NSS&1.07&NSS\\
\enddata
\tablecomments{The multivariate ($\Phi(nR_e),nR_e$) fit with the (1) gravitational potential at radius R and columns (2) - (11) as in Table~\ref{Tab19} columns (1) - (10).}
\end{deluxetable}

 \clearpage

\begin{deluxetable}{lllccccccc}
\tabletypesize{\scriptsize}
\tablecaption{Fitting Parameters: $\log{M_{\bullet}} = a_1\log{{M({10R_e})} \over {M({10R_e})}_0}+a_2\log{10R_e \over {10R_e}_0}+d$ \label{Tab21}}
\tablewidth{0pt}
\tablehead{
\colhead{Sample} & \colhead{$a_1$}&\colhead{$a_2$}  & \colhead{d} &
\colhead{$\epsilon_{yin}$} &
\colhead{$\sigma_{fy}$} &
\colhead{$F_{\sigma_y}$} & \colhead{$sig_{\sigma}$} & \colhead{$F_{1}$} & \colhead{$sig_1$}
}
\startdata
15E&1.54$\pm$0.23&-1.17$\pm$0.35&8.11$\pm$0.07&0.18&0.26&1.19&NSS&1.71&75SS\\
23gal&1.09$\pm$0.19&-0.45$\pm$0.30&8.08$\pm$0.07&0.28&0.33&0.82&NSS&1.06&NSS\\
8B&0.82$\pm$0.33&0.26$\pm$0.80&8.14$\pm$0.23&0.39&0.43&0.78&NSS&0.87&NSS\\
Cap8&1.61$\pm$0.29&-1.20$\pm$0.42&8.17$\pm$0.10&0.20&0.27&1.83&NSS&2.19&75SS\\
\enddata
\tablecomments{The multivariate ($M(10R_e),10R_e$) fit with columns (1) - (10) as in Table~\ref{Tab19}.}
\end{deluxetable}
\clearpage

\clearpage

\clearpage

\clearpage

\clearpage

\clearpage

\clearpage

\clearpage

\clearpage

\clearpage

\clearpage

\clearpage

\clearpage

\clearpage

\end{document}